\documentclass[]{aastex631}
\usepackage{graphicx}
\usepackage{longtable}\LTcapwidth=\textwidth
\usepackage{multirow}
\usepackage{subfigure}
\usepackage{ulem}

\usepackage{CJK}
\pdfoutput=1

\graphicspath{{./}{figures/}}

\begin{document}
\begin{CJK*}{UTF8}{gbsn}

\title{ALMA Survey of Orion Planck Galactic Cold Clumps (ALMASOP): How do dense core properties affect the multiplicity of protostars?}


\author[0000-0003-4506-3171]{Qiu-yi Luo(罗秋怡)}\thanks{E-mail:lqy@shao.ac.cn}
\affiliation{Shanghai Astronomical Observatory, Chinese Academy of Sciences, 80 Nandan Road, Shanghai 200030, People’s Republic of China}
\affiliation{School of Astronomy and Space Sciences, University of Chinese Academy of Sciences, No. 19A Yuquan Road, Beijing 100049, People’s Republic of China}

\author[0000-0002-5286-2564]{Tie Liu(刘铁)}\thanks{E-mail:liutie@shao.ac.cn}
\affiliation{Shanghai Astronomical Observatory, Chinese Academy of Sciences, 80 Nandan Road, Shanghai 200030, People’s Republic of China}

\author[0000-0002-8149-8546]{Ken'ichi Tatematsu}
\affil{Nobeyama Radio Observatory, National Astronomical Observatory of Japan,
National Institutes of Natural Sciences,
Nobeyama, Minamimaki, Minamisaku, Nagano 384-1305, Japan}
\affiliation{Department of Astronomical Science,
The Graduate University for Advanced Studies, SOKENDAI,
2-21-1 Osawa, Mitaka, Tokyo 181-8588, Japan}

\author{Sheng-Yuan Liu}
\affiliation{Institute of Astronomy and Astrophysics, Academia Sinica, Roosevelt Road, Taipei 10617, Taiwan (R.O.C)}

\author{Pak Shing Li}
\affiliation{Astronomy Department, University of California, Berkeley, CA 94720-3411, USA}

\author{James di Francesco }
\affiliation{NRC Herzberg Astronomy and Astrophysics,
5071 West Saanich Road,
Victoria, BC, V9E 2E7, Canada}
\affiliation{Department of Physics and Astronomy, University of Victoria,
3800 Finnerty Road, Elliot Building,
Victoria, BC, V8P 5C2, Canada}

\author[0000-0002-6773-459X]{Doug Johnstone}
\affiliation{NRC Herzberg Astronomy and Astrophysics,
5071 West Saanich Road,
Victoria, BC, V9E 2E7, Canada}
\affiliation{Department of Physics and Astronomy, University of Victoria,
3800 Finnerty Road, Elliot Building,
Victoria, BC, V8P 5C2, Canada}

\author{Paul F. Goldsmith}
\affiliation{Jet Propulsion Laboratory, California Institute of Technology, 4800 Oak Grove Drive, Pasadena CA 91109, USA}

\author[0000-0002-2338-4583]{Somnath Dutta}
\affiliation{Institute of Astronomy and Astrophysics, Academia Sinica, Roosevelt Rd, Taipei 10617, Taiwan (R.O.C)}

\author{Naomi Hirano}
\affiliation{Institute of Astronomy and Astrophysics, Academia Sinica, Roosevelt Rd, Taipei 10617, Taiwan (R.O.C)}

\author{Chin-Fei Lee}
\affiliation{Institute of Astronomy and Astrophysics, Academia Sinica, Roosevelt Rd, Taipei 10617, Taiwan (R.O.C)}

\author{Di Li}
\affiliation{National Astronomical Observatories of China, Chinese Academy of Sciences, Beijing, 100012, China}

\author{Kee-Tae Kim}
\affiliation{Korea Astronomy and Space Science Institute, 776 Daedeokdae-ro, Yuseong-gu, Daejeon 34055, Republic of Korea}

\author{Chang Won Lee}
\affiliation{Korea Astronomy and Space Science Institute, 776 Daedeokdae-ro, Yuseong-gu, Daejeon 34055, Republic of Korea}

\author{Jeong-Eun Lee}
\affiliation{The School of Space Research, Kyung Hee University, 1732 Deogyeong-daero, Giheung-gu, Yongin-si, Gyeonggi-do, Republic of Korea}

\author{Xun-chuan Liu}
\affiliation{Shanghai Astronomical Observatory, Chinese Academy of Sciences, 80 Nandan Road, Shanghai 200030, People’s Republic of China}

\author[0000-0002-5809-4834]{Mika Juvela}
\affiliation{Department of Physics, P.O.Box 64, FI-00014,
University of Helsinki, Finland}

\author{Jinhua He}
\affiliation{Yunnan Observatories, Chinese Academy of Sciences, 396 Yangfangwang, Guandu District, Kunming, 650216, P. R. China
}
\affiliation{Chinese Academy of Sciences South America Center for Astronomy, National Astronomical Observatories, CAS, Beijing 100101, China
}
\affiliation{Departamento de Astronomía, Universidad de Chile, Las Condes, 7591245 Santiago, Chile}

\author{Sheng-Li Qin}
\affiliation{Department of Astronomy, Yunnan University, Kunming 650091,People’s Republic of China}

\author{Hong-Li Liu}
\affiliation{Department of Astronomy, Yunnan University, Kunming 650091,People’s Republic of China}

\author{David Eden}  
\affiliation{Astrophysics Research Institute, Liverpool John Moores University, IC2, Liverpool Science Park, 146 Brownlow Hill, Liverpool L3 5RF, UK}
\affiliation{Armagh Observatory and Planetarium, College Hill, Armagh, BT61 9DB, UK}

\author{Woojin Kwon}  
\affiliation{Department of Earth Science Education, Seoul National University, 1 Gwanak-ro, Gwanak-gu, Seoul 08826, Republic of Korea}
\affiliation{SNU Astronomy Research Center, Seoul National University, 1 Gwanak-ro, Gwanak-gu, Seoul 08826, Republic of Korea}

\author[0000-0002-4393-3463]{Dipen Sahu}
\affiliation{Academia Sinica Institute of Astronomy and Astrophysics, 11F of AS/NTU Astronomy-Mathematics Building, No.1, Sec. 4, Roosevelt Rd, Taipei 10617, Taiwan, R.O.C.}

\author[0000-0003-1275-5251]{Shanghuo Li}
\affiliation{Korea Astronomy and Space Science Institute, 776 Daedeokdae-ro, Yuseong-gu, Daejeon 34055, Republic of Korea}

\author[0000-0001-5950-1932]{Feng-Wei Xu}
\affiliation{Department of Astronomy, School of Physics, Peking University, Beijing 100871, Peopleʼs Republic of China}
\affiliation{Kavli Institute for Astronomy and Astrophysics, Peking University, Haidian District, Beijing 100871, Peopleʼs Republic of China}

\author{Si-ju Zhang}
\affiliation{Kavli Institute for Astronomy and Astrophysics, Peking University, Haidian District, Beijing 100871, Peopleʼs Republic of China}

\author[0000-0002-1369-1563]{Shih-Ying Hsu}
\affiliation{National Taiwan University (NTU), No. 1, Section 4, Roosevelt Rd, Taipei 10617, Taiwan (R.O.C.);seansyhsu@gmail.com}
\affiliation{Institute of Astronomy and Astrophysics, Academia Sinica, Roosevelt Rd, Taipei 10617, Taiwan (R.O.C)}

\author{Leonardo Bronfman}
\affiliation{Departamento de Astronomía, Universidad de Chile, Las Condes, 7591245 Santiago, Chile}

\author[0000-0002-7125-7685]{Patricio Sanhueza} %
\affiliation{National Astronomical Observatory of Japan, National Institutes of Natural Sciences, 2-21-1 Osawa, Mitaka, Tokyo 181-8588, Japan}
\affil{Department of Astronomical Science, SOKENDAI (The Graduate University for Advanced Studies), 2-21-1 Osawa, Mitaka, Tokyo 181-8588, Japan}

\author[0000-0002-8898-1047]{Veli-Matti Pelkonen}
\affiliation{Institut de Ci\`{e}ncies del Cosmos, Universitat de Barcelona, IEEC-UB, Mart\'{i} i Franqu\`{e}s 1, E08028 Barcelona, Spain}

\author{Jian-wen Zhou}
\affiliation{National Astronomical Observatories of China, Chinese Academy of Sciences, Beijing, 100012, China}
\affiliation{School of Astronomy and Space Sciences, University of Chinese Academy of Sciences, No. 19A Yuquan Road, Beijing 100049, People’s Republic of China}

\author{Rong Liu}
\affiliation{National Astronomical Observatories of China, Chinese Academy of Sciences, Beijing, 100012, China}
\affiliation{School of Astronomy and Space Sciences, University of Chinese Academy of Sciences, No. 19A Yuquan Road, Beijing 100049, People’s Republic of China}

\author{Qi-lao Gu}
\affiliation{Shanghai Astronomical Observatory, Chinese Academy of Sciences, 80 Nandan Road, Shanghai 200030, People’s Republic of China}

\author{Yue-fang Wu}
\affiliation{Department of Astronomy, School of Physics, Peking University, Beijing 100871, Peopleʼs Republic of China}
\affiliation{Kavli Institute for Astronomy and Astrophysics, Peking University, Haidian District, Beijing 100871, Peopleʼs Republic of China}

\author{Xiao-feng Mai(麦晓枫)}
\affiliation{Shanghai Astronomical Observatory, Chinese Academy of Sciences, 80 Nandan Road, Shanghai 200030, People’s Republic of China}
\affiliation{School of Astronomy and Space Sciences, University of Chinese Academy of Sciences, No. 19A Yuquan Road, Beijing 100049, People’s Republic of China}

\author{Edith Falgarone}  
\affiliation{LERMA, Observatoire de Paris, PSL Research University, CNRS, Sorbonne Universit\'{e}s, UPMC Univ. Paris 06, Ecole normale sup\'{e}rieure, F-75005 Paris, France}

\author{Zhi-Qiang Shen}
\affiliation{Shanghai Astronomical Observatory, Chinese Academy of Sciences, 80 Nandan Road, Shanghai 200030, People’s Republic of China}

\begin{abstract}
During the transition phase from a prestellar to a protostellar cloud core, one or several protostars can form within a single gas core. The detailed physical processes of this transition, however, still remain unclear. We present 1.3\,mm dust continuum and molecular line observations with the Atacama Large Millimeter/submillimeter Array (ALMA) toward 43 protostellar cores in the Orion Molecular Cloud Complex ($\lambda$ Orionis, Orion B, and Orion A) with an angular resolution of $\sim$ 0$\farcs$35 ($\sim$ 140 au). In total, we detect 13 binary/multiple systems. We derive an overall multiplicity frequency (MF) of 28$\%$ $\pm$ 4$\%$ and a companion star fraction (CSF) of 51$\%$ $\pm$ 6$\%$, over a separation range of 300-8900 au. The median separation of companions is about 2100 au. The occurrence of stellar multiplicity may depend on the physical characteristics of the dense cores. Notably, those containing binary/multiple systems tend to show higher gas density and Mach number than cores forming single stars. The integral-shaped filament (ISF) of Orion A giant molecular cloud (GMC), which has the highest gas density and hosts high-mass star formation in its central region (the Orion Nebula cluster), shows the highest MF and CSF among the Orion
GMCs. In contrast, the $\lambda$ Orionis Giant Molecular Cloud (GMC) has a lower MF and CSF than the Orion B and Orion A GMCs, indicating that feedback from H{\sc ii} regions may suppress the formation of multiple systems. We also find that the protostars comprising a binary/multiple system are usually at different evolutionary stages.

\end{abstract}


\section{Introduction} \label{sec:intro}

Molecular clouds exhibit hierarchical structures at different levels from large to small scales. Dense cores lie at the terminus where stars are born through gravitational fragmentation \citep{heggie_binary_1975,1979Cohen,1987ARA&A..25...23S,2007ARA&ABergin,kraus_multiple_2012,duchene_stellar_2013,2014Reipurth}. 
Isolated star formation has been investigated for decades in the context of star formation
\citep{di_francesco_observational_2006}. The formation of binary/multiple star systems, however, has been not as well studied, though these systems are commonly seen in star associations and clusters \citep{kraus_multiple_2012,lomax_simulations_2015,2020lee,dutta_alma_2020-1,2014Reipurth}.

Binary/multiple systems in the main sequence have been studied for decades. Indeed, star systems that have more than one star are as common as single star systems in the Milky Way \citep{1978SS,1989MS,1991AAD,2009Connelley,raghavan_survey_2010-1}.  Previous near-infrared observational studies have concentrated on the multiplicity statistics for young stellar objects (YSOs) \citep{haisch_jr_near-infrared_2004,Duchene2004,Duchene2007,2008Connelley,kraus_role_2007,Kraus2011,2015Daemgen,kounkel_hst_2016,Ma2019}. The high occurrence rates of multiple YSO systems revealed in these studies imply that stellar multiplicity is determined in the star formation process. In these surveys, the fraction of stars having companions (CSF) ranges from 16$\%$ $\pm$ 4$\%$ to 62$\%$ $\pm$ 14$\%$, and the fraction of systems having multiple YSOs (MF) ranges from 18$\%$ $\pm$ 4$\%$ to $\sim$75$\%$. In the past 10 years, high resolution interferometric observations with the Submillimeter Array(SMA), Very Large array (VLA), and Atacama Large Millimeter/submillimeter Array (ALMA) of nearby clouds have systematically revealed the multiplicity of low-mass protostars in still earlier phases, notably the Class 0/I phase \citep{2013Chen,2015leemass,hatziminaoglou_multiplicity_2018}. 
These observations indicate that the CSF ranges from $\sim$71$\%$ to 91$\%$ $\pm$ 5$\%$ while MF ranges from the 40$\%$ to 60$\%$ for young protostars with separations below 10,000 au. Recently, higher angular resolution ALMA and VLA observations by \cite{Tobin2022} revealed the multiplicity of Orion protostars that were previously identified in the Herschel Orion Protostar Survey (HOPS): the CSF is 44$\%$ $\pm$ 3$\%$ and the MF is 30$\%$ $\pm$ 3$\%$.

Although different theoretical models of binary/multiple systems at various evolutionary stages have been proposed, e.g, turbulent fragmentation \citep{goodwin_simulating_2004,fisher_turbulent_2004} and disk fragmentation \citep{1989Adams,1994Bonnell}, the most relevant scenario remains unclear. Nevertheless, some predictions can be tested. For example, turbulent fragmentation suggests a non-linear gravitational collapse, resulting in wider-separation binary/multiple systems (above 1000 au). On the other hand, disk fragmentation caused by gravitational instability is more likely to generate closer (below 600 au) binary/multiple systems \citep{1986Beichman,raghavan_survey_2010-1,reipurth_visual_2007-1,1979Cohen,2014Reipurth}.

In a multiplicity study of all known protostars (94) in the Perseus molecular cloud, \cite{tobin_vla_2016} found a bimodal distribution of separation among binary/multiple systems, with peaks at $\sim$75 au and $\sim$3000 au. The bimodal distribution was recently confirmed by a more detailed analysis of protostar separations in the Orion and Perseus molecular clouds \citep{Tobin2022}.  Based on these results, \cite{Tobin2022} suggest that multiples with small separations ($<$500 au) are likely produced by both disk fragmentation and turbulent fragmentation with migration, and those with wide separation ($>10^3$ au) result primarily from turbulent fragmentation. \cite{2017NatLee} found a low-mass binary system in the earliest stages of formation with the rotation axes of its disks misaligned, suggesting that this binary system is likely formed due to turbulent fragmentation. This system, however, has small separation ($\sim$860 au), which has probably been decreased by the migration of the protostars \citep{2020lee}. Further, \citet{2016TobinN}
detected a close triple system that was likely formed out of a protostellar disk undergoing gravitational instability. 

However, previous investigations have rarely addressed how the properties of the host dense cores affect stellar multiplicity, even though they may play an important role. In this work, we explore the relation between the physical characteristics of dense cores and the formation of single or multiple stellar systems. This paper is structured as follows: Section \ref{sec:floats} introduces our sample, and Section \ref{sec:Obs} describes the observational data from the surveys we use in the paper. In Section \ref{sec:results}, we report the results of multiplicity analysis in binary/multiple systems. Section \ref{sec:discussion} discusses the origin of multiplicity, environmental effects, and the physical and chemical differences among protostars. Section \ref{sec:summary} provides a summary.

\section{The Sample} \label{sec:floats}

The Planck survey detected 13,188 Planck Galactic cold clumps (PGCCs) across the whole sky that exhibit extremely low temperatures (T $\sim$ 14 K) \citep{planck_catalog2016}. The PGCCs are therefore excellent targets for studying the very initial conditions of star formation \citep{Planck2011,Juvela2010,Wu2012,Montillaud2015,planck_catalog2016,liu_top-scope_2018,Eden2019,Xu2020,2021xfw}.

In follow-up observations, the James Clerk Maxwell Telescope (JCMT) large program SCOPE observed $\sim$1,300 PGCCs in the 850 $\mu$m continuum to study the early evolution of dense cores, which targets high-column-density ( $>$5$\times$10$^{20}$cm$^{-2}$ for a 5$\arcmin$ beam of the Planck telescope) clumps  \citep[][]{liu_top-scope_2018,Eden2019}. The SCOPE sample is important for statistically investigating evolution between the starless and protostellar star formation phases. 

The Orion Molecular Cloud Complex is about 380-420 pc away from us, and it has been extensively studied with many observations \citep{kim_molecular_2020,SahuLiu2021}. As a part of SCOPE, \cite{yi_planck_2018-1} turned to the Orion giant molecular cloud (GMC) complex, and observed 58 PGCCs in the three GMCs (Orion A, Orion B and $\lambda$ Orionis) with the JCMT in the 850 $\mu$m continuum. Beyond those they included 38 other PGCCs, for which archival 850 $\mu$m continuum data were available in the JCMT Science Archive. The sample of \cite{yi_planck_2018-1} is complete for PGCCs with column densities higher than 5$\times$10$^{20}$cm$^{-2}$ (for a 5$\arcmin$ beam of the Planck telescope) in the Orion GMCs. In total, 119 dense cores were identified from JCMT SCUBA-2 observations of these Orion PGCCs, forming a unique sample of cold cores at similar distances for further studies. 

As a follow-up observation of these Orion PGCCs, the ALMA Survey of Orion Planck Galactic Cold Clumps (ALMASOP) subsequently observed 72 out of the 119 dense Orion PGCCs at high resolution ($\sim$140 au). This resolution is high enough to resolve close binary/multiple systems with separations of a few hundred astronomical units. These 72 cores are among the densest in these clouds and are arguably the closest to the onset of star formation \citep[][]{dutta_alma_2020-1}, i.e., those that have started to collapse or have already formed protostars. In previous work with ALMASOP survey data, we focused on the chemical evolution of hot corino sources \citep{hsu_alma_2020,Hsu2022}, the fragmentation of prestellar cores \citep{sahu_alma_2021}, and outflow jets \citep{dutta_alma_2020-1}. 

\cite{dutta_alma_2020-1} presented the 1.3\,mm continuum emission maps in ALMASOP observations and identified protostars in 43 cores. Multiple protostars were frequently seen within a single core, but the multiplicity of these protostellar systems was not discussed in \cite{dutta_alma_2020-1}. The present paper investigates binary/multiple systems in ALMASOP, and focuses on how dense core properties relate to the multiplicity of protostars. The major limitation of this work is that very close binary systems with separation $\le$ 140 au are not resolvable due to the angular resolution limits of the ALMASOP observation. 

\section{Observations and data}\label{sec:Obs}
\subsection{ALMA Observations}

ALMASOP (project ID: 2018.1.00302.S.; PI: Tie Liu), observed 72 extremely cold young dense cores in the Orion molecular clouds, including 23 starless core candidates and 49 protostellar core candidates, with ALMA Band\,6 during 2018 October to 2019 January. The observations were performed in four spectral windows centered at 216.6, 218.9, 231.0 and 233.0 GHz each with a 1.875 GHz bandwidth, with a velocity resolution of $\sim$ 1.4 km/s. Three array configurations were used for the observations: 12m C43-5 (TM1), 12m C43-2 (TM2), and 7m Atacama Compact Array (ACA). Additional details of the observations are presented in \cite{dutta_alma_2020-1}. Several molecular line transitions were observed: CO(J=2-1), C$^{18}$O(J=2-1), N$_2$D$^+$(J=3-2), DCO$^+$(J=3-2), DCN(J=3-2), SiO(J=5-4). In this paper, we utilize the results from the 1.3\,mm continuum and CO(J=2-1) data.

The acquired visibilities were calibrated with the standard pipeline in CASA \citep{2007McMullin}. The 1.3\,mm continuum images were generated using the TCLEAN task of CASA with a threshold of 3 $\sigma$ theoretical sensitivity in the three sets of continuum images with the ACA (beam size $\sim$ 5.8 arcsec) FWHM, TM2+ACA (beam size $\sim$ 1.2 arcsec) FWHM, and TM1+TM2+ACA (beam size $\sim$ 0.35 arcsec) FWHM combination. We applied the “hogbom” deconvolver, and Briggs weighting with a robust value of +0.5 to obtain high-resolution maps, which allow us to identify multiple components more precisely. The TM1+TM2+ACA combination provides the best resolution to distinguish multiple components. The typical sensitivity of 1.3 mm continuum emission ranges from 0.01 to 0.2 mJy~beam$^{-1}$ in the TM1+TM2+ACA data \citep{dutta_alma_2020-1}. The corresponding 3$\sigma$ mass sensitivity is better than 0.002 M$_{\sun}$, assuming a dust temperature of 25 K and a distance of 400 pc.

\subsection{JCMT SCUBA-2 Observations}

The natal cores of the ALMASOP sources were observed in the 850\,$\mu$m continuum with JCMT/SCUBA-2, as a part of the JCMT legacy survey SCOPE, "SCUBA-2 Continuum Observations of Pre-protostellar Evolution" (project ID: M16AL003, PI: Tie Liu). The beam size of JCMT at 850$\mu$m is about 14$\arcsec$. An analysis of the 850\,$\mu$m continuum emission of these dense cores was presented in \cite{yi_planck_2018-1}. The dense core parameters including the sizes, masses and H$_2$ mean densities derived by \cite{yi_planck_2018-1} are listed in Table \ref{tab:binary} and Table \ref{tab:single}. It is worth mentioning that we updated these parameters following new distance measurements as used in \cite{kim_molecular_2020}, which are 380 pc for $\lambda$ Orionis, 390 pc and 430 pc for Orion A, 390 pc and 420 pc for Orion B.

\subsection{No: 45m observations}

The ALMASOP sources were also observed in eight molecular lines (CCS(J$_N$=8$_7$-7$_6$), CCS(J$_N$=7$_6$-6$_5$), HC$_3$N(J=9-8), N$_2$H$^+$(J=1-0), DNC(J=1-0), HN$^{13}$C(J=1-0), N$_2$D$^+$(J=1-0), c-C$_3$H$_2$ J$_{K_{a}K_{c}}$(2$_{12}$-1${01}$), NH$_3$(J,K)=(1,1)) toward the 850 $\mu$m intensity peak positions of the SCUBA-2 cores with the 45 m radio telescope at Nobeyama Radio Observatory (No:45m, project IDs: CG161004, LP177001; PI: K. Tatematsu) from 2015 December to 2019 May \citep{tatematsu_astrochemical_2017,kim_molecular_2020,Tatematsu2021}. The observations were conducted with the receivers TZ1, T70 and FOREST which are dual-polarization and two-sideband superconductor–insulator–superconductor receivers. The velocity resolution is 0.05-0.06 km/s for both TZ1 and T70, 0.1 km/s for FOREST and 0.05 km/s for H22. The details of No: 45m observations are described in \cite{kim_molecular_2020}. Among the observed molecular lines, the N$_2$H$^+$ line emission exhibits a spatial distribution similar to that of the SCUBA-2 850\,$\mu$m dust continuum emission \citep{tatematsu_astrochemical_2017,Tatematsu2021}. Therefore, we adopt the Mach number, system velocity, and line width (FWHM) of N$_2$H$^+$(J=1-0) from \cite{kim_molecular_2020} for further analysis in this work, which are compiled in Table \ref{tab:binary} and Table \ref{tab:single}. In addition, there are 16 dense cores in the ALMASOP survey that were also mapped in N$_2$H$^+$(J=1-0) line emission \citep{tatematsu_astrochemical_2017,Tatematsu2021}. We also present the N$_2$H$^+$(J=1-0) maps for these cores in this work.

\subsection{Infrared data}

We used the Spitzer Enhanced Imaging Products (SEIP) from the Spitzer Heritage Archive \citep{2012Megeath} and Wide-field Infrared Survey Explorer (WISE) data \citep{Wright2010} to help classify protostars distributed throughout the region of the cores. The SEIP includes data from the four channels of the IRAC instrument (3.6, 4.5, 5.8 and 8 $\mu$m) and the 24 $\mu$m channel of the MIPS instrument. The WISE data were also observed in 3.4, 4.6, 12, and 22 $\mu$m.
In this paper, we mainly use the highest-resolution 3.6/4.5/8 $\mu$m data in SEIP data observations or the highest-resolution 3.4/4.6/12 $\mu$m data in the WISE observations.

\section{Results}
\label{sec:results}

\subsection{Identification of binary/multiple systems}

\begin{figure*}[ht!]
    \centering
    \includegraphics[width=18cm]{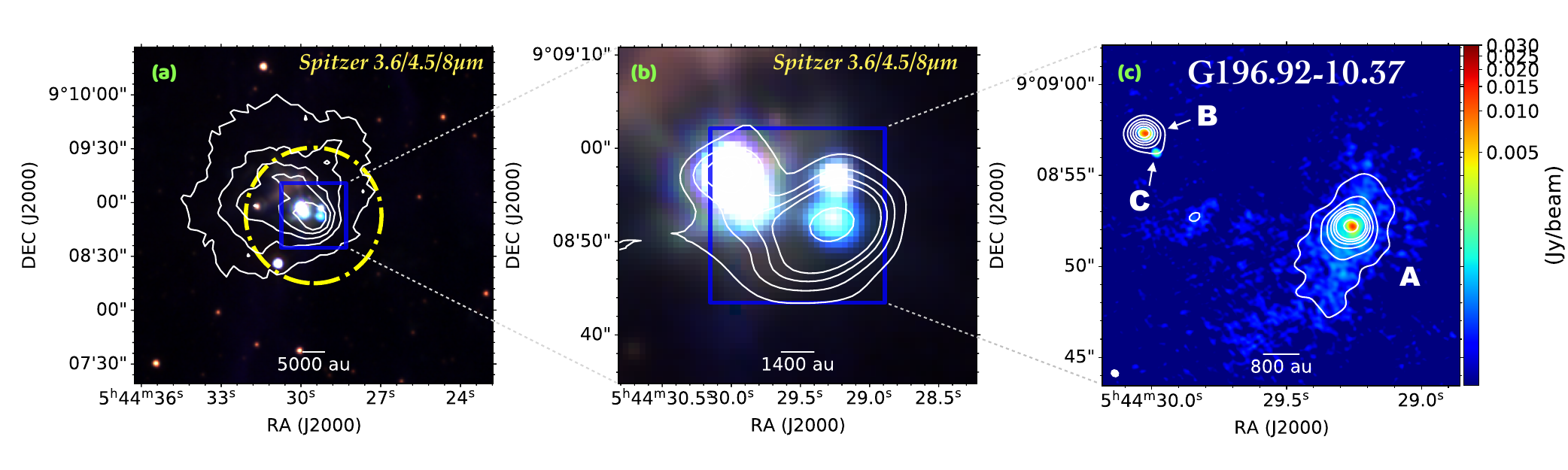}
    \caption{Images of an exemplar core G196.92-10.37. (a) JCMT SCUBA-2 850 $\mu$m contours superimposed with RGB image of Spitzer 3.6/4.5/8 $\mu$m data. The yellow circle represents the field of view (FOV) of the ALMA ACA image shown in panel (b). The contour levels are from  6$\sigma$ to 30$\sigma$ with steps of 6$\sigma$\, (1$\sigma$ is 13.8 mJy beam$^{-1})$. (b) Zoom into the infrared data, with contours of the ALMA ACA 1.3\,mm dust continuum data. The contour levels are 3$\sigma$, 6$\sigma$, 9$\sigma$, 12$\sigma$ and 30$\sigma$ (1$\sigma$ is 5 mJy beam$^{-1})$. (c) Color image showing ALMA 1.3\,mm dust continuum emission from combined TM1+TM2+ACA data, and contours showing the 1.3\,mm dust continuum emission from lower-resolution TM2+ACA data. The contours correspond to 5$\sigma$, 10$\sigma$, 15$\sigma$, 20$\sigma$, 25$\sigma$, 30$\sigma$, 50$\sigma$, 70$\sigma$ and 90$\sigma$ (1$\sigma$ is 4 mJy beam$^{-1})$.}
    \label{Figure 1}
\end{figure*}

We focus on the 43 protostellar cores among 72 cores that were detected in 1.3\,mm continuum emission in the ALMASOP survey \citep{dutta_alma_2020-1}. 
The 1.3\,mm continuum emission from these cores is relatively bright, and indicates that one or more compact objects, previously identified as young protostars \citep{dutta_alma_2020-1}, have formed within them. Figure \ref{Figure 1} presents the infrared and (sub)millimeter continuum images obtained with various instruments for an exemplar core G196.92-10.37. Multiple point sources are seen in the Spitzer image but some may be contamination from background/foreground stars or very evolved protostars because they do not show strong 1.3\,mm continuum emission  \citep{Tobin2022}. We will only consider young protostars with disks and envelopes, which should have strong 1.3 mm continuum emission, in further multiplicity analysis. From the highest-resolution ALMA 1.3\,mm continuum image shown in the right panel, one can identify a triple protostellar system formed in the exemplar core G196.92-10.37. The image descriptions of the other cores are compiled in the Appendix~\ref{AppA}.

We identify binary/multiple systems within the SCUBA-2 cores, whose sizes are smaller than or comparable to the FOV of the ALMA 12 m array, following the same strategy used in previous works \citep[e.g.,][]{tobin_vla_2016,Tobin2022}. The core sizes are listed in Table \ref{tab:binary} and \ref{tab:single}. In total, we identify 13 binary/multiple systems and 29 single systems, and one binary-system candidate (one of its members is a prestellar core). Here we note again that our observations cannot resolve systems with below 140 au separation, thus some single star systems could also be very close binary/multiple systems. Some ALMASOP sources were also observed in the VANDAM survey \citep[see Table \ref{tab:tobin};][]{Tobin2022}. We find that two single-protostar systems (G206.12-15.76 and G211.47-19.27S) in the ALMASOP survey were further resolved to very close binary/multiple systems in the VANDAM survey. However, the very close binary/multiple systems in  G206.12-15.76 and G211.47-19.27S are apparently formed due to disk fragmentation, and hence their properties are not determined by the fragmentation of their natal dense cores. Therefore, they are treated as single stars in the below analysis of the relations between core properties and protostellar multiplicity. The other single stars in ALMASOP survey remain to be single in the VANDAM survey.

\begin{figure}[ht!]
    \centering
    \includegraphics[width=12cm]{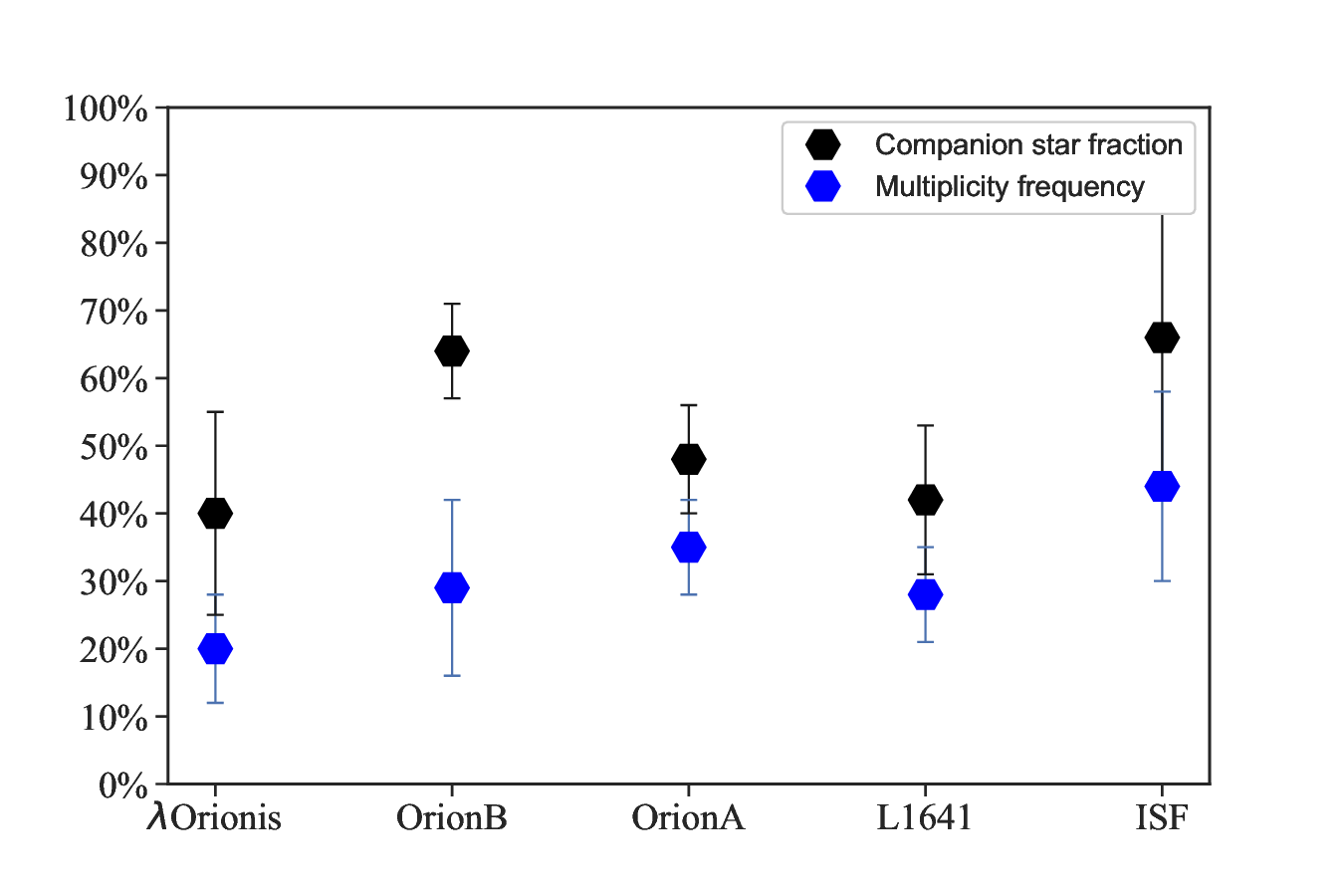}
    \caption{MF (blue hexagons) and CSF (black hexagons) in three GMCs ($\lambda$ Orionis, Orion A, and Orion B) in the Orion region, and in two subregions of Orion A, L1641 and ISF.}
    \label{fig:mfcsf}
\end{figure}

To quantify the multiplicity in each Orion GMC, we adopt the statistical parameters of MF and CSF from Batten (1973). They give a good indication of the proportion of binary/multiple systems and the average number of companions. The two quantities associated with the number of star systems and companions are given by
\begin{equation}
    MF=\frac{B + T + Q + ...}{S + B + T + Q ...}
\end{equation}

\begin{equation}
    CSF = \frac{B + 2T + 3Q + ...}{S + B + T + Q ...}
\end{equation}

where \textit{S}, \textit{B}, \textit{T}, and \textit{Q} are the numbers of single, binary, triple, and quadruple systems in the sample, respectively.  
Figure \ref{fig:mfcsf} shows the MF and CSF of each of the three Orion GMCs, and the actual quantities are listed in Table \ref{tab:mfcsf}. As the two subregions (separated by a  -6$\arcdeg$ decl.) of the Orion A, the integral-shaped filament (ISF) and L1641, have very different physical environments and levels of star formation activity \citep{Megeath2016}, we also calculate their MF and CSF separately. The overall MF and CSF for the ALMASOP sample are 28$\%$ $\pm$ 4$\%$ and 51$\%$ $\pm$ 6$\%$, respectively. For comparison, \citet{Tobin2022} found that the overall MF and CF for the Orion HOPS protostars are $30\%\pm3\%$ and $44\%\pm3\%$, respectively, which are broadly consistent with our statistical results. However, we note that their sample contains very close binary/multiple systems with separations smaller than 100 au, which cannot be resolved in our observations.

In our sample, the $\lambda$ Orionis cloud shows the lowest MF and CSF, and the ISF region in Orion A shows the highest MF and CSF.
When comparing the two subregions of Orion A, the ISF region, which contains the Orion Nebula cluster (ONC) has much higher multiplicity than L1641. The different occurrences of stellar multiplicity in the Orion GMCs may be caused by environmental effects such as stellar feedback, gas density, and turbulence levels, which we explore later in Section \ref{sec:discussion}.

\subsection{Properties of the natal dense cores}

We are interested in how the properties of dense cores affect the multiplicity of the protostars that form within them. To explore this possible connection, we summarize detailed information of the 43 protostellar cores from previous studies and 
divide the cores into two groups: dense cores forming single stars and dense cores forming more than one protostar. As the assumptions and calculation procedures for many of the parameters have already been described in detail in previous works \citep{yi_planck_2018-1,kim_molecular_2020,dutta_alma_2020-1}, we simply describe the core parameters here. The core parameters of these two groups are listed in Table \ref{tab:binary} and Table \ref{tab:single}. 

\cite{yi_planck_2018-1} used the ClumpFind task in the STARLINK package to identify dense cores. The method was to fit the SCUBA-2 850\,$\mu$m continuum emission maps in a 2D Gaussian profiles with the threshold of 5 $\sigma$. ClumpFind limited the edge of the core according to the threshold, and extracted the sizes and fluxes relying on the full width half max(FWHM) brightness and the integrated fluxes of the core.

The flux density produced by the SCUBA-2 850\,$\mu$m dust emission can be expressed as:  
\begin{equation}
    S^{\prime}_{\nu}=\Omega \tau_\lambda B_{\nu}(T_{dust}) =  \Omega \mu m_H \kappa_{\nu}B_{\nu}(T_{dust})
\end{equation}

The H$_2$ column density of the cores can be therefore be derived from the flux density. The calculation of the following parameters, with flux density involved, follows the optically thin assumption:
\begin{equation}
    N({\rm H}_2)=\frac{S^{\prime}_{\nu}}{\Omega \mu m_H \kappa_{\nu} B_{\nu}(T_{dust})}
\end{equation}
where S$^{\prime}_{\nu}$ is the beam-averaged flux density, 
 $\tau_\lambda$ is the dust optical depth, $\Omega$ is the solid angle of the source, $\mu$ is the mean molecular weight and its value is 2.8, $m_{\rm H}$ is the mass unit of atomic hydrogen, and $B_{\mu}(T_{\rm dust})$ is the Planck function of dust temperature $T_{\rm dust}$, which is from the PGCCs catalog. The dust opacity per gram of gas is calculated following the equation from \citet{1990Beckwith}: $\kappa_{\nu}$ =$0.1\left(\nu/10^{12}\rm {Hz}\right)^\beta$ cm$^2$g$^{-1}$, where $\beta$ is the dust
emissivity spectral index from \citet{planck_catalog2016}. The column densities of the 43 protostellar cores in the ALMASOP sample range from 0.4 $\pm$ 0.1 $\times$ 10$^{23}$ cm$^{-2}$ to 11.7 $\pm$ 2.5 $\times$ 10$^{23}$ cm$^{-2}$, with a median value of 2.5 $\times$ 10$^{23}$ cm$^{-2}$.

The mean H$_2$ number densities of the cores are estimated from the H$_2$ column densities based on the SCUBA-2 850 $\mu$m dust continuum:
\begin{equation}
    n_{{\rm H}_2}=\frac{N({\rm H}_2)}{R}
\end{equation}
where $R$ is the core diameter and is calculated using $R = \sqrt{a\cdot b}$, \textit{a} and \textit{b} being the major and minor axes of FWHM obtained with ClumpFind and converted to linear distances assuming source distances of \citet{kim_molecular_2020}. The mean density of the cores ranges from 0.7 $\times$ 10$^{5}$ cm$^{-3}$ to 40.8 $\times$ 10$^{5}$ cm$^{-3}$, with a median value of 5.4 $\times$ 10$^{5}$ cm$^{-3}$.

The core masses are inferred from the SCUBA-2 850 $\mu$m continuum fluxes compiled in \cite{yi_planck_2018-1}:
\begin{equation}
    M_{core}=\frac{S^{\prime}_{\nu}D^2}{\kappa_{\nu}B_{\nu}(T_{\rm dust})}
\end{equation}
likewise, where S$^{\prime}_{\nu}$ is the flux density of cores from the SCUBA-2 850\,$\mu$m observations, and $D$ is the distance adopted by \cite{kim_molecular_2020}. The dust opacity $\kappa_{\nu}$ is adopted from \citet{1990Beckwith}.
As listed in Table \ref{tab:binary} and Table \ref{tab:single}, the mass of the cores ranges from 0.14 $\pm$ 0.04 $\textit{M}_{\odot}$ to 11.64 $\pm$ 2.18 $\textit{M}_{\odot}$, with a median mass of 2.27 $\textit{M}_{\odot}$. 

The high-resolution ALMA 1.3 mm continuum emission traces the total gas mass of the envelope and disk, $\textit{M}_{enve+disk}$, for each protostar. $\textit{M}_{enve+disk}$ can be roughly estimated based on the integrated 1.3\,mm continuum flux density under an optically thin assumption, which was already calculated in \cite{dutta_alma_2020-1}. The smallest gas mass of protostars in the whole sample is 0.009 $\textit{M}_{\odot}$ and the largest mass is 2.312 $\textit{M}_{\odot}$, with an median mass of 0.3 $\textit{M}_{\odot}$. Firstly, we estimate the mass of the envelope+disk for individual protostars ($M_{\rm enve+disk}^{*}$). Secondly, we estimate the total envelope+disk masses ($M_{enve+disk}$) of all protostars within each core.  
We then calculate a 'compact gas fraction' (CGF) to describe the fraction of gas that has been accumulated from the dense core into the protostellar envelope and disk:
\begin{equation}
 CGF = M_{enve+disk} / M_{core}
\end{equation}
The CGF is a ratio calculated after collecting all the $\textit{M}_{enve+disk}$ in each core. The CGF is shown in the Column (14) of Table \ref{tab:binary} and \ref{tab:single}. The minimum and maximum values of the CGF for whole cores are 0.2$\%$ and 80$\%$, and the median value is 11$\%$. 

Once the perturbations caused by self-gravity inside a molecular cloud are larger than the Jeans length, they can lead the region to become unstable. Hence, Jeans fragmentation may lead to substructures that become binary/multiple systems. The Jeans lengths of dense cores can be calculated via the following equations from \citet{2014Wang}:

\begin{equation}
 L_{Jeans}=c_s\left(\frac{\pi}{G\rho}\right)^{1/2}=0.066\left(\frac{T}{10{\rm\,K}}\right)^{1/2}\left(\frac{n}{10^5\,{\rm cm}^{-3}}\right)^{-1/2}{\rm pc}
\end{equation}
where $T$ is the dust temperature of the dense core from \citet{2017Kounkel}, and $n$ is the mean column density of H$_2$ from \citet{yi_planck_2018-1}.  The Jeans lengths of dense cores in our sample range from 2500 au to 19,300 au with a median value of 7000 au.

The Mach number ($\mathcal{M}$) is a measurement of turbulence within of dense cores. We adopt the Mach numbers from \cite{kim_molecular_2020}, which are calculated using
\begin{equation}
    \mathcal{M}=\frac{\sigma_{NT}}{c_s}
\end{equation}
where $\sigma_{NT}$ is the nonthermal velocity dispersion derived from $\sigma_{NT}$ = $\sqrt{\frac{\triangle v}{8ln2} - \frac{k_{b}T_{k}}{m}}$. Here, $k_{B}$ is the Boltzmann constant, $T_k$ the kinetic temperature, $\triangle v$ the N$_2$H$^+$ (J=1-0) line width (FWHM), $m$ the mass of N$_2$H$^+$ molecule, and $c_s$ the sound speed, which depends on the temperature assumed.
The Mach number of the cores ranges from 0.6 $\pm$ 0.1 to 2.4 $\pm$ 0.1, with a median value of 1.1.

In order to examine whether the statistical properties of groups of dense cores harboring different types of protostellar systems differ from one another, we calculate the median and mean values of the above-derived core parameters in the two groups of cores, and list them in Table \ref{tab:my-table}. Statistically, the $N({\rm H}_2)$, $n({\rm H}_2)$, $\mathcal{M}$, and $\textit{M}_{core}$ of cores containing binary/multiple systems are larger than those of cores forming single stars. The $\textit{L}_{Jeans}$ and $M_{\rm enve+disk}^{*}$ of individual protostars in those multiple systems are smaller than those of single systems.  The core sizes of the two groups, however, do not show any noticeable difference.

\begin{figure*}[ht!]
    \centering
    \includegraphics[width=16cm]{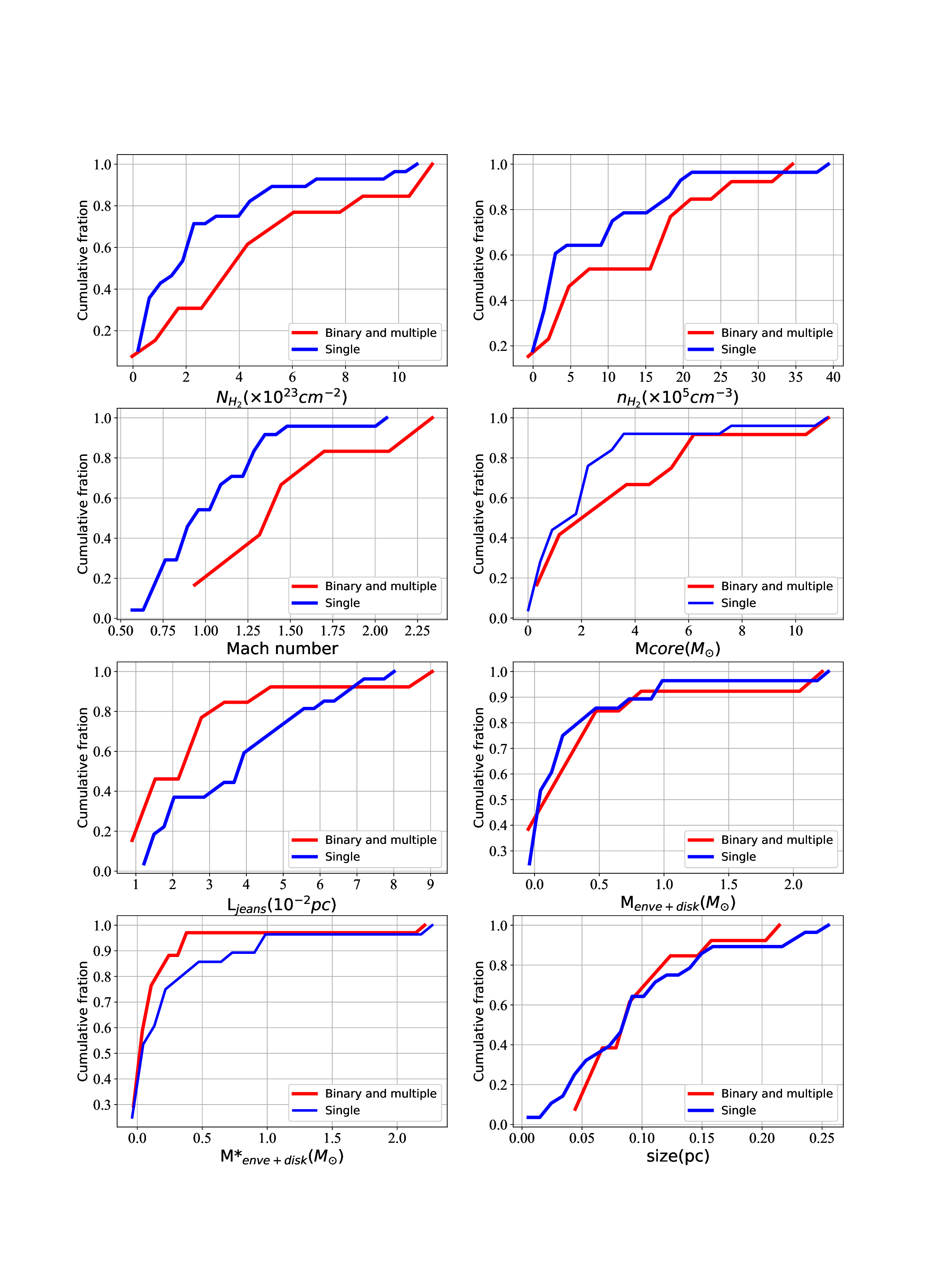}
    \caption{Cumulative distribution functions of the physical parameters in the two groups of dense cores. The compared core parameters are listed in Table \ref{tab:my-table}: H$_2$ column density, H$_2$ number density, Mach number, Jeans length, core mass, mass of envelope and disk, and size of core. In each panel, the blue line stands for single-star systems and the red line stands for binary/multiple systems. 
    $M_{\rm enve+disk}$ stands for the total envelope+disk mass of a core. $M_{\rm enve+disk}^{*}$ stands for the each protostar's envelope+disk mass.}
    \label{Cumulative}
\end{figure*}

We present the cumulative distribution functions of the core parameters of the two groups of cores in Figure \ref{Cumulative}. In addition, we use the Kolmogorov–Smirnov (KS) test to gauge whether or not these two groups of cores have the same underlying core parameter distributions. Overall, the distributions of H$_2$ column density, H$_2$ number density and Mach number of the two groups appear to differ substantially with p-value smaller than 10$\%$. In particular, the two groups show significant differences in H$_2$ column density and Mach number with very low p-values ($<5\%$) in the KS test of their distributions. In contrast, the distributions in $\textit{M}_{core}$, size, and $\textit{M}_{enve+disk}$ of the two groups of cores are statistically similar, as indicated by the very high p-values ($>28\%$) in the KS test. We find that the proportion of envelope+disk masses below 1 $\textit{M}_{\odot}$ for protostars in binary/multiple systems is larger than that in single systems.
We discuss these results more thoroughly in Section \ref{sec:discussion} below.

\begin{deluxetable}{lcccccccccccccc}
\rotate
\tabletypesize{\tiny}
\tablecaption{Physics and chemical properties of binary/multiple systems\label{tab:binary}}
\tablehead
{
 & & & & & & & &\colhead{ N$_2$H$^+$} & & & & & & \\ \cline{8-10}
\colhead{Source} & \colhead{R.A.(J2000)} &\colhead{Decl.(J2000)}&  \colhead{size} &  \colhead{$T_d$} &\colhead{$N(H_2)$}& \colhead{$n_H$$_2$}& \colhead{V$_{lsr}$}
&  \colhead{$\Delta$ V$_{HFS}$} &  \colhead{$\Delta$ V$_{GA}$} &  \colhead{$\mathcal{M}$ }&  \colhead{M$_{enve+disk}$}&  \colhead{ M$_{core}$}&  \colhead{CGF }&  \colhead{L$_{jeans}$} \\
 & \colhead{(h:m:s)}   & \colhead{(d:m:s)}   &  \colhead{(pc)} &  \colhead{(K)} &  \colhead{($\times$ 10$^{23}$cm$^{-2}$)}&   \colhead{($\times$ 10$^{5}$cm$^{-3}$)}
&  \colhead{(km s$^{-1}$)} &  \colhead{(km s$^{-1}$)} &  \colhead{(km s$^{-1}$)}&  \colhead{}&  \colhead{(M$_\odot$)} &  \colhead{(M$_\odot$)}
&\colhead{} & \colhead{(au)}
}
\decimalcolnumbers
\startdata
G196.92-10.37 & 05:44:29.56 & +09:08:50.20 & 0.22 & 14.8 ± 0.4 & 1.8 $\pm$ 0.1 & 5.6 $\pm$ 0.1  & 11.68 & 0.84 & 1.02 & 1.5 $\pm$ 0.2  & 0.116 $\pm$ 0.034 & 5.69 $\pm$ 0.32 & 0.02 & 7000   \\
G205.46-14.56M1$^{a}$ & 05:46:08.06 & -00:10:43.71 & 0.09 & 12.5 $\pm$ 0.9 & 8.7 $\pm$ 0.8 & 35.9 $\pm$ 1.7& 9.99 & 0.73 & 1.18 & 1.4 $\pm$ 0.1 &  2.309 $\pm$ 0.96 & 6.36 $\pm$ 1.33 & 0.36 & 2500   \\
G205.46-14.56M2$^{a}$ & 05:46:07.89 & -00:10:01.82 & 0.06 & 12.5 $\pm$ 0.9 & 5.5 $\pm$ 0.5 & 22.6 $\pm$ 1.5&10.05 & 0.88 & 1.26 & 1.7 $\pm$ 0.1 &  0.291 $\pm$ 0.063& 1.83 $\pm$ 0.38 & 0.16 & 3200    \\
G205.46-14.56S1 & 05:46:07.05 & -00:13:37.78 & 0.13 & 12.5 $\pm$ 0.9 & 6.7 $\pm$ 0.6 & 27.4 $\pm$ 1.7 & 10.31 & 1.18 & 1.23 & 2.4 $\pm$ 0.1 &  0.542 $\pm$ 0.179& 11.64 $\pm$ 2.18 & 0.04 & 2900    \\
G206.93-16.61E2 & 05:41:37.31 & -02:17:18.13 & 0.10 & 16.8 $\pm$ 5.3 & 4.5 $\pm$ 1.5 & 18.8 $\pm$ 2.4 & 9.76 & 0.60 & 0.58 & 1.0 $\pm$ 0.1 &  0.863 $\pm$ 0.195 & 4.08 $\pm$ 0.16 & 0.21 & 4000    \\
G207.36-19.82N1 & 05:30:50.94 & -04:10:35.60 & 0.06 & 11.9 $\pm$ 1.4 & 3.5 $\pm$ 0.5 & 5.7 $\pm$ 1.1 &10.52 & 1.13 & 1.48 & 2.3 $\pm$ 0.1 &  0.124 $\pm$ 0.048 & 1.15 $\pm$ 0.46 & 0.10 & 6200    \\
G208.68-19.20N2 & 05:35:20.45 & -05:00:50.39 & 0.05 & 19.7 $\pm$ 3.8 & 11.6 $\pm$ 3.0 & 18.9 $\pm$ 3.6 & 11.14 & 0.44 & 0.67 & 0.6 $\pm$ 0.1 &  4.786$\pm$ 2.046 & 2.27 $\pm$ 1.15 &  & 4400 \\
G208.68-19.20N3 & 05:35:18.02 & -05:00:20.70 & 0.05 & 19.7 $\pm$ 3.8 & 11.7 $\pm$ 2.5 & 18.9 $\pm$ 3.7 & 11.12 & 0.69 & 0.96 & 1.1 $\pm$ 0.1 &  0.607 $\pm$ 0.195 & 2.65 $\pm$ 1.50 & 0.23 & 4400    \\
G208.68-19.20S & 05:35:26.32 & -05:03:54.39 & 0.09 & 19.7 $\pm$ 3.8 & 4.6 $\pm$ 1.0 & 7.6 $\pm$ 1.4 & 10.35 & 0.95 & 0.94 & 1.5 $\pm$ 0.1 &  0.464 $\pm$ 0.190& 3.55 $\pm$ 1.03 & 0.13 & 6900    \\
G209.55-19.68N1 & 05:35:08.90 & -05:55:54.40 & 0.16 & 14.1 $\pm$ 4.0 & 0.4 $\pm$ 0.1 &  0.7 $\pm$ 0.1 & 7.20 & 0.84 & 1.30 & 1.6 $\pm$ 0.1 &  0.334 $\pm$ 0.087& 0.75 $\pm$ 0.04 & 0.44 & 19300    \\
G210.49-19.79W & 05:36:18.86 & -06:45:28.03 & 0.11 & 11.8 $\pm$ 2.7 & 11.3 $\pm$ 2.9 & 18.4 $\pm$ 0.2& 9.01 & 0.60 & 0.77 &  1.2 $\pm$ 0.1 & 0.209 $\pm$ 0.086 & 6.62 $\pm$ 0.25 & 0.03 & 3400   \\
G210.97-19.33S2 & 05:38:45.30 & -07:01:04.41 & 0.07 & 12.8 $\pm$ 1.1 & 1.2 $\pm$ 0.3 & 2.0 $\pm$ 0.1 & & & 0.86 &  &  0.038 $\pm$ 0.012 & 0.37 $\pm$ 0.09 & 0.10 & 10900    \\
G211.47-19.27N & 05:39:57.18 & -07:29:36.07 & 0.07 & 12.4 $\pm$ 1.1 & 4.1 $\pm$ 0.4 & 6.6 $\pm$ 0.3 & 3.99 & 0.52 & 0.97 & 1.0 $\pm$ 0.2 &  0.105 $\pm$ 0.035 & 1.54 $\pm$ 0.43 & 0.06 & 5900    \\
G212.10-19.15N2 & 05:41:24.03 & -07:53:47.51 & 0.12 & 10.8 $\pm$ 1.4 & 1.8 $\pm$ 0.3 & 2.9 $\pm$ 0.3& 4.45 & 0.69 & 1.04 & 1.5 $\pm$ 0.1 &  0.048 $\pm$ 0.016 & 1.53 $\pm$ 0.72 & 0.03 & 8300  \\
\hline
\enddata
\tablecomments{Column(1): ALMASOP core name. Note that the marked $^a$ is different from the JCMT dense core name and those marked with * are dense core detected only in the ALMASOP survey.
(2)-(3): coordinates in equatorial system (J2000) from \citep{yi_planck_2018-1} and \citep{dutta_alma_2020-1}. (4): Core size from \citet{yi_planck_2018-1}. (5): Dust temperature comes from \citet{2017Kounkel}. (6): H$_2$ column density cited from \citet{kim_molecular_2020}. (7): H$_2$ number density from \citet{yi_planck_2018-1}. (8)-(10): N$_2$H$^+$(J=1-0) of systemic velocity, FWHM, and FWHM inferred by Gaussian fitting from \citet{kim_molecular_2020}. (11): Mach number cited from \citet{kim_molecular_2020}. (12): Gas mass of envelope and disk from \citep{dutta_alma_2020-1}.}
\end{deluxetable}

\begin{deluxetable}{lcccccccccccccc}
\rotate
\tabletypesize{\tiny}
\tablecaption{Physics and chemical properties of single systems\label{tab:single}}
\tablehead
{
 & & & & & & & &\colhead{ N$_2$H$^+$} & & & & & & \\ \cline{8-10}
\colhead{Source} & \colhead{R.A.(J2000)} &\colhead{Decl.(J2000)}&  \colhead{size} &  \colhead{$T_d$} &\colhead{$N(H_2)$}& \colhead{$n_H$$_2$}& \colhead{V$_{lsr}$}
&  \colhead{$\Delta$ V$_{HFS}$} &  \colhead{$\Delta$ V$_{GA}$} &  \colhead{$\mathcal{M}$ }&  \colhead{M$_{enve+disk}$}&  \colhead{ M$_{core}$}&  \colhead{CGF }&  \colhead{L$_{jeans}$} \\
 & \colhead{(h:m:s)}   & \colhead{(d:m:s)}   &  \colhead{(pc)} &  \colhead{(K)} &  \colhead{($\times$ 10$^{23}$cm$^{-2}$)}&   \colhead{($\times$ 10$^{5}$cm$^{-3}$)}
&  \colhead{(km s$^{-1}$)} &  \colhead{(km s$^{-1}$)} &  \colhead{(km s$^{-1}$)}&  \colhead{}&  \colhead{(M$_\odot$)} &  \colhead{(M$_\odot$)}
&\colhead{} & \colhead{(au)}
}
\decimalcolnumbers
\startdata
G191.90-11.21S & 05:31:31.73 & +12:56:14.99 & 0.10 & 14.7 $\pm$ 4.1 & 0.7 $\pm$ 0.2 & 2.2 $\pm$ 0.5 &  10.53 & 0.41 & 0.38 & 0.7 $\pm$ 0.1 & 0.079 $\pm$ 0.034& 0.93 $\pm$ 0.09 & 0.085 & 11100   \\
G192.12-11.10 & 05:32:19.54 & +12:49:40.19 & 0.12 & 13.3 $\pm$ 3.0 & 1.8 $\pm$ 0.4 & 5.4 $\pm$ 0.2 & 10.02 & 0.73 & 0.35 & 1.4 $\pm$ 0.2 & 0.340 $\pm$ 0.145&  2.32 $\pm$ 0.33 & 0.147 & 6800\\
G192.32-11.88N & 05:29:54.47 & +12:16:56.00 & 0.06 & 17.3 $\pm$ 6.0 & 1.1 $\pm$ 0.4 & 3.4 $\pm$ 0.4 &12.12 & 0.66 & 0.59 & 1.1 $\pm$ 0.2 &0.408 $\pm$ 0.174 & 0.51 $\pm$ 0.05 & 0.800 &9700  \\
G192.32-11.88S & 05:29:54.74 & +12:16:32.00 & 0.05 & 17.3 $\pm$ 6.0 & 0.9 $\pm$ 0.4 & 2.9 $\pm$ 0.2 & 12.08 & 0.54 & 0.58 & 0.9 $\pm$ 0.1 & 0.100 $\pm$ 0.043&0.23 $\pm$ 0.02 & 0.435 &10500\\
G200.34-10.97N & 05:49:03.71 & +05:57:55.74 & 0.09 & 13.5 $\pm$ 0.9 & 0.8 $\pm$ 0.1 & 2.5 $\pm$ 0.6 & 13.36 & 0.45 & 0.51 & 0.8 $\pm$ 0.1 & 0.068 $\pm$ 0.029 &0.81 $\pm$ 0.06 & 0.084 &10000 \\
G201.52-11.08 & 05:50:59.01 & +04:53:53.10 & 0.05 & 13.6 $\pm$ 1.3 & 0.8 $\pm$ 0.1 & 3.3 $\pm$ 0.8  &  &  & & & 0.060 $\pm$ 0.026&0.14 $\pm$ 0.04 & 0.429 & 8800  \\
G203.21-11.20W1 & 05:53:42.83 & +03:22:32.90 & 0.12 & 11.2 $\pm$ 0.7 & 2.7 $\pm$ 0.3 & 11.1 $\pm$ 1.9 &10.70 & 0.50 & 1.02 & 1.0 $\pm$ 0.1 &  0.091 $\pm$ 0.039 & 2.88 $\pm$ 0.13 & 0.032 & 4300   \\
G203.21-11.20W2 & 05:53:39.62 & +03:22:24.90 & 0.10 & 11.2 $\pm$ 0.7 & 3.2 $\pm$ 0.4 & 13.4 $\pm$ 1.5 & 10.11 & 0.50 & 0.72 & 1.0 $\pm$ 0.1 & 0.034 $\pm$ 0.015 &2.57 $\pm$ 0.19 & 0.013 & 3900  \\
G205.46-14.56N1$^{a}$ & 05:46:09.65 & -00:12:16.45 & 0.05 & 12.5 $\pm$ 0.9 & 5.5 $\pm$ 0.5 & 22.5 $\pm$ 1.1 & 9.92 &  & 1.34 &  & 0.475 $\pm$ 0.203&1.06 $\pm$ 0.16 & 0.448 & 3200     \\
G205.46-14.56N2$^{a}$ & 05:46:07.49 & -00:12:22.42 & 0.03 & 12.5 $\pm$ 0.9 & 4.7 $\pm$ 0.4 & 19.2 $\pm$ 1.9 &10.29 & 0.70 & 0.75 & 1.4 $\pm$ 0.1 & 0.223 $\pm$ 0.095 & 0.49 $\pm$ 0.09 & 0.475 & 3500\\
G205.46-14.56S2 & 05:46:04.49 & -00:14:18.87 & 0.03 & 12.5 $\pm$ 0.9 & 4.8 $\pm$ 0.5 & 19.8 $\pm$ 2.9& 10.44 & 0.46 & 0.54 & 0.9 $\pm$ 0.1 &0.069 $\pm$ 0.029 &0.47 $\pm$ 0.10 & 0.147 & 3400   \\
G205.46-14.56S3 & 05:46:03.54 & -00:14:49.34 & 0.04 & 12.5 $\pm$ 0.9 & 5.0 $\pm$ 0.5 & 20.4 $\pm$ 1.9 & 10.36 & 0.61 & 0.81 & 1.2 $\pm$ 0.1 &0.167 $\pm$ 0.072 &0.88 $\pm$ 0.16 &0.190 &3300 \\
G206.12-15.76 & 05:42:45.26 & -01:16:11.37 & 0.14 & 11.9 $\pm$ 1.6 & 2.6 $\pm$ 0.4 & 10.8 $\pm$ 1.3 &  &  &  &  & 1.035 $\pm$ 0.442 &3.95 $\pm$ 1.73 & 0.262 & 4500  \\
G206.93-16.61W2$^{a}$ & 05:41:25.04 & -02:18:08.11 & 0.06 & 16.8 $\pm$ 5.3 & 9.9 $\pm$ 3.1 & 40.08 $\pm$ 4.7 & 9.25 & 0.65 & 1.02 & 1.1 $\pm$ 0.1 &0.771 $\pm$ 0.333 & 3.27 $\pm$ 0.15 & 0.235 &2800  \\
G208.68-19.20N1 & 05:35:23.37 & -05:01:28.70 & 0.09 & 19.7 $\pm$ 3.8 & 10.9 $\pm$ 2.4 & 17.7 $\pm$ 3.4 & 11.13 & 0.72 & 0.96 & 1.1 $\pm$ 0.1 &2.312 $\pm$ 0.988 &7.79 $\pm$ 2.53 & 0.297 &4500 \\
G208.68-20.04E & 05:32:48.40 & -05:34:47.14 & 0.13 & 12.8 $\pm$ 4.2 & 2.6 $\pm$ 0.9 & 4.3 $\pm$ 0.3 & 8.74 & 0.38 & 0.52 & 0.7 $\pm$ 0.1 & 0.073 $\pm$ 0.031 &3.95 $\pm$ 0.29 & 0.018  & 7400 \\
G208.89-20.04Walma* & 05:32:28.03 & -05:34:26.69 &  &  &  & &  &  & & &0.028 $\pm$ 0.012  &  &  &     \\
G209.55-19.68S1 & 05:35:13.25 & -05:57:58.65 & 0.23 & 14.1 $\pm$ 4.0 & 0.8 $\pm$ 0.2 & 1.2 $\pm$ 0.2& 7.35 & 0.68 & 0.94 & 1.3 $\pm$ 0.1 &0.264 $\pm$ 0.113 & 2.31 $\pm$ 0.97 & 0.114 &14800 \\
G209.55-19.68S2$^a$ & 05:35:08.96 & -05:58:26.38 & 0.16 & 14.1 $\pm$ 4.0 & 1.0 $\pm$ 0.3 & 1.6 $\pm$ 0.5 & 8.11 & 0.45 & 0.51 & 0.8 $\pm$ 0.1 & 0.084 $\pm$ 0.036 & 1.52 $\pm$ 0.26 & 0.054 &12800  \\
G210.37-19.53S & 05:37:00.55 & -06:37:10.16 & 0.15 & 14.0 $\pm$ 4.1 & 1.1 $\pm$ 0.3 & 1.8 $\pm$ 0.7 & 5.61 & 0.72 & 0.69 & 1.3 $\pm$ 0.1 &0.133 $\pm$ 0.050& 2.27 $\pm$ 0.23 & 0.059 & 12000   \\
G210.82-19.47S & 05:38:03.67 & -06:58:24.141 & 0.08 &  & 0.4 $\pm$ 0.1  & 0.7 $\pm$ 0.8 & &  &  &  & 0.010 $\pm$ 0.004&0.22$\pm$ 0.125 & 0.600     \\
G211.01-19.54N & 05:37:57.23 & -07:06:56.72 & 0.10 & 14.7 $\pm$ 8.4 & 2.5 $\pm$ 1.4 &  4.1 $\pm$ 0.2 & 6.03 & 0.83 & 0.79 & 1.5 $\pm$ 0.1 &0.130 $\pm$ 0.056 &2.16 $\pm$ 0.36 & 0.060 &8200  \\
G211.01-19.54S & 05:37:59.04 & -07:07:24.14 & 0.07 & 14.7 $\pm$ 8.4 & 2.5 $\pm$ 1.4 & 4.0 $\pm$ 0.1 & 5.72 & 0.70 & 1.01 & 1.3 $\pm$ 0.1 &0.021 $\pm$ 0.009 &1.05 $\pm$ 0.32 & 0.020&8300  \\
G211.16-19.33N2 & 05:39:05.83 & -07:10:41.52 & 0.10 & 12.5 $\pm$ 1.8 & 0.8 $\pm$ 0.1 & 1.2 $\pm$ 0.1 & 3.49 & 0.46 & 0.70 & 0.9 $\pm$ 0.1 &0.016 $\pm$ 0.007& 0.48 $\pm$ 0.20 & 0.033&13900  \\
G211.47-19.27S & 05:39:56.10 & -07:30:28.40 & 0.01 & 12.4 $\pm$ 1.1 & 7.2 $\pm$ 0.8 & 11.7 $\pm$ 0.3 & 5.52 & 1.03 & 0.93 & 2.1 $\pm$ 0.1 &0.990 $\pm$ 0.424 &11.39 $\pm$ 3.18 & 0.087 &4500  \\
G212.10-19.15S & 05:41:26.39 & -07:56:51.81 & 0.15 & 10.8 $\pm$ 1.4 & 1.9 $\pm$ 0.3 & 3.0 $\pm$ 0.4 & 3.78 & 0.38 & 0.74 & 0.6 $\pm$ 0.1 & 0.237 $\pm$ 0.101& 2.41 $\pm$ 1.00 & 0.098 & 8200  \\
G212.84-19.45N & 05:41:32.07 & -08:40:10.94 & 0.10 & 11.7 $\pm$ 1.1 & 2.0 $\pm$ 0.2 & 3.2 $\pm$ 0.5 & 4.31 & 0.34 & 0.42 & 0.8 $\pm$ 0.1 &0.273 $\pm$ 0.117& 1.26 $\pm$ 0.50 & 0.217 & 8300   \\
G215.87-17.62M & 05:53:32.52 & -10:25:05.99 & 0.24 & 12.2 $\pm$ 1.2 & 0.5 $\pm$ 0.1 & 0.8 $\pm$ 0.1 & 8.96 & 0.35 & 0.46 & 0.7 $\pm$ 0.1 &0.065 $\pm$ 0.028& 2.50 $\pm$ 0.64 & 0.026 & 16800   \\
G215.87-17.62N & 05:53:41.91 & -10:24:02.00 & 0.26 & 12.2 $\pm$ 1.2 & 0.6 $\pm$ 0.1 & 1.0 $\pm$ 0.2  & 9.28 & 0.46 & 0.89 & 0.9 $\pm$ 0.4 &0.0090 $\pm$ 004& 3.96 $\pm$ 0.84 & 0.002& 15000    \\
\hline
\enddata
\tablecomments{Column(1): ALMASOP core name. Note that the marked $^a$ is different from the JCMT dense core name and those marked with * are dense core detected only in the ALMASOP survey.
(2)-(3): coordinates in equatorial system (J2000) from \citep{yi_planck_2018-1} and \citep{dutta_alma_2020-1}. (4): Core size from \citet{yi_planck_2018-1}. (5): Dust temperature comes from \citet{2017Kounkel}. (6): H$_2$ column density cited from \citet{kim_molecular_2020}. (7): H$_2$ number density from \citet{yi_planck_2018-1}. (8)-(10): N$_2$H$^+$(J=1-0) of systemic velocity, FWHM, and FWHM inferred by Gaussian fitting from \citet{kim_molecular_2020}. (11): Mach number cited from \citet{kim_molecular_2020}. (12): Gas mass of envelope and disk from \citep{dutta_alma_2020-1}.}
\end{deluxetable}

\begin{deluxetable}{cccccccc}
\tablecaption{Stellar multiplicity in Orion Molecular Cloud Complex\label{tab:mfcsf}}
\tabletypesize{\footnotesize}
\tablehead{
\colhead{ Subregions} & \colhead{MF } &  \colhead{CSF} &
\colhead{Sample Number} & \colhead{Sample Number}& \colhead{Sample Number}& \colhead{Sample Number}& \colhead{Sample Number} \\
\colhead{} & \colhead{} & \colhead{} & \colhead{(Single System)}& \colhead{(Binary System)} & \colhead{(Triple System)} &\colhead{(Quadruple System)} & \colhead{(Five-star System)}
}
\decimalcolnumbers
\startdata
 $\lambda$ Orionis & 20$\%$ $\pm$ 8$\%$ & 40$\%$ $\pm$ 15$\%$ & 4& 0 &1&0&0 \\
Orion B & 29$\%$ $\pm$ 7$\%$ & 64$\%$ $\pm$ 13$\%$ & 10 & 2 &0&1&1 \\
Orion A & 35$\%$ $\pm$ 7$\%$ & 48$\%$ $\pm$ 8$\%$  & 15 & 5 &2&1&0 \\
L1641 & 28$\%$ $\pm$ 7$\%$ & 42$\%$ $\pm$ 11$\%$  & 10 & 3 &0&1&0 \\
ISF & 44$\%$ $\pm$ 14$\%$ & 66$\%$ $\pm$ 22$\%$  & 5 & 2 &2&0&0  \\
\enddata
\end{deluxetable}

\begin{table}
\begin{longtable}[c]{cc|clcl|clll|ccl}
\caption{Comparison of physical parameters of cores}
\label{tab:my-table}\\
\hline
\hline
 & {\textbf{Parameter}} & \multicolumn{4}{c|}{\textbf{Core (Single Systems)}} & \multicolumn{4}{c|}{\textbf{Core (Binary/Multiple Systems)}} & \multicolumn{2}{c}{\textbf{KS Test}} &  \\ \cline{3-12}
 &  & \textbf{Number} & \textbf{Mean} & \textbf{Median} & \textbf{Sigma} & \textbf{Number} & \textbf{Mean} & \textbf{Median} & \textbf{Sigma} & \textbf{Statistic} & \textbf{\textit{p}-value} &  \\ \cline{2-12}
\endfirsthead
\multicolumn{13}{c}%
{{\bfseries Table \thetable\ continued from previous page}} \\
\hline
 & \multirow{2}{*}{\textbf{Parameter}} & \multicolumn{4}{c|}{\textbf{Single system}} & \multicolumn{4}{c|}{\textbf{Binary/Multiple system}} & \multicolumn{2}{c}{\textbf{KS-test}} &  \\ \cline{3-12}
 &  & \textbf{Number} & \textbf{Mean} & \textbf{Median} & \textbf{sigma} & \textbf{N} & \textbf{Mean} & \textbf{Median} & \textbf{sigma} & \textbf{statistic} & \textbf{p-value} &  \\ \cline{2-12}
\endhead
\cline{2-12}
\endfoot
\endlastfoot
 & \textbf{$N(H_2)$($\times$10$^{23}$ cm$^{-2}$)} & 28 & 2.76 & 1.90 & 2.69 & 13 & 5.06 & 4.50 & 3.52 & 0.45 & 4.3$\%$   \\
 & \textbf{$n_H$$_2$($\times$10$^{5}$ cm$^{-3}$)} & 28 & 8.33 & 3.70 & 9.19 & 13 & 13.31 & 7.60 & 10.64 & 0.41 & 7.0$\%$   \\
 & \textbf {$\mathcal{M}$} & 24 & 1.06 & 1.00 & 0.34 & 12 & 1.51 & 1.50 & 0.43 & 0.50 & 3.0$\%$   \\
 & \textbf{$L_{jeans}$(10$^{-2}$ pc)} & 27 & 3.95 & 3.97 & 1.98 & 13 & 3.16 & 2.86 & 2.11 & 0.36 & 15.0$\%$  \\
 & \textbf{$M_{core}$(M$_\odot$)} & 28 & 2.27 & 1.84 & 2.39 & 13 & 3.67 & 2.65 & 3.08 & 0.31 & 28.0$\%$  \\
 & \textbf{$M_{enve+disk}$(M$_\odot$)} & 28 & 0.30 & 0.15 & 0.47 & 13 & 0.47 & 0.29 & 0.58 & 0.35 & 18.9$\%$   \\
 & \textbf{$M_{\rm enve+disk}^{*}$(M$_\odot$)} & 28 & 0.30 & 0.15 & 0.47 & 34 & 0.18 & 0.11 & 0.38 & 0.22 & 40.1$\%$   \\
 & \textbf{$Size$($\times$10$^{-1} $pc)} & 28 & 1.02 & 1.00 & 0.60& 13 & 1.02 & 0.90 & 0.45 & 0.22 & 89.7$\%$  \\
 & \textbf{$Velocity~Gradient$ (km/s)} & 11 & 3.93 & 3.60 & 1.55 & 5 & 4.05 & 3.62 & 1.70 & 0.25 & 92.6$\%$  \\ 
\hline
\end{longtable}
\textbf{Note}: G208.68-19.20N2 is not included in the statistics because it is a protobinary system candidate.
\end{table}


\subsection{N$_2$H$^+$ maps of 16 protostellar cores}

\begin{figure*}[ht!]
    \centering
    \includegraphics[width=19cm]{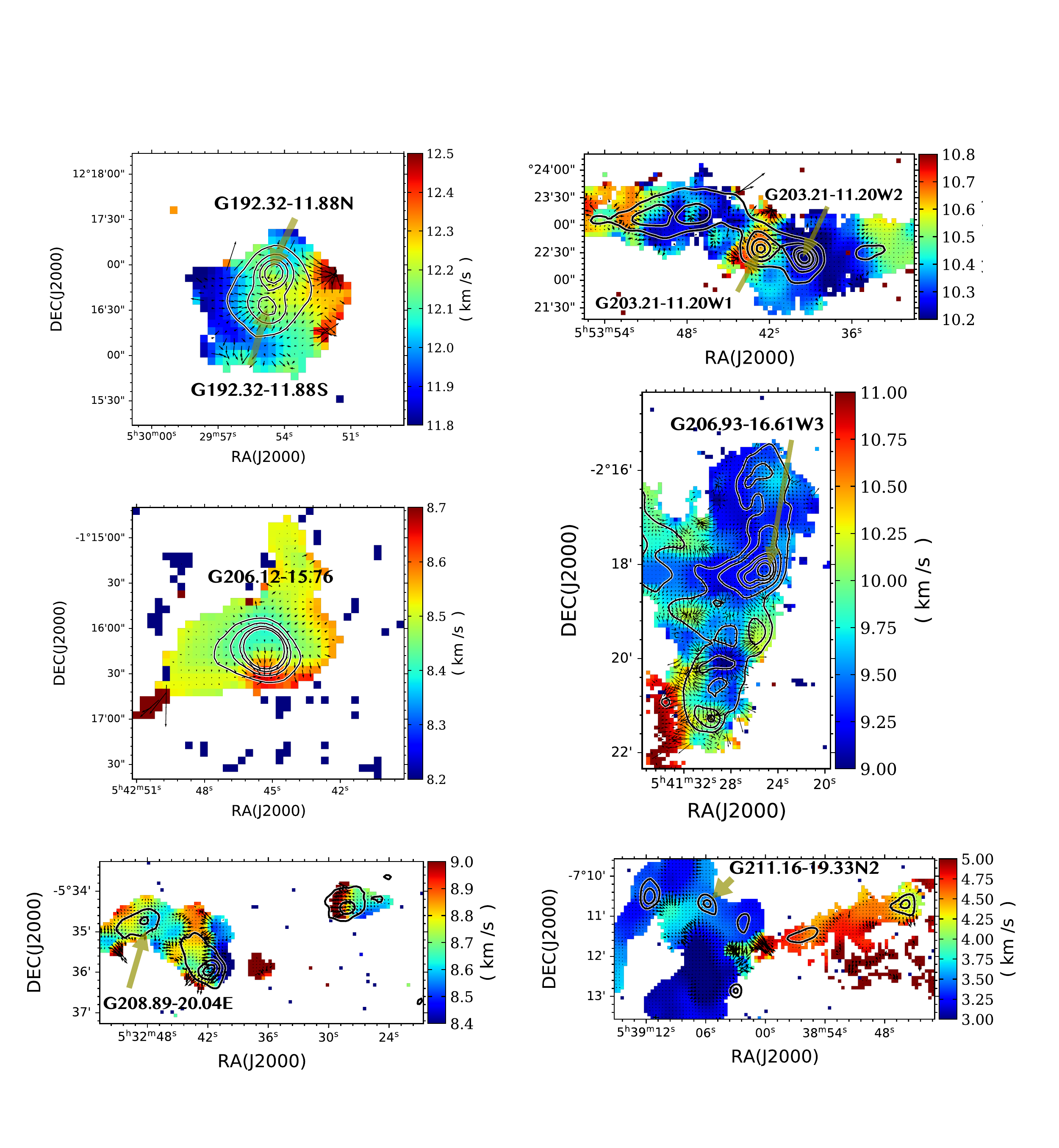}
    \caption{N$_2$H$^+$ velocity maps with contours of the JCMT 850 $\mu$m data. The contour levels are from 5 $\sigma$ to 25 $\sigma$ with steps of 5 $\sigma$. The single-system ALMASOP cores are labeled in black and binary/multiple systems in orange. The arrows show the directions of the local velocity gradients, with the lengths indicating their magnitudes.}
    \label{n2h+_s}
\end{figure*}

\begin{figure*}[ht!]
    \centering
    \includegraphics[width=19cm]{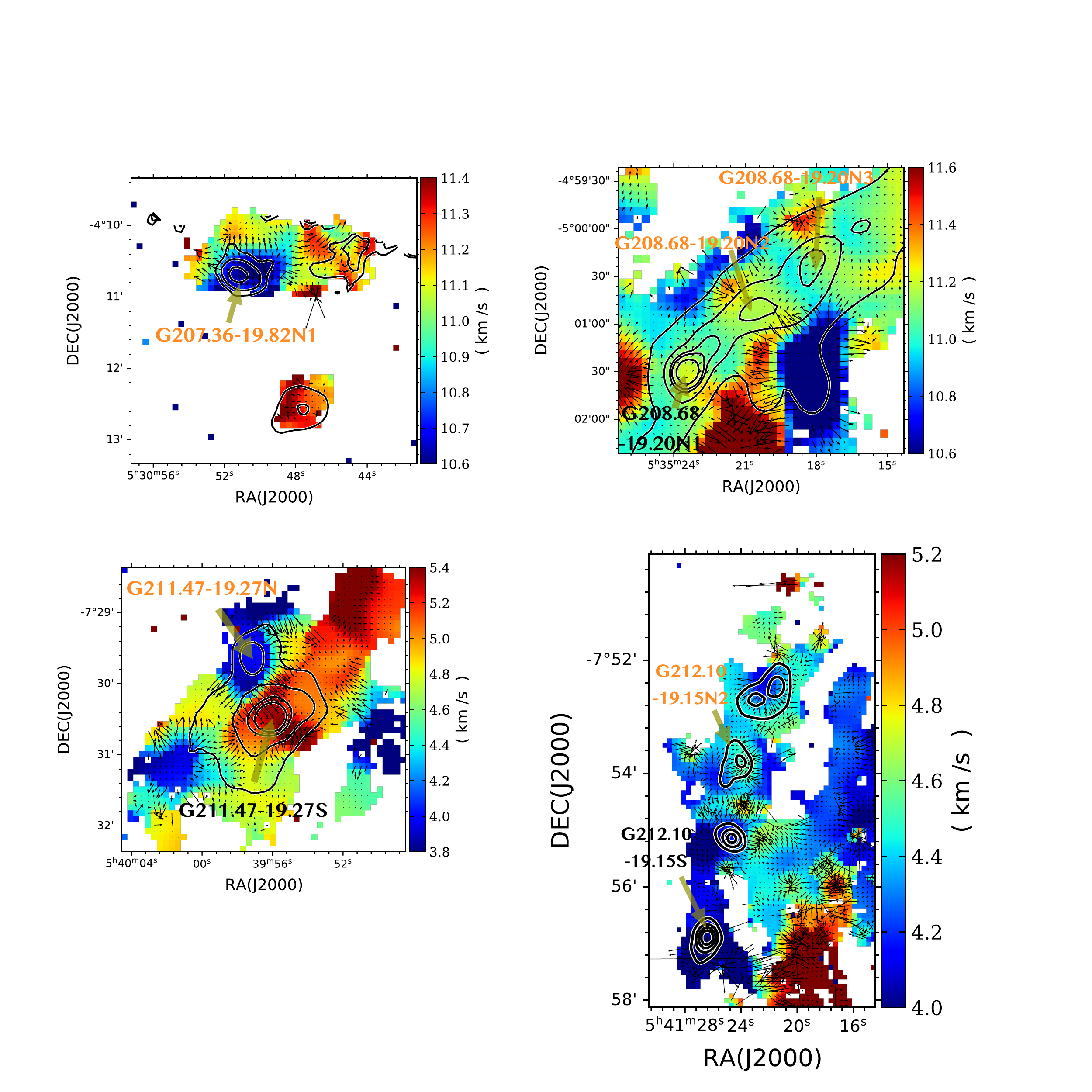}
    \caption{N$_2$H$^+$ velocity maps with contours of the JCMT 850 $\mu$m data. The contour levels are from 5 $\sigma$ to 25 $\sigma$ with steps of 5 $\sigma$. The single-system ALMASOP cores are labeled in black and binary/multiple systems in orange. The arrows show the directions of the local velocity gradients, with the lengths indicating their magnitudes.}
    \label{n2h+_b}
\end{figure*}

There are 16 protostellar cores in the ALMASOP survey that have been mapped in N$_2$H$^+$ J=1-0 line emission with the No: 45\,m telescope \citep{Tatematsu2021}. Among the 16 cores, five are forming binary/multiple systems and the other 11 contain single protostars. We note that the fraction of binary/multiple systems in this small sample is similar to that of the whole ALMASOP sample. With these data, we can investigate whether or not there are significant differences in gas kinematics between the two groups of cores. We fit the hyperfine spectra to all N$_2$H$^+$ emission lines with signal-to-noise ratios (SNR) higher than 3, and finally derive the centroid velocity field maps of these cores. The velocity field maps of dense cores that are forming single systems are shown in Figure \ref{n2h+_s}, while the maps of cores forming binary/multiple systems are shown in Figure \ref{n2h+_b}. Most of the cores show velocity gradients in the N$_2$H$^+$ emission maps, no matter how many protostars have formed within them. 

To evaluate the gas kinematics quantitatively, we derive local velocity gradients across these maps in steps of 10$\arcsec$. The local velocity gradients are shown as arrows on the velocity field maps. In Table \ref{tab:my-table}, we present the maximum, minimum, mean, and standard deviation of the local velocity gradients in these 16 dense cores. The median velocity gradient of the single-system cores ranges from 1.92 to 7.21 km~s$^{-1}$~pc$^{-1}$, with a mean value of 3.94 km~s$^{-1}$~pc$^{-1}$. For comparison, the median velocity gradient of the binary/multiple cores ranges from 1.90 to 7.08 km~s$^{-1}$~pc$^{-1}$, with a mean value of 4.05 km~s$^{-1}$~pc$^{-1}$. We also list the statistics of the velocity gradients in the Table \ref{tab:my-table}. 
These statistics indicate that there is no significant difference in velocity gradient between the two groups of cores with a \textit{p}-value in KS test of 92$\%$. These results, however, need to be tested with a much larger sample of cores from higher angular resolution observations, which may separate more clearly the core kinematics from those of the surroundings. The ALMASOP data, unfortunately, do not have high enough spectral resolution or suitable molecular line tracers for gas kinematics studies on the core scale.

\begin{deluxetable}{lccccc}

\tablecaption{Velocity gradients in dense cores \label{tab:vs}}
\tabletypesize{\footnotesize}
\tablehead{
\colhead{Core Name} & \colhead{Mean } &\colhead{Median }&  \colhead{Min}   & \colhead{Max} & \colhead{Std}\\
 \colhead{} & \colhead{( km~s$^{-1}$~pc$^{-1}$)} &  \colhead{(km~s$^{-1}$~pc$^{-1}$)} &  \colhead{( km~s$^{-1}$~pc$^{-1}$)}  & \colhead{(km~s$^{-1}$~pc$^{-1}$)} & \colhead{(km~s$^{-1}$~pc$^{-1}$)}}
\decimalcolnumbers
\startdata
G192.32-11.88N     & 2.49 & 2.51 & 0.39 & 4.14 & 1.10 \\
G192.32-11.88S     & 1.87 & 1.92 & 1.55& 2.33 & 0.29 \\
G203.21-11.20W1    & 2.42 & 2.07 & 0.48 & 5.57 & 1.57 \\
G203.21-11.20W2    & 3.40 & 2.75 & 1.25 & 7.44 & 1.67 \\
G206.12-15.76      & 3.89 & 4.60 & 0.22 & 8.09 & 2.45\\
G206.93-16.61W3    & 4.04 & 3.60 & 0.27 & 10.99 & 2.53 \\
G208.68-19.20N1    & 5.14 & 5.20 & 1.52 & 9.80 & 2.01 \\
G208.89-20.04E     & 3.36 & 3.50 & 0.99 & 5.23 & 1.02 \\
G211.16-19.33N2    & 4.42 & 4.58 & 2.82 & 5.84 & 0.73 \\
G211.47-19.27S     & 10.80 & 7.21 & 0.13 & 67.64 & 12.05 \\
G212.10-19.15S     & 5.39 & 5.39 & 2.02 & 10.08 & 2.43 \\
\hline
G207.36-19.82N1    & 8.20 & 7.08 & 1.60 & 16.39 & 4.41 \\
G208.68-19.20N2    & 2.12 & 1.90 & 0.49 & 4.67 & 1.22 \\
G208.68-19.20N3    & 4.17 & 4.32 & 0.63 & 8.26 & 1.97 \\
G211.47-19.27N     & 8.57 & 3.62  & 1.02 & 27.99 & 8.86 \\
G212.10-19.15N2    & 3.55 & 3.32 & 0.95 & 8.44 & 1.69 \\
\hline
\enddata
\tablecomments{The eleven cores above the dividing line are those containing only one protostar, and the remaining five binary/multiple cores are those containing binary/multiple systems.}
\end{deluxetable}

\subsection{Separation of Protostars in Binary/Multiple Systems}

\begin{figure}[ht!]
    \centering
    \includegraphics[width=12cm]{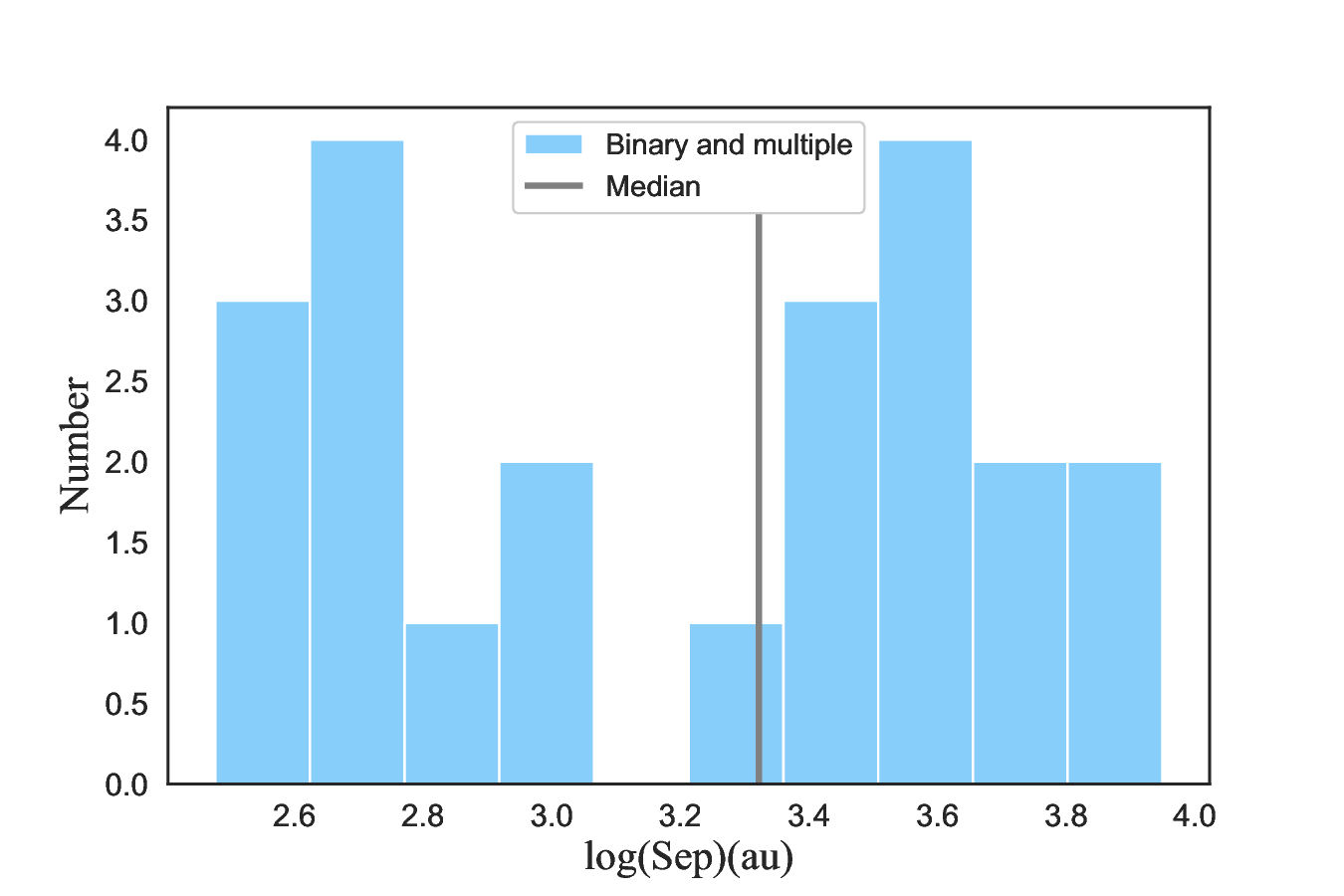}
    \caption{The separations of companion stars are shown as a histogram. The gray vertical line represents the median separation. }
    \label{Sep}
\end{figure}

We derive the projected separations among member protostars, for 36 protostars in the 13 binary/multiple-system cores. For high-order systems, we use the MiSTree (minimum spanning tree) package \footnote{https://joss.theoj.org/papers/10.21105/joss.01721.pdf} to get the distance or separation between protostars. Figure \ref{Sep} shows the distribution of separation in the binary/multiple systems. The projected separation between companion protostars of the whole sample ranges from 300 au to 8900 au, and the median projected separation is about 2100 au. Our result highlights a bimodal behavior in the projected separation distribution with one peak around 500 au and the other peak around 3500 au, similar in character to the bimodal distribution of projected separations presented in \cite{tobin_vla_2016} and \cite{Tobin2022}.

\section{Discussion}\label{sec:discussion}
\subsection{Origin of the multiplicity of protostars in dense cores}

\textbf{
\begin{figure}[ht!]
    \centering
    \includegraphics[width=12cm]{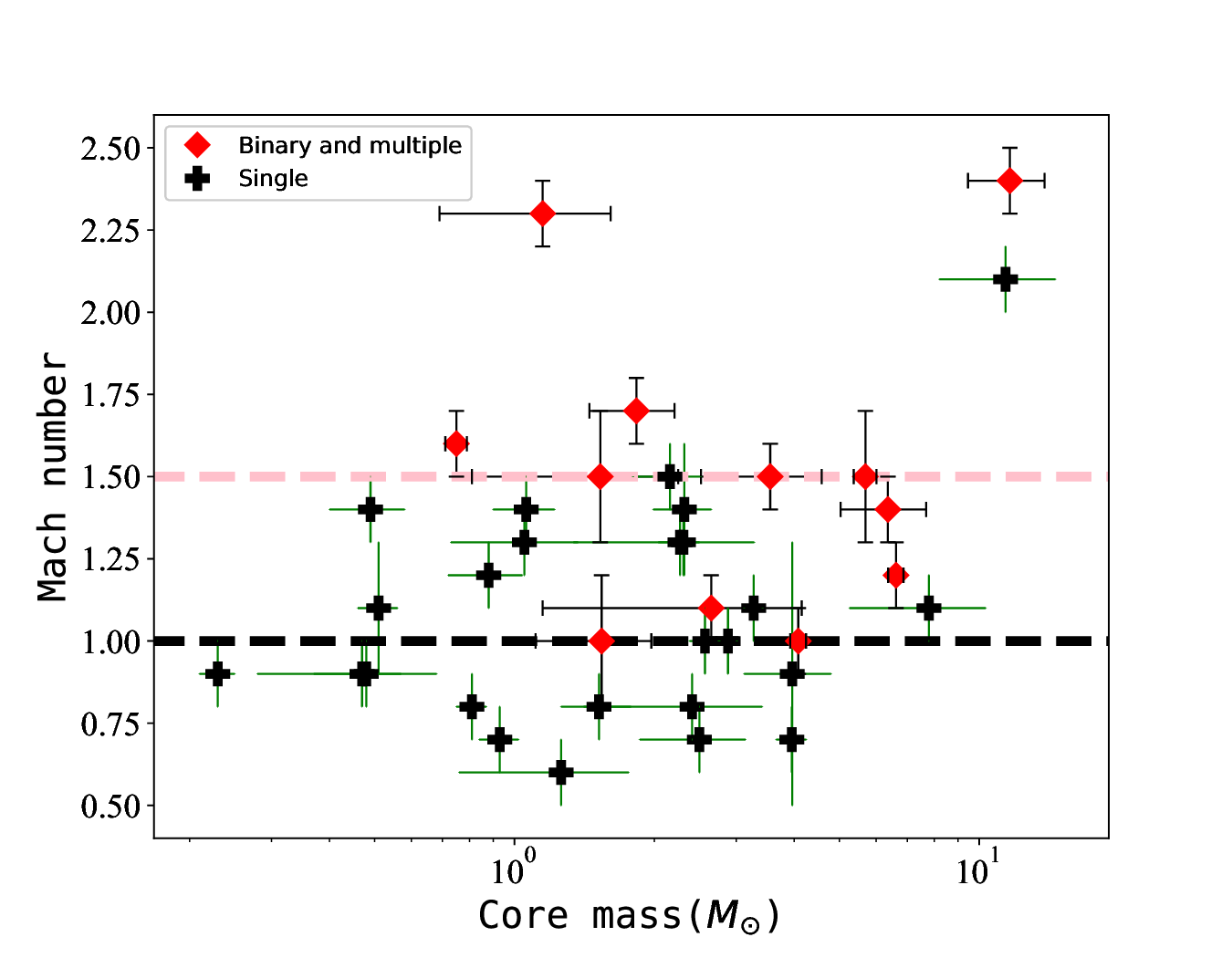}
    \caption{Comparison of core mass and Mach number. The red dots represent binary/multiple systems and the black dots represents single systems. The two dashed lines represent the median Mach numbers of the two core groups.}
    \label{fig:mm}
\end{figure}
}

In Section 4.2, we statistically compare the physical properties of dense cores that form different systems. We find that dense cores forming binary/multiple systems are statistically denser and have higher $\textit{N}(H_2)$ and $n_{H_2}$, and greater turbulence (higher $\mathcal{M}$) than single-system cores. Dense cores with higher densities also show smaller Jeans lengths, and thus are more likely to fragment into substructures that could form a binary/multiple systems.

The higher Mach number of binary/multiple-system cores suggests that their turbulent motions are more prevalent in them than in single-system cores. Higher Mach numbers, however, may not reflect the initial turbulence levels before core fragmentation, because outflows from protostars may have injected energy to induce the observed core turbulence \citep{2021Yun}. However, as shown in Figure \ref{figA2} in the appendix, in most cases, only one protostar in a binary/multiple system is drives a highly collimated and energetic bipolar outflow. In addition, the outflow properties (e.g., outflow velocities, masses, momenta, and energies) are not significantly different in various dense cores forming either single stars or binary/multiple systems \citep[see S. Dutta et al. 2022, in preparation; and][]{dutta_alma_2020-1}. Therefore, the higher Mach numbers in dense cores forming binary/multiple systems may not be caused by outflow activity.

Furthermore, a higher Mach number may also reflect internal core accretion. Statistically, binary/multiple-system cores are more massive than single-system cores while the sizes of the two core groups are indistinguishable. This behavior may indicate that during gravitational collapse, binary/multiple-system cores accumulate more mass through the accretion of gas. The large accretion flows of dense cores may also induce more turbulence inside the cores. However, as shown in Figure~\ref{fig:mm}, Mach numbers are not well correlated with core masses. In general, for dense cores with similar core masses, those cores forming binary/multiple systems have larger Mach numbers than those forming single stars, further suggesting that larger Mach numbers are not associated with larger core masses.

Higher Mach numbers of binary/multiple-system cores could also be related to other bulk motions such as core rotation. Indeed, rotation has been widely revealed in cores of the Orion GMCs \citep{Tatematsu2016,Xu2020}. In particular, hierarchical protostellar systems may form by rotation-driven fragmentation, as demonstrated in early numerical simulations \citep{Boss1987,1991BossN,1992Chapman,Bonnell1992,Whitworth1995}. From the single-dish N$_2$H$^+$ maps shown in Section 4.3, we do see clear velocity gradients across cores. Whether or not these velocity gradients are caused by core rotation, however, needs to be tested with higher spatial and spectral resolution observations. Here at least, we find no significant differences in velocity gradients among cores forming various numbers of stars.

Therefore, higher Mach numbers in dense cores likely arise from their initial turbulence levels, and binary/multiple systems tend to form in more turbulent cores. Significant levels of turbulence in cores can be a key trigger of fragmentation because turbulent fluctuations within a bound core can produce multiple nonlinear perturbations in density that exceed the local Jeans mass and causes the overdense region to collapse faster than the background core \citep{goodwin_simulating_2004,fisher_turbulent_2004,offner_formation_2010}. In addition, turbulent fragmentation has been suggested to be the dominant channel for binary/multiple-system formation of low-mass stars in numerical simulations \citep{offner_formation_2010,Tobin2022}.

To summarize, we find that the multiplicity of protostars that form in dense cores may be determined by the core gas density and Mach number. However, correlations between these parameters and time evolution need more refined analyses in future. We find that the protostars are at different evolutionary stages even in the same system, which makes the comparison more complicated (for more details on the evolutionary stages, see Section 5.3 below).

\subsection{Environmental effects on the multiplicity of protostars}

Figure \ref{fig:mfcsf} shows the multiplicity of the three GMCs: $\lambda$ Orionis, Orion B, and Orion A, of which $\lambda$ Orionis has the lowest MF and CSF. Several previous observations indicated that the physical properties of the three GMCs are markedly diverse. \cite{yi_planck_2018-1} showed that the mean column densities of dense cores in these clouds can be ranked as $\lambda$ Orionis, Orion A, and Orion B, from lowest to highest. Furthermore, cores in $\lambda$ Orionis have the highest dust gas temperatures of the three clouds. Orion A has the most numerous young stars, while Orion B is probably an even younger cloud than Orion A \citep{yi_planck_2018-1,2008ballyb}. The median core volume density of Orion B is greater than that of Orion A, although the temperatures are similar in the two subregions \citep{yi_planck_2018-1}. The higher core volume density reduces the Jeans length in the Orion B cloud, resulting in an overall higher CSF compared to the Orion A cloud. The northern part of Orion A, the ISF, however, stands out in the Orion complex due to its intense star formation \citep{Megeath2016} and extremely high gas density \citep{Hacar2018,Schuller2021,2021Yun}. It is the most massive filament among the Gould Belt clouds \citep{Bally1987,Schuller2021}. The ISF region, with high-mass star formation in its dense center (the ONC), shows the highest MF and CSF among Orion GMCs. In contrast, the southern part of Orion A, L1641, containing only low-mass star formation, shows a comparable MF and smaller CSF when compared with Orion B GMC. These results indicate that binary/multiple systems form more frequently in high-density clouds. 

Meanwhile, the $\lambda$ Orionis cloud, which has a unique ringlike structure, is composed of an H{\sc ii} region and OB associations in the center. The $\lambda$ Orionis dense cores that we observed in the ALMASOP survey are exposed to the strong UV radiation from the H{\sc ii} region, where the ionization front heats the cores. In addition, the shock and energy input to the environment from supernova explosion feedback and stellar feedback from the OB association, for example, UV radiation and photoevaporation, can also influence the star formation process in these bright-rimmed dense cores, consequently limiting the star formation rates therein \citep{cunha1996,maddalena1987,tie2016,kounkel_supernovae_2020,Yi2021}. In simulations, the radiative feedback of OB stars or an H{\sc ii} regions has been shown to be efficient in suppressing fragmentation in dense gas \citep{Myers2013}. However, we note that binary/multiple systems could also form in some dense cores (such as G196.92-10.37 in the B35 cloud) in the $\lambda$ Orionis GMC because these cores are mostly shielded from the UV radiation due to their high density and large mass. Such cores, however, are very rare in the $\lambda$ Orionis GMC \citep{yi_planck_2018-1}, which may explain the overall low formation rate of binary/multiple stars in the $\lambda$ Orionis GMC. In contrast, although the ISF region in Orion A is also exposed to strong UV radiation from OB clusters, the ISF shows a very high MF and CSF because of its extremely high density.

\subsection{Different evolutionary stages of protostars in Binary/Multiple Systems}

The physical properties of protostars in binary/multiple-system cores are discussed in the appendix. We find that the protostars in most multiple systems are not at identical stages of evolution, as judged by their infrared emission. Interestingly, we find that in some systems (such as G196.92-10.37, G205.46-14.56M1, G205.46-14.56S1, G208.68-19.20N2, G208.68-19.20N3 and G210.49-19.79W) infrared-brighter member protostars, which are more evolved, tend to be located farther from the SCUBA-2 core centers than these less evolved member protostars. This may indicate that protostars will migrate out of their natal cores as they evolve. These wide binary/multiple systems are very likely to become gravitationally unbound as they evolve. Previous statistical studies have also found that the MF and CSF decrease from Class 0 to Class I, and to Class II protostars and field stars \citep{2013Chen,Tobin2022}, indicating that a large fraction of binary/multiple systems formed in early evolutionary stages will become unbound finally. The ALMASOP data, however, do not have high enough spectral resolution or suitable molecular line tracers for investigating the kinematics and dynamics of member protostars. Therefore, with the present data, we cannot judge whether the binary/multiple systems are gravitationally bound or not. Future more sensitive and higher spectral resolution molecular line observations with ALMA will help determine the gravitational stability or even orbital motions of these systems. 

Figure \ref{figA2} in the appendix presents the CO outflow maps for these systems. In most cases, only one member protostar in a binary/multiple system drives a highly collimated bipolar outflow, which is commonly seen for Class 0 protostars. The other protostars, however, are not associated with energetic outflows, indicating that they could be at a more evolved phase. This difference also implies that protostars in binary/multiple systems could have very different accretion histories or they could have competed for gas to accrete from their natal cores in their formation process. 

In the turbulent fragmentation model, multiple protostars can form in the same core with unevenly sized substructures created by turbulent fluctuation. In this scenario, the evolution timescale of each member protostar in the same core could also be very different. Therefore, the turbulent fragmentation model can well explain the different evolutionary stages of member protostars in binary/multiple systems, as witnessed here.

\section{Summary}\label{sec:summary}

We studied 43 low-mass protostellar cores in the Orion Molecular Cloud Complex that were observed in ALMASOP. We used 1.3\,mm continuum emission to detect protostars. We investigated the stellar multiplicity of these cores, and analyzed how the different properties of the dense cores may influence the formation of single systems or binary/multiple systems. Our main results are summarized as follows:

(1) From 1.3 mm continuum emission, we identified 13 binary/multiple systems, 1 binary-system candidate, and 29 single-star systems. The overall MF and the CSF of the ALMASOP sample in the Orion complex ($\lambda$ Orionis, Orion B, and Orion A) are 28$\%\pm4\%$ and $51\%\pm6\%$, respectively. Among the Orion GMCs, $\lambda$ Orionis has the lowest multiplicity (MF and CSF), while the northern part of Orion A (the ISF) has the highest MF and CSF.

(2) We have determined the separation of companions in each dense core. The median separation is about 2100 au for the 13 binary/multiple systems. A bimodal distribution is seen in companion separation with one peak at $\sim$500 au and the another peak at $\sim$3500 au.

(3) We found evidence that the properties of the dense core have an effect on binary/multiple-system formation.
Specifically, we suggest that the gas density and Mach number of cores may be key factors that promote binary/multiple-star formation in the early stages of dense core to stellar system evolution.

(4) The low multiplicity in $\lambda$ Orionis may be due to further environmental effects. There, strong UV radiative feedback from the giant H{\sc ii} region in $\lambda$ Orionis may destroy dense cores and suppress the formation of binary/multiple stellar systems.

(5) We found that protostars in each binary/multiple system are usually at very different evolutionary stages. Some protostars drive highly collimated CO outflows and are likely at the Class 0 stage, while other protostars without strong outflows are likely at later stages. More evolved member protostars tend to be located farther from the dense core centers than these less evolved member protostars, indicating that protostars will migrate out of their natal cores as they evolve.

We note that our studies are limited by the small sample, and also limited by the relatively poor spatial and spectral resolution. Our results could be further tested using future higher spatial and spectral resolution observations toward a more complete dense core sample in various molecular clouds that are in widely different environments.

\begin{acknowledgments}

This paper makes use of the following ALMA data: ADS/
JAO.ALMA\#2018.1.00302.S. ALMA is a partnership of ESO
(representing its member states), NSF (USA) and NINS
(Japan), together with NRC (Canada), MOST and ASIAA
(Taiwan), and KASI (Republic of Korea), in cooperation with
the Republic of Chile. The Joint ALMA Observatory is
operated by ESO, auI/NRAO, and NAOJ.

T.L. acknowledges support from the National Natural Science Foundation of China (NSFC) through grants No. 12073061 and No. 12122307, the International Partnership Program of the Chinese Academy of Sciences (CAS) through grant No. 114231KYSB20200009, the Shanghai Pujiang Program (20PJ1415500), and science research grants from the China Manned Space Project with no. CMS-CSST-2021-B06.

K.T. was supported by Japan Society for the Promotion of Science (JSPS) KAKENHI (grant No. 20H05645).

D.J. and J.d.F. are supported by NRC Canada and by NSERC Discovery Grants.

C.-F.L. acknowledge grants from the Ministry of Science and Technology of Taiwan (MoST  107-2119-M-001-040-MY3 and 110-2112-M-001-021-MY3) and Academia Sinica (Investigator Award AS-IA-108-M01).

This research was carried out in part at the Jet Propulsion Laboratory, which is operated by the California Institute of Technology under a contract with the National Aeronautics and Space Administration (80NM0018D0004).

J.-E.L. was supported by a National Research Foundation of Korean (NRF) grant funded by the Korea government (MSIT) (grant No.2021R1A2C1011718).

J.H. acknowledges the support of NSFC projects 11873086 and U1631237. This work is sponsored (in part) by the CAS, through a grant to the CAS South America Center for Astronomy in Santiago, Chile.

S.-L.Q. is supported by the NSFC with grant No. 12033005.

S.Z. acknowledges the support of the China Postdoctoral Science Foundation through grant No. 2021M700248.

L.B. gratefully acknowledges support by the ANID BASAL projects ACE210002 and FB210003.

P.S. was supported by a Grant-in-Aid for Scientific Research (KAKENHI No. 18H01259) of JSPS. 

V.-M.P. acknowledges support by the grant PID2020-115892GB-I00 funded by MCIN/AEI/10.13039/501100011033.

\end{acknowledgments}

\vspace{5mm}
\facilities{ALMA, JCMT, No: 45m}

\software{astropy \citep{2013Astropy},  
        CASA \citep{2007McMullin},
        Starlink \citep{Currie2014},
        MiSTree \citep{Naidoo2019}
          }



\appendix

\section{APPENDIX: Description of individual Binary/Multiple systems  } \label{AppA}

In this appendix, we provide maps for the 14 dense cores containing binary/multiple systems and discuss the properties of member protostars based on near- and mid-infrared data from Spitzer (3.6 $\mu$m, 4.5 $\mu$m and 8 $\mu$m) or WISE (3.4 $\mu$m, 4.6 $\mu$m and 12 $\mu$m) data, JCMT 850 $\mu$m data, ALMA 1.3\,mm  dust continuum data, and CO outflows. Figure \ref{figA1} shows the infrared and dust continuum images of these systems, and each row of Figure \ref{figA1} shows one source. Figure \ref{figA2} presents their CO outflow maps. Table \ref{tab:tobin} lists the ALMASOP sources that were also observed in the VANDAM survey.

\begin{deluxetable*}{llc}

\tablecaption{ALMASOP Sources observed in the VANDAM survey\label{tab:tobin}}
\tablehead{
\colhead{ALMASOP sources} & \colhead{HOPS}  & \colhead{Class of HOPS}\\
}
\decimalcolnumbers
\startdata
& Binary/Multiple Systems &\\
\hline
G205.46-14.56M1$\_$A & HOPS-317-A & 0\\
G205.46-14.56M1$\_$B & HOPS-317-B& 0\\
G205.46-14.56M2$\_$A & HOPS-387-B& I\\
G205.46-14.56M2$\_$B & HOPS-387-A& I \\
G205.46-14.56M2$\_$C & HOPS-386-A& I\\
G205.46-14.56M2$\_$D & HOPS-386-B& I\\
G205.46-14.56M2$\_$E & HOPS-386-C& I\\
G205.46-14.56S1$\_$A & HOPS-358-A& 0\\
G205.46-14.56S1$\_$B & HOPS-358-B& 0\\
G206.93-16.61E2$\_$A & HOPS-298-A& I\\
G206.93-16.61E2$\_$B & HOPS-298-B& I\\
G208.68-19.20N2$\_$B & HOPS-89& Flat\\
G208.68-19.20N3$\_$B & HOPS-92-A-A& Flat\\
G208.68-19.20N3$\_$C & HOPS-92-B& Flat\\
G208.68-19.20S$\_$A & HOPS-84-A& I\\
G208.68-19.20S$\_$B & HOPS-84-B& I\\
G209.55-19.68N1$\_$B & HOPS-12-B-A& 0\\
G209.55-19.68N1$\_$C & HOPS-12-A& 0\\
G210.49-19.79W$\_$A & HOPS-168-A & 0\\
G210.97-19.33S2$\_$A & HOPS-377& 0 \\
G210.97-19.33S2$\_$B & HOPS-144 & I\\
G211.47-19.27N$\_$A & HOPS-290-B& 0 \\
G211.47-19.27N$\_$B & HOPS-290-A& 0 \\
G212.10-19.15N2$\_$A& HOPS-263& I \\
G212.10-19.15N2$\_$B& HOPS-262& Flat\\
\hline
& Single-star Systems & \\
\hline
G205.46-14.56N1 & HOPS-402& 0\\
G205.46-14.56N2 & HOPS-401& 0\\
G205.46-14.56S2 & HOPS-385& Flat\\
G205.46-14.56S3 & HOPS-315& I\\
G206.12-15.76*& HOPS-400-B& 0\\
G206.93-16.61W2& HOPS-399& 0\\
G209.55-19.68S1& HOPS-11& 0\\
G209.55-19.68S2& HOPS-10& 0\\
G210.37-19.53S& HOPS-164& 0\\
G210.83-19.47S& HOPS-156& I\\
G211.01-19.54N& HOPS-153& 0\\
G211.01-19.54S& HOPS-152& 0\\
G211.16-19.33N2& HOPS-133& I\\
G211.47-19.27S*& HOPS-288-A-A& 0\\
G212.10-19.15S& HOPS-247& 0\\
G212.84-19.45N& HOPS-244& 0\\
\hline
\enddata
\tablecomments{The two sources marked with * are identified in the VANDAM survey as binary/multiple systems. HOP-400 contains two protostars, A and B, are separated by 180 au. HOPS-288 contains three protostars; the remaining two protostars are 60 au and 250 au away from HOPS-288-A-A.}
\end{deluxetable*}

\subsection{G196.92-10.37}

G196.92-10.37 is located in an isolated dark nebula named B35A (LDN 1594), which is part of the $\lambda$ Orionis molecular cloud. Several YSOs were found in B35A through the 37\,m telescope of the Haystack Observatory and the 43-m telescope of NRAO \citep{1989Benson}. \cite{2008Connelley} discovered a protobinary system using the UH 2.2\,m infrared telescope near the dense core. 
The Two Micron All Sky Survey (2MASS) and the Spitzer Legacy Program found two YSO candidates, which are identified as protostars B and C in this dense core in our observation \citep{20032mass,2003Evans,2013Nobel}.

As shown in Figure \ref{Figure 1} (a), G196.92-10.37 is a cometary bright-rimmed core that is forming a small cluster. From the zoomed-in image in Figure \ref{Figure 1} (b), we find three infrared point sources in the ALMA FOV. The eastern infrared point source is further resolved into a close binary (B and C) in ALMA observations, as shown in Figure \ref{Figure 1} (c). The southwest infrared point source is also detected in the ALMA 1.3 mm continuum, and is labeled as A in Figure \ref{Figure 1} (c). The infrared point source that is located to the north of A has not been detected in 1.3\,mm continuum emission, indicating that it may be a foreground/background star or a very evolved protostar.

The three protostars detected by ALMA are at very different phases of evolution \citep{dutta_alma_2020-1}. 
G196.92-10.37$\_$A is a Class 0 source with an envelope+disk mass of 0.069 $\pm$ 0.029 $\textit{M}_\odot$. The ALMA 1.3 mm dust continuum observations show an extended structure around this protostar.
The G196.92-10.37$\_$B protostar is in the Class I phase, and its envelope+disk mass is 0.042 $\pm$ 0.018 $\textit{M}_\odot$. The separation between A and B is $\sim$4700 au. 
G196.92-10.37$\_$C is a Class I protostar with a low thermal luminosity and an envelope+disk mass of 0.005 $\pm$ 0.002 $\textit{M}_\odot$. B and C form a close binary with a projected separation of only $\sim$300 au. The two Class I protostars B and C are located away from the center of the SCUBA-2 core. In contrast, the youngest protostar A is located at the center.

As shown in Figure \ref{figA2}, G196.92-10.37$\_$A drives a strong and large-scale wide-angle outflow, while the outflows detected in the vicinity of 196.92-10.37$\_$B and G196.92-10.37$\_$C are very weak.

\subsection{G205.46-14.56M1}

G205.46-14.56M1 is located in the L1630 region of Orion B. Many Herbig-Haro objects such as HH 24-26 YSOs have been found near the dense core \citep{2001Phillips}. 
Among them, HH 24 MMS is identified as a binary system driving a bipolar jet discovered by VLA. The 6.9\,mm continuum emission reveals that the protostellar system is still in the early phase \citep{2008Kang}. The separation of the companions is about 360 au. 
In addition, the VISTA/VIRCAM near-infrared survey and JCMT SCUBA-2 survey have found some additional YSOs around the core \citep{2015Spezzi,kirk_jcmt_2016-1}.

Figure \ref{figA1} presents maps for G205.46-14.56M1. Two protostars (A and B) are identified in our ALMA observation, as shown in panel (c). 
The G205.46-14.56M$\_$B, located at the center of the SCUBA-2 core, is not detected at infrared wavelengths, indicating that it is a prestellar core or in an extremely early stage of evolution. As shown in Figure \ref{figA2}, this object is not associated with an apparent outflow, further suggesting that it is more likely a prestellar core. Its total gas mass is 2.245 $\pm$ 0.960 $\textit{M}_\odot$.
G205.46-14.56M1$\_$A is a Class I protostar with an envelope+disk mass of 0.064 $\pm$ 0.028 $\textit{M}_\odot$. It is near the edge of the SCUBA-2 core. However, it is still drives a very collimated outflow as shown in Figure \ref{figA1}. The distance between these two objects is $\sim$2300 au.

\subsection{G205.46-14.56M2}

G205.46-14.56M2 is also located in the L1630 region of Orion B. Some YSOs have been previously identified by 2MASS, Spitzer, and JCMT SCUBA surveys \citep{20032mass,2012Megeath,kirk_jcmt_2016-1}.
This dense core is surrounded by a well-known HH 24 complex and the two Class 0 protostars SSV63E and SSV63W \citep{ozawa_detection_1999}.

In G205.46-14.56M2, we find a rare multiple system containing five protostars. Among them, A$/$B and C$/$D seem to form a pair of twin binary systems. From the ALMA 1.3\,mm dust continuum image, it seems that the four protostars are not connected to protostar E, as shown in Figure \ref{figA1} (c). G205.46-14.56M2$\_$A and G205.46-14.56M2$\_$B are the closest companions of all, with a separation of only $\sim$700 au. G205.46-14.56M2$\_A$ is in unknown stage and G205.46-14.56M2$\_$B is in the Class I phase with envelope+mass masses of 0.037 $\pm$ 0.016 $\textit{M}_\odot$ and 0.124 $\pm$ 0.053 $\textit{M}_\odot$, respectively. 
G205.46-14.56M2$\_$C and G205.46-14.56M2$\_$D are relatively close, and their separation is $\sim$1000 au. G205.46-14.56M2$\_$C is in unknown stage and G205.46-14.56M2$\_$D is in the Class I phase with envelope+disk masses of 0.091 $\pm$ 0.039 $\textit{M}_\odot$ and 0.028 $\pm$ 0.01 2 $\textit{M}_\odot$. The distance between these two pairs is $\sim$3800 au.
In our analysis, we find that G205.46-14.56M2$\_$E, located $\sim$5000 au away from the other four protostars, has an unknown evolutionary stage with an envelope+disk mass of 0.011 $\pm$ 0.005  $\textit{M}_\odot$.

As shown in Figure \ref{figA2}, G205.46-14.56M2$\_B$ and G205.46-14.56M2$\_$D drive weak but well-collimated outflows. G205.46-14.56M2$\_$A and G205.46-14.56M2$\_$C are also associated with weak outflows. 

This multiple system will be discussed in a detailed paper by T. Liu et al. (2022, in preparation)

\subsection{G205.46-14.56S1}

G205.46-14.56S1 is located in the HH 24-26 region of Orion B. HH 25MMS, a Class 0 source studied extensively by the JCMT, VLA and IRAM, is situated near this dense core \citep{1998Gibb,1995Bontemps,1999Lis,2001Phillips,2004Gibb,2013Chen}.
In addition, several YSOs were discovered in the vicinity by Spitzer, AKARI, and JCMT SCUBA surveys \citep{2012Megeath,2013Nobel,2015kang,furlan_herschel_2016,kirk_jcmt_2016-1}.

As shown in Figure \ref{figA1} (a), no infrared sources are found at the center of the 850 $\mu$m continuum emission map. Two protostars, however, are identified in the our ALMA observation. G205.46-14.56S1$\_$A and G205.46-14.56S1$\_$B are in the Class 0 phase with a separation of $\sim$5200 au and their envelope+disk masses are 0.151 $\pm$ 0.065  $\textit{M}_\odot$ and 0.391 $\pm$ 0.167 $\textit{M}_\odot$, respectively.

Figure \ref{figA2} shows a very collimated outflow traced by $^{12}$CO that is driven by G205.46-14.56S1$\_$B. On the other hand, G205.46-14.56S1$\_$A is drives a much weaker outflow, which seems to overlay the outflow from G205.46-14.56S1$\_$B.

\subsection{G206.93-16.61E2}

G206.93-16.61E2 in Orion B is found close to the reflection nebula NGC 2023.
Several YSOs have been discovered in JCMT SCUBA-2, Spitzer, and 2MASS all-sky surveys near G206.93-16.61E2 \citep{20032mass,2012Megeath,kirk_jcmt_2016-1}.

As shown in Figure \ref{figA1} (a), a bright infrared source is detected in the ALMA FOV. In the ALMA image, as shown in Figure \ref{figA1} (c), the infrared source is resolved into four protostars in this system. Since these four protostars cannot be separated in infrared band, their evolutionary stages are not well determined. Their envelope+disk masses are 0.280 $\pm$ 0.120  $\textit{M}_\odot$, 0.112 $\pm$ 0.048  $\textit{M}_\odot$, 0.219 $\pm$ 0.097  $\textit{M}_\odot$ and 0.252 $\pm$ 0.110 $\textit{M}_\odot$, respectively.
Three protostars (A, C and D) are close to one another with an average separation of $\sim$450 au, while the other protostar G206.93-16.61E2$\_$B is approximately 1000 au away from them. From Figure \ref{figA2}, one can see that the outflows of this core are very complicated, which seems to be caused by the orbital rotation of the member protostars in the multiple system. More details of this system will be presented in a following paper (Q. Luo et al. 2022, in preparatioin).

\subsection{G207.36-19.82N1}

G207.36-19.82N1 is located in Orion A. In the VISTA Orion A survey, a protostar candidate named 2MASS J05305129-0410322 was identified, which we regard as G207.36-19.82N1$\_$B ALMA data observation \citep{2018Meingast}. An HH 58 object was also discovered near the core \citep{tatematsu_astrochemical_2017}.

Figure \ref{figA1} (a) shows an infrared source slightly offset from the center of the 850 $\mu$m continuum, implying that the evolutionary stage of G207.36-19.82N1$\_$B may be similar to that of G207.36-19.82N1$\_$A. In our ALMA observation, we find a binary system in this dense core. The two protostars have envelope+disk masses of 0.113 $\pm$ 0.048  $\textit{M}_\odot$ and 0.011 $\pm$ 0.005  $\textit{M}_\odot$ and the separation between them is $\sim$5400 au. As shown in Figure \ref{figA1} (c), G207.36-19.82N1$\_$A shows a flattened disklike structure.

Figure \ref{figA2} reveals a high-velocity outflow with asymmetric structures in the vicinity of G207.36-19.82N1$\_$A, while G207.36-19.82N1$\_$B is not associated with an outflow. Both of them are likely at the Class II phase or an even later phase.

\subsection{G208.68-19.20N2}

G208.68-19.20N2 is located in Orion A and some YSOs were discovered to be associated with it in JCMT SCUBA and Green Bank Telescope 3.3\,mm continuum emission observation \citep{2008DiFrancesco,2014Schnee}. 

A protobinary system candidate containing a starless core and a protostar is found in this dense core based on our ALMA 1.3\,mm dust continuum observations.
The starless core G208.68-19.20N2$\_$A is very dense and will collapse to form a new protostar \citep{sahu_alma_2021}.
The protostar G208.68-19.20N2$\_$B is not associated with an outflow, indicating that it is likely an evolved Class I protostar. The envelope+disk mass of this protostar is just 0.009 $\pm$ 0.001 $\textit{M}_\odot$.

\subsection{G208.68-19.20N3}

G208.68-19.20N3 is located in a filament in Orion A. As shown in Figure \ref{figA1} (b), our ACA 1.3\,mm continuum observations resolve the core into two subcores. The southern subcore is associated with an infrared object, which is further resolved into a close binary in our high-resolution ALMA observations shown in Figure \ref{figA1} (c). In total, the ALMA 1.3\,mm emission continuum reveals three protostars in this core. G208.68-19.20N3$\_$A is a Class 0 protostar and its distance from the other two protostars is $\sim$6000 au. G208.68-19.20N$\_$B and G208.68-19.20N$\_$C are only $\sim$600 au apart. Both of them are Class I protostars, and located offset from the center of the SCUBA-2 core. These three protostars have the envelope+disk masses of 0.436 $\pm$ 0.189  $\textit{M}_\odot$, 0.078 $\pm$ 0.033  $\textit{M}_\odot$, 0.093 $\pm$ 0.040  $\textit{M}_\odot$, respectively. Figure \ref{figA2} shows a collimated outflow from G208.68-19.20N$\_$A that is perpendicular to its extended disklike structure. The two remaining protostars are also associated with outflows but most of their outflows are located beyond our FOV.

\subsection{G208.68-19.20S}

G208.68-19.20S is in Orion A, and several protostars have been discovered there in the AzTEC 1.1\,mm survey \citep{shimajiri_catalog_2015}.

As shown in Figure \ref{figA1} (c), two Class I protostars are found in our ALMA observations, and they are close to each other with a separation of only $\sim$300 au. Their envelope+disk masses are 0.421 $\pm$ 0.180 $\textit{M}_\odot$ for the protostar A and 0.043 $\pm$ 0.061  $\textit{M}_\odot$ for the protostar B. As shown in Figure \ref{figA2}, the binary system drives an outflow along the east-west direction and the outflow is much weaker than other Class 0 outflows in the ALMASOP sample. Moreover, there is no obvious red-shift emission of the high-speed component of $^{12}$CO around the protostars. 

\subsection{G209.55-19.68N1}

G209.55-19.68N1 is in Orion A, and a young star named HOPS-12 has been discovered near this dense core \citep{furlan_herschel_2016}. 


From ALMA data, three protostars are detected in this dense core. Figure \ref{figA1} (c) shows two substructures encasing the three protostars.
G209.55-19.68N1$\_$B and G209.55-19.68N1$\_$C reside in the same substructure and are about 400 au apart. G209.55-19.68N1$\_$A lies in another substructure and is separated from G209.55-19.68N1$\_$B by $\sim$4800 au.
Their envelope+disk masses are 0.141 $\pm$ 0.060  $\textit{M}_\odot$, 0.061 $\pm$ 0.027  $\textit{M}_\odot$, 0.132 $\pm$ 0.057 \textbf{M$_\odot$} for components A, B and C, respectively.

As shown in Figure \ref{figA2}, G209.55-19.68N1$\_$A drives a very collimated outflow. The red lobe of outflow from G209.55-19.68N1$\_$B, however, is distinctly distorted. Moreover, the CO emission near G209.55-19.68N1$\_$C is weak and may be contaminated by the outflow from G209.55-19.68N1$\_$B.

\subsection{G210.49-19.79W}

G210.49-19.79W is in Orion A.
In the vicinity of the dense core, past VLA observations have detected the VLA 4 source with an H$_2$O maser \citep{Rodrguez2000}.

There are two protostars in different evolutionary stages in the dense core.
G210.49-19.79W$\_$A is proved to be in Class 0 and its envelope+disk mass is 0.201 $\pm$ 0.086 $\textit{M}_\odot$. G210.49-19.79W$\_$B is at a more evolved stage and its envelope+disk mass is 0.008 $\pm$ 0.003 $\textit{M}_\odot$. The separation of one from the other is $\sim$2800 au. The younger protostar G210.49-19.79W$\_$A is located at the center of the SCUBA-2 core, while the older protostar is clearly offset from the core center. 

As shown in Figure \ref{figA2}, G210.49-19.79W$\_$A drives a collimated wide-angle outflow. We do not detect $^{12}$CO emission in the surroundings of G210.49-19.79W$\_$B, which may imply that G210.49-19.79W$\_$B has stopped accretion.

\subsection{G210.97-19.33S2}

G210.97-19.33S2 is located in Orion A, and we have identified a protobinary system in its center. G210.97-19.33S2$\_$A is proved be an Class 0 protostar and G210.97-19.33S2$\_$B is an Class I protostar. Its envelope+disk masses are 0.020 $\pm$ 0.009 $\textit{M}_\odot$ and 0.018 $\pm$ 0.008 $\textit{M}_\odot$ respectively. The separation between the two protostars (A and B) is $\sim$3400 au. Furthermore, we have discovered two additional protostars in the outer part of the core: G210.97-19.33S2$\_$C and G210.97-19.33S2$\_$D.
Figure \ref{figA1} (c) shows that G210.97-19.33S2C and G210.97-19.33S2$\_$D are more widely separated from the core center. Their position is near the edge of the FOV, and their emission is weak.

According to the high-velocity $^{12}$CO emission map in Figure \ref{figA2}, we have only detected outflows near G210.97-19.33S2$\_$A. The CO emission around G210.97-19.33S2$\_$B is very faint. G210.97-19.33S2$\_$C and G210.97-19.33S2$\_$D do not exhibit any line emission.

\subsection{G211.47-19.27N}

G211.47-19.27N is a faint SCUBA-2 850 $\mu$m source to the north of a star cluster in Orion A. Several YSOs have been found in its surroundings by Spitzer observations and the Bolocam Galactic Plane Survey \citep{2012Megeath,2015MErello}. The infrared source associated with G211.47-19.27N is very faint and diffuse, as shown in Figure \ref{figA1} (a) and Figure \ref{figA1} (b). As shown in Figure \ref{figA1} (c), two protostars are identified in the ALMA data, and they share a common envelope. The distance between the two protostars is $\sim$300 au, indicating that they are forming a close binary system. These two protostars are in the Class 0 phase, and their envelope+disk masses are 0.067 $\pm$ 0.030 $\textit{M}_\odot$ and 0.039 $\pm$ 0.019 $\textit{M}_\odot$, respectively.

From Figure \ref{figA2}, we can see that the two Class 0 protostars drive powerful outflows, but it is impossible to identify the outflow-driving source.

\subsection{G212.10-19.15N2}

G212.10-19.15N2 is in Orion A. Two YSOs were detected in this core in Spitzer and 2MASS all-sky observations \citep{20032mass,2012Megeath}.

We have identified two Class I protostars in this dense core. G212.10-19.15N2$\_$A has an envelope+disk mass of 0.034 $\pm$ 0.014 $\textit{M}_\odot$, and G212.10-19.15N2$\_$B has an envelope+disk mass of 0.014 $\pm$ 0.006 $\textit{M}_\odot$. They are $\sim$2800 au apart. According to Figure \ref{figA2}, neither of these two protostars exhibits a clear outflow.

\clearpage

\begin{figure*}
\begin{minipage}[t]{0.9\linewidth}
  \vspace{5pt}
  \centerline{\includegraphics[width=1.085\linewidth]{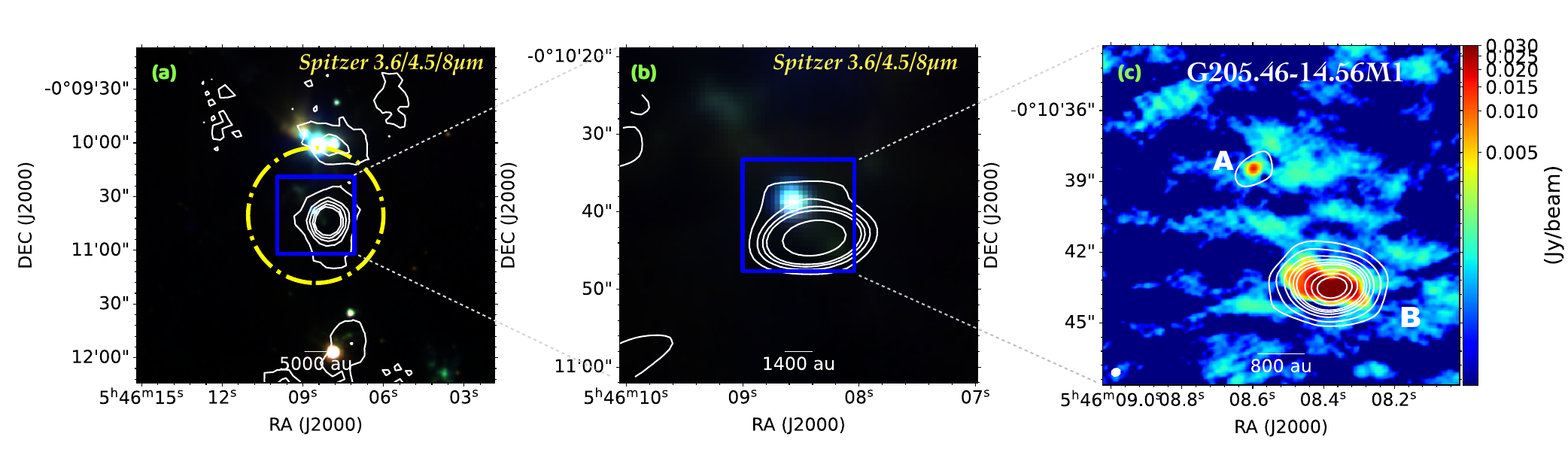}}
  \vspace{5pt}
  \centerline{\includegraphics[width=1.085\linewidth]{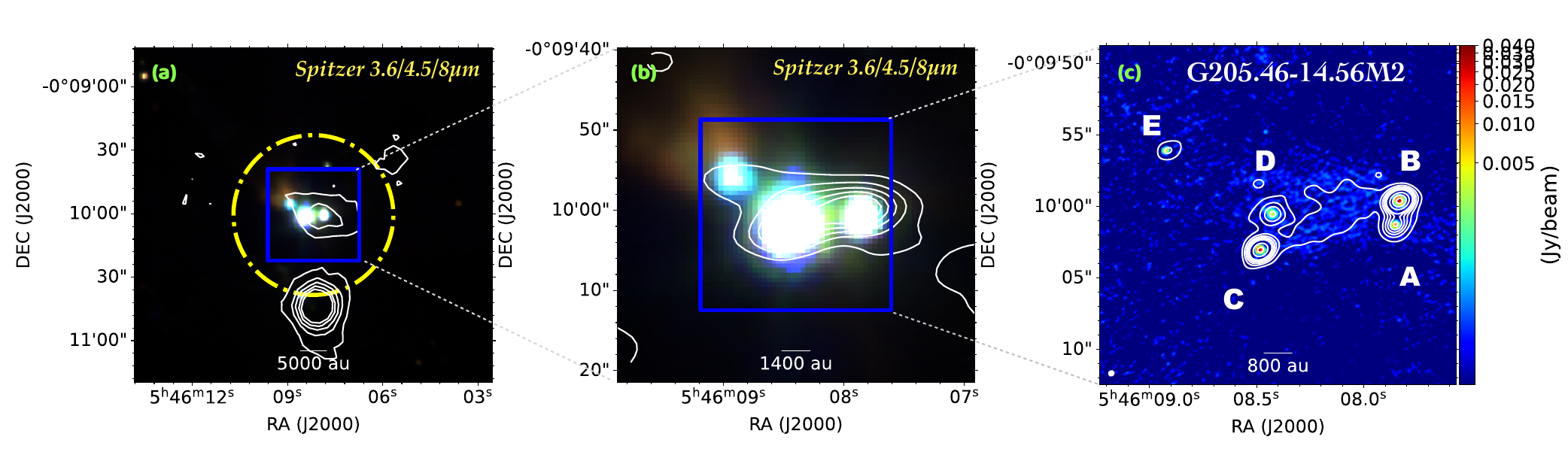}}
  \vspace{5pt}
  \centerline{\includegraphics[width=1.085\linewidth]{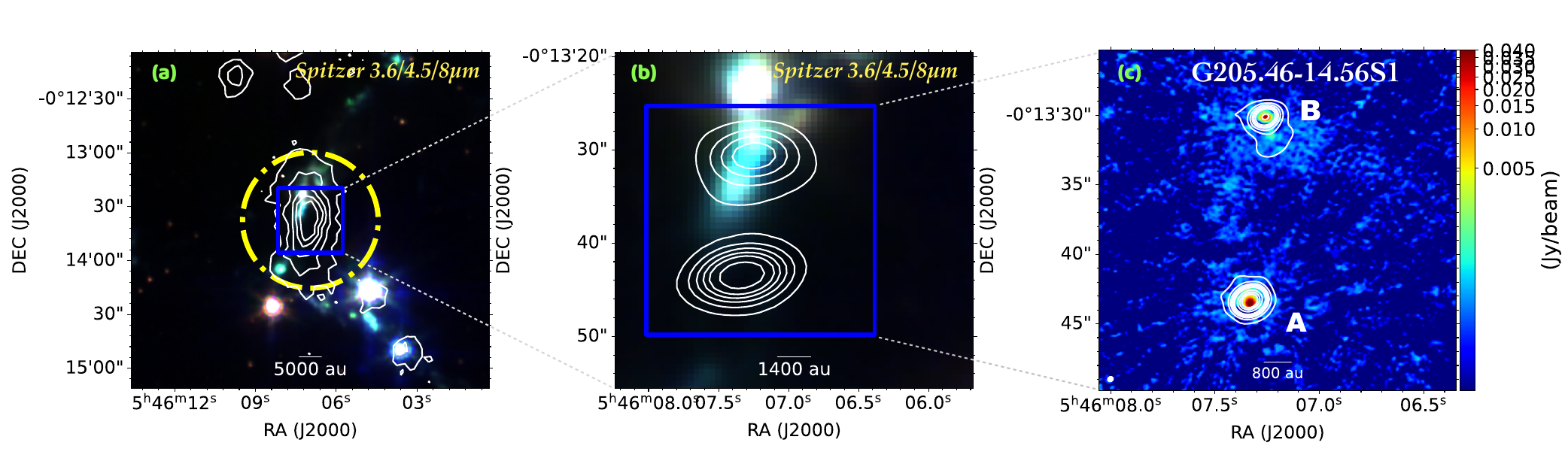}}
      \vspace{5pt}
  \centerline{\includegraphics[width=1.085\linewidth]{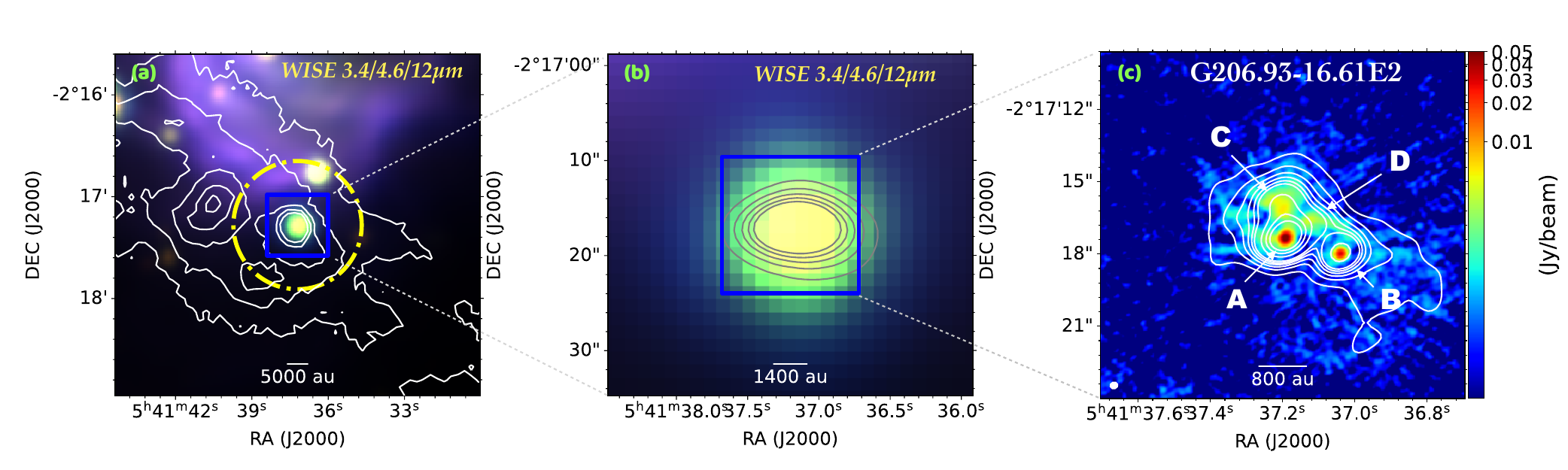}}
     \end{minipage}
  \caption{Same as in Figure 1 but for other sources. Each row presents maps for one source, and the source name is labeled in panel (c).}
\label{figA1}
\end{figure*}

\setcounter{figure}{7}

\begin{figure*}
\begin{minipage}[t]{0.9\linewidth}
  \vspace{5pt}
  \centerline{\includegraphics[width=1.085\linewidth]{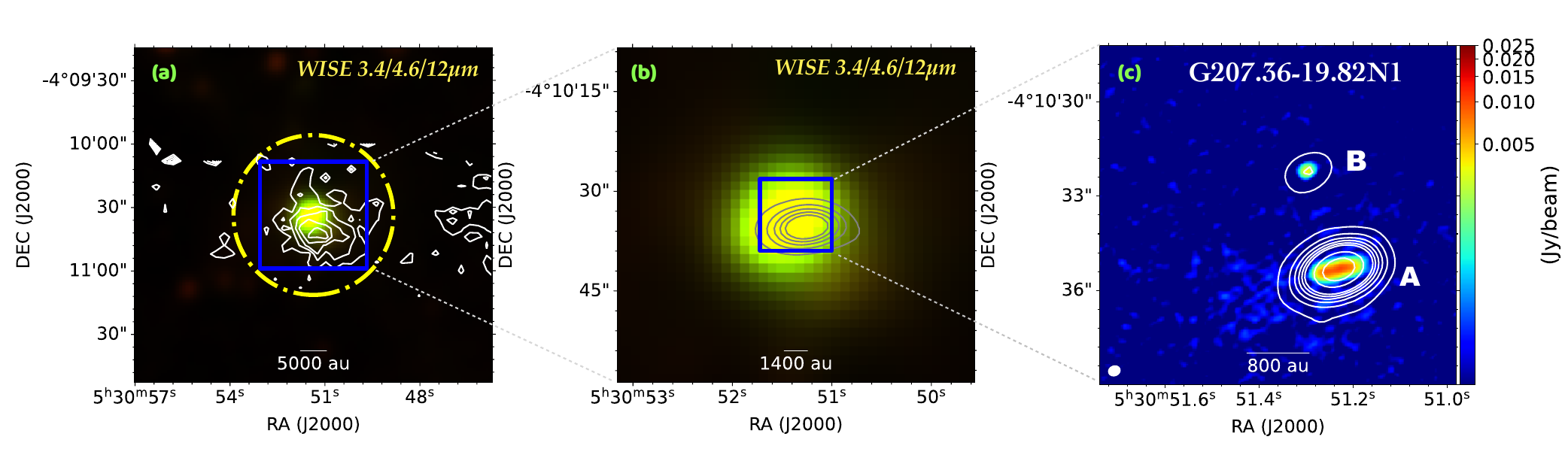}}
  \vspace{5pt}
  \centerline{\includegraphics[width=1.085\linewidth]{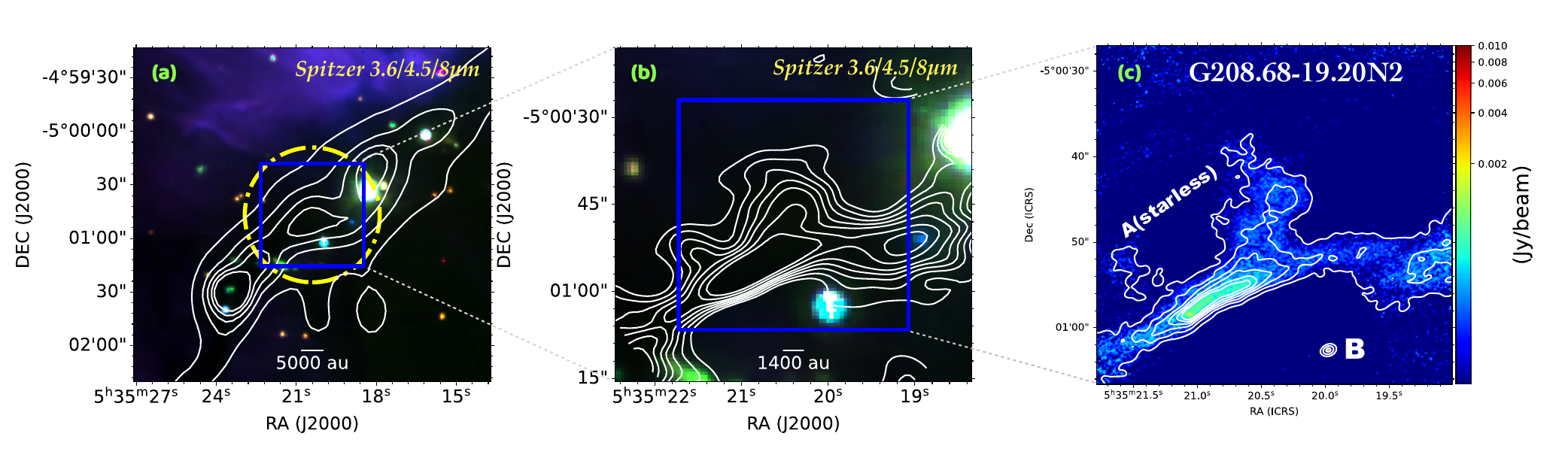}}
      \vspace{5pt}
  \centerline{\includegraphics[width=1.085\linewidth]{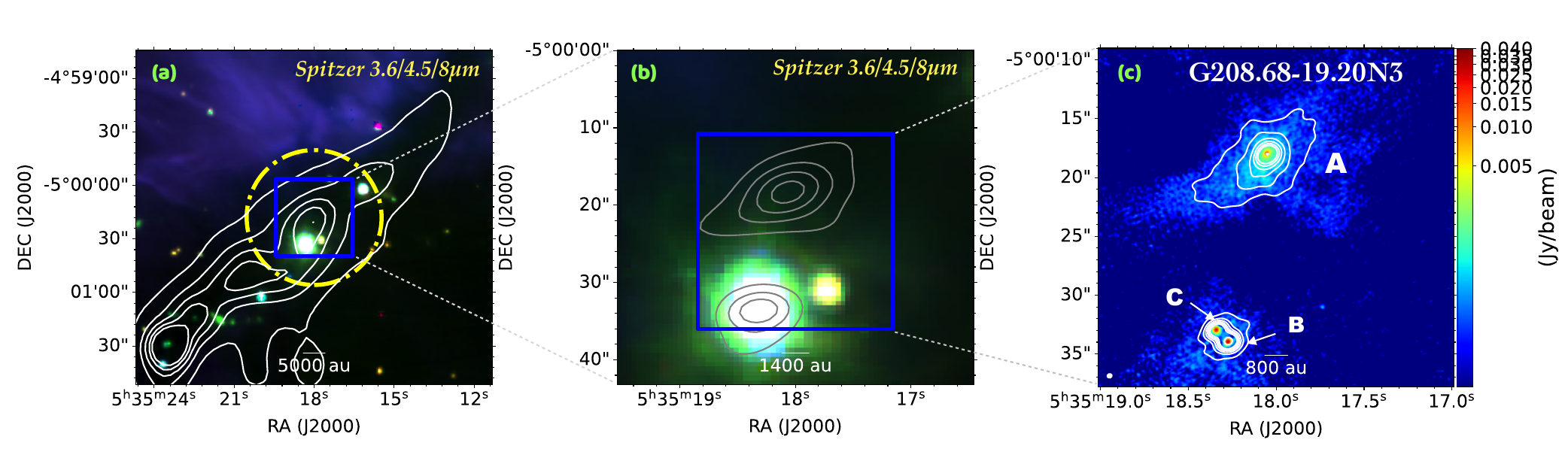}}
      \vspace{5pt}
  \centerline{\includegraphics[width=1.085\linewidth]{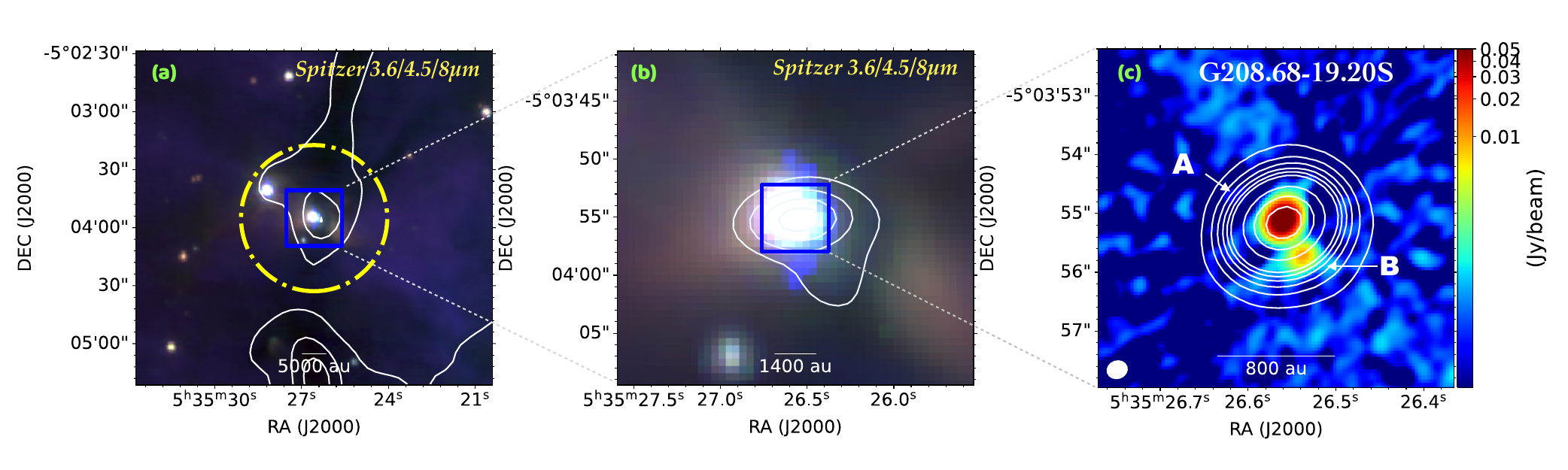}}
  \end{minipage}
  \caption{Continued}
\label{figA1}
\end{figure*}

\setcounter{figure}{7}

\begin{figure*}
\begin{minipage}[t]{0.9\linewidth}
  \vspace{5pt}
  \centerline{\includegraphics[width=1.085\linewidth]{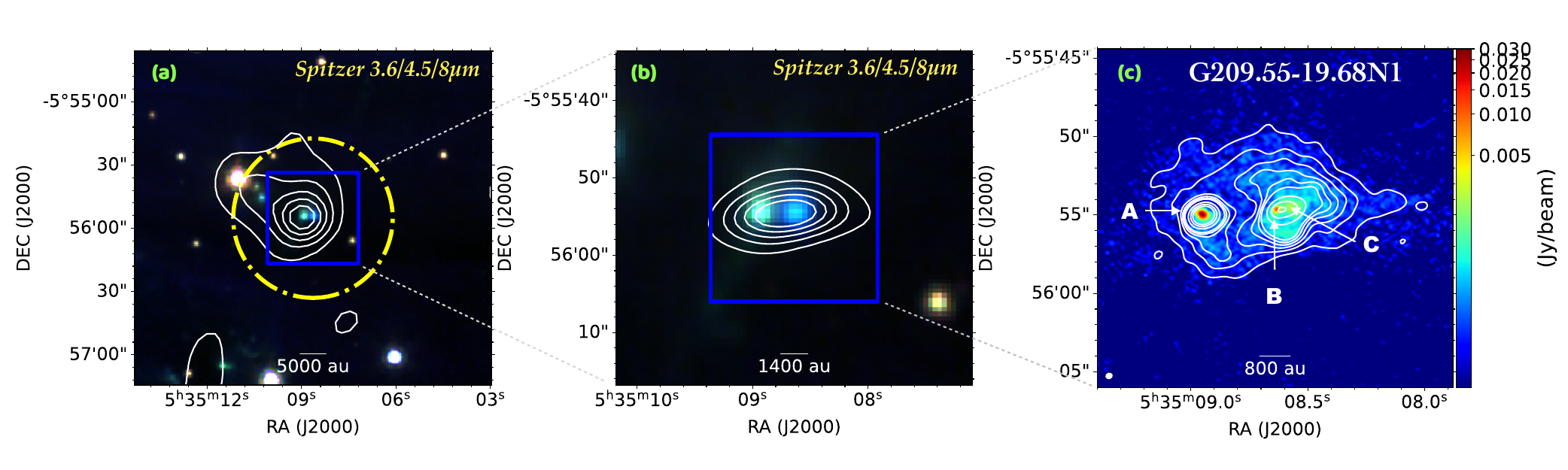}}
  \vspace{5pt}
  \centerline{\includegraphics[width=1.085\linewidth]{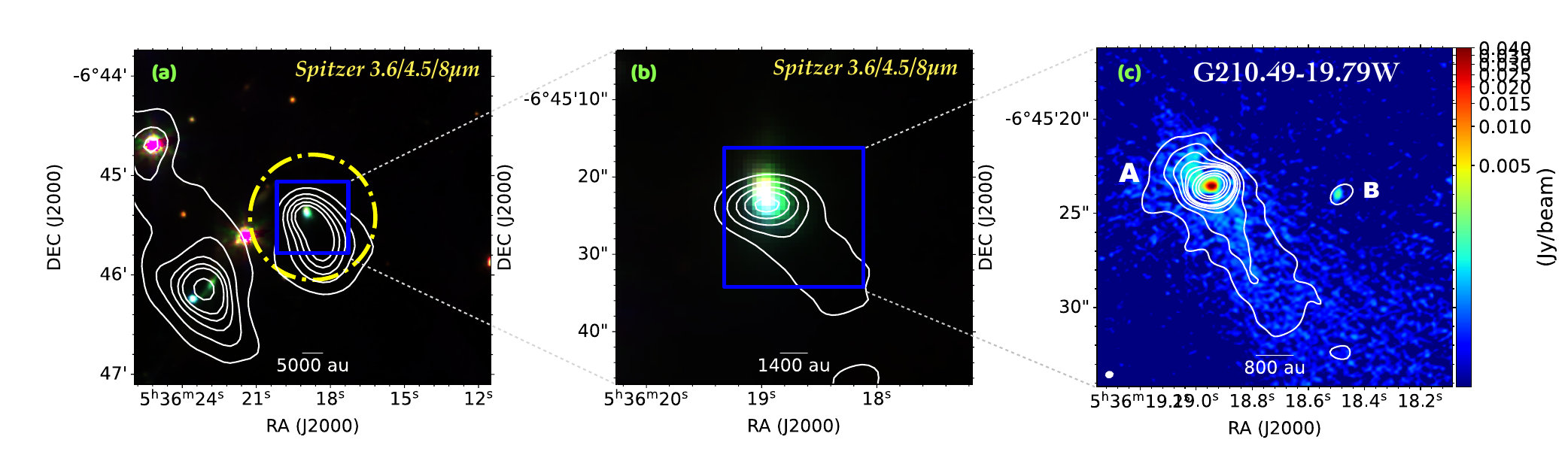}}
  \vspace{5pt}
  \centerline{\includegraphics[width=1.085\linewidth]{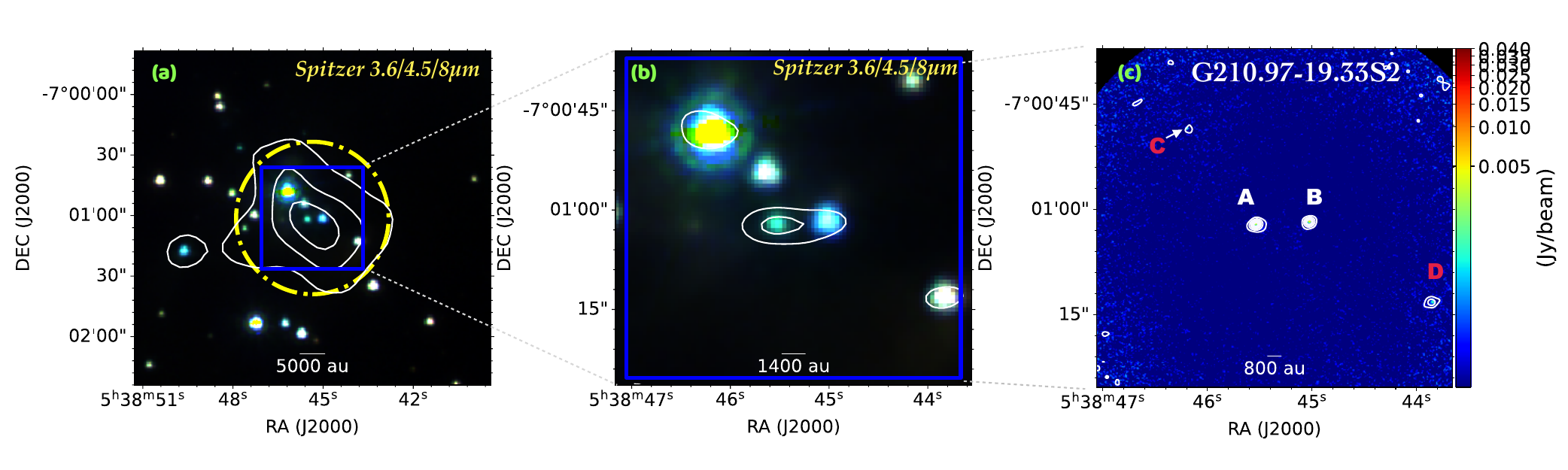}}
  \vspace{5pt}
  \centerline{\includegraphics[width=1.085\linewidth]{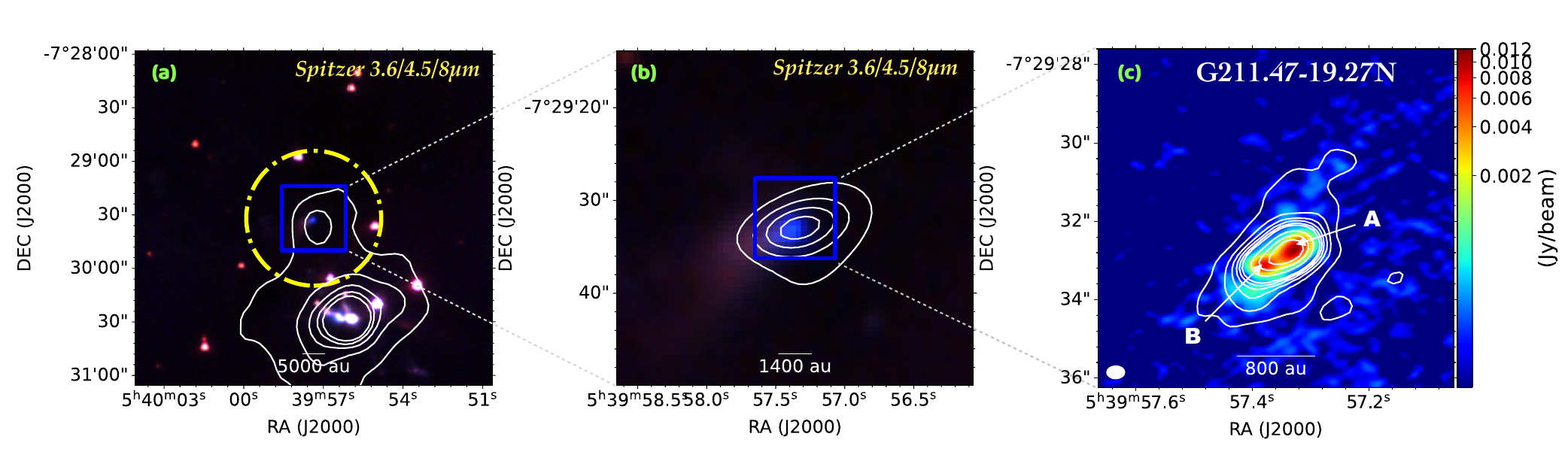}}
  \end{minipage}
  \caption{Continued}
\label{figA1}
\end{figure*}

\setcounter{figure}{7}

\begin{figure*}
\begin{minipage}[t]{0.9\linewidth}
  \vspace{5pt}
  \centerline{\includegraphics[width=1.085\linewidth]{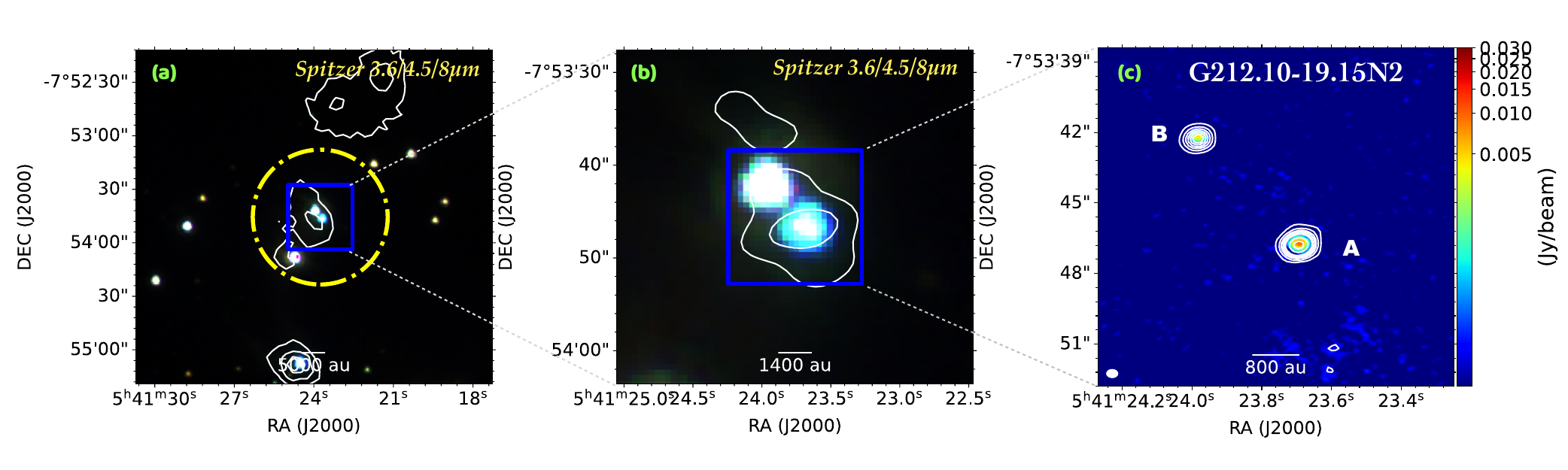}}
  \end{minipage}
  \caption{Continued}
\label{figA1}
\end{figure*}

\begin{figure*}[htbp]
\centering
{
    \begin{minipage}[b]{0.3\linewidth}
         \centering
         \includegraphics[width=1\linewidth]{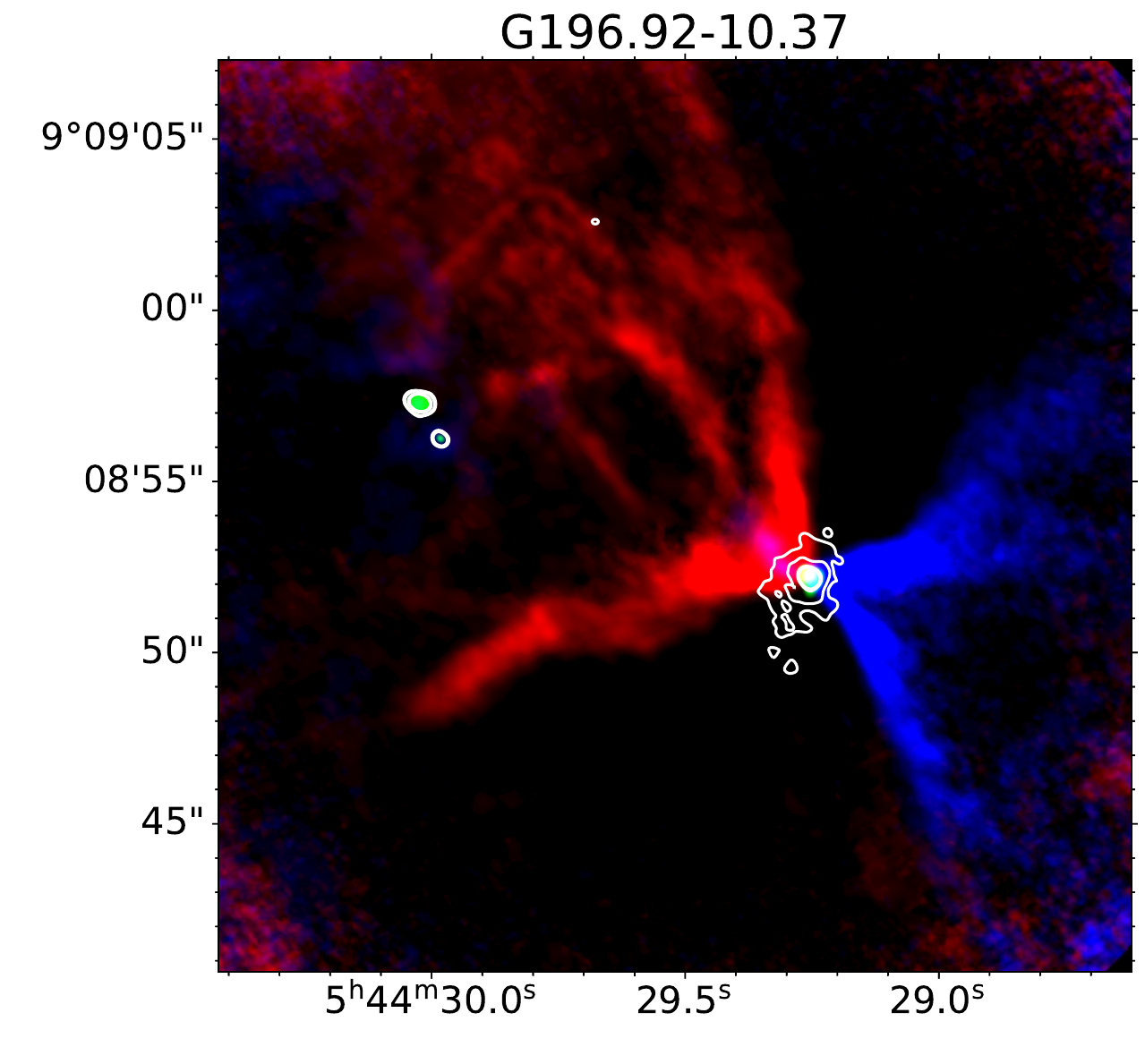}
     
    \end{minipage}
}
{
    \begin{minipage}[b]{0.3\linewidth}
         \centering
         \includegraphics[width=1\linewidth]{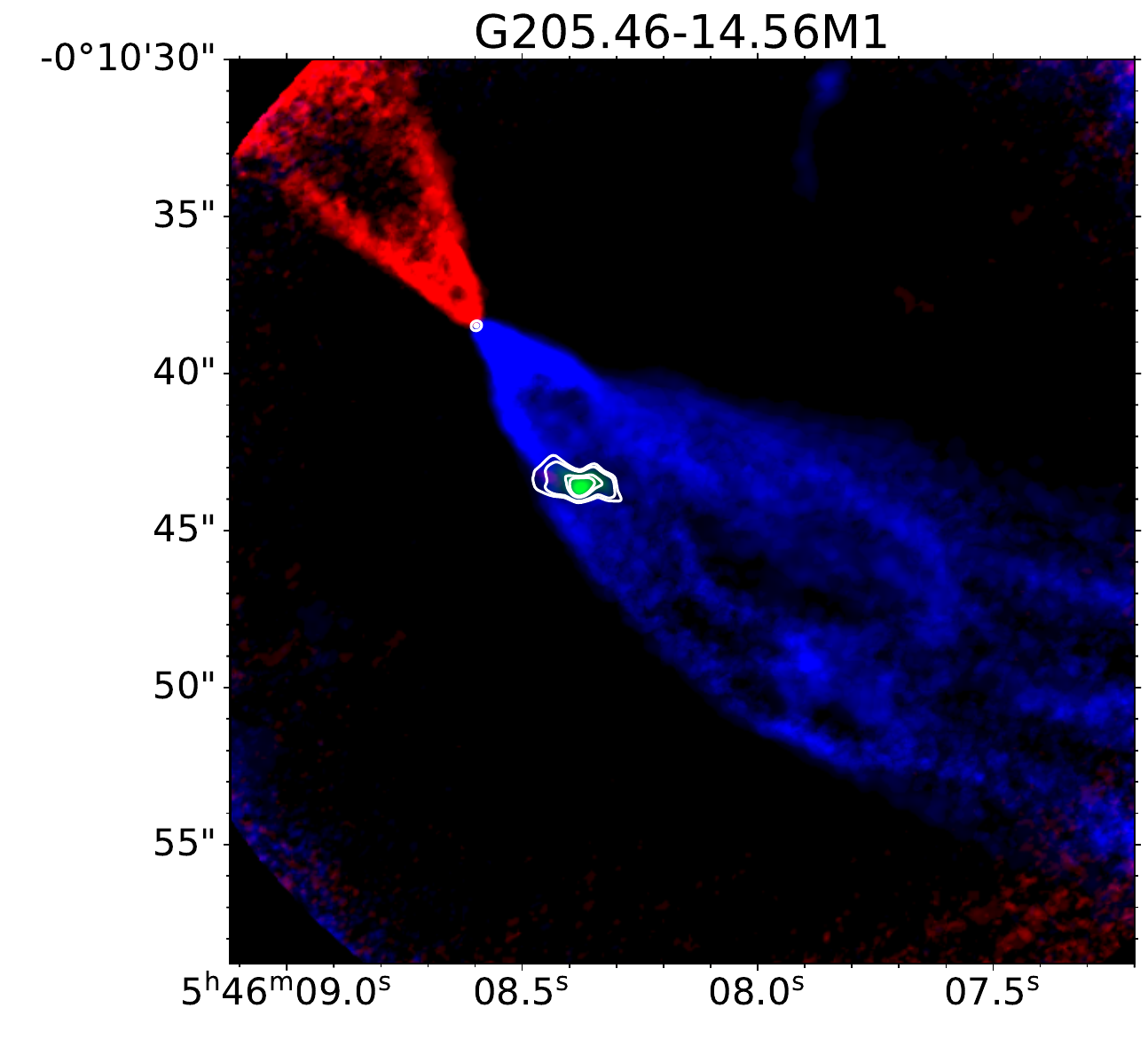}
     
    \end{minipage}
}
{
    \begin{minipage}[b]{0.3\linewidth}
         \centering
         \includegraphics[width=1\linewidth]{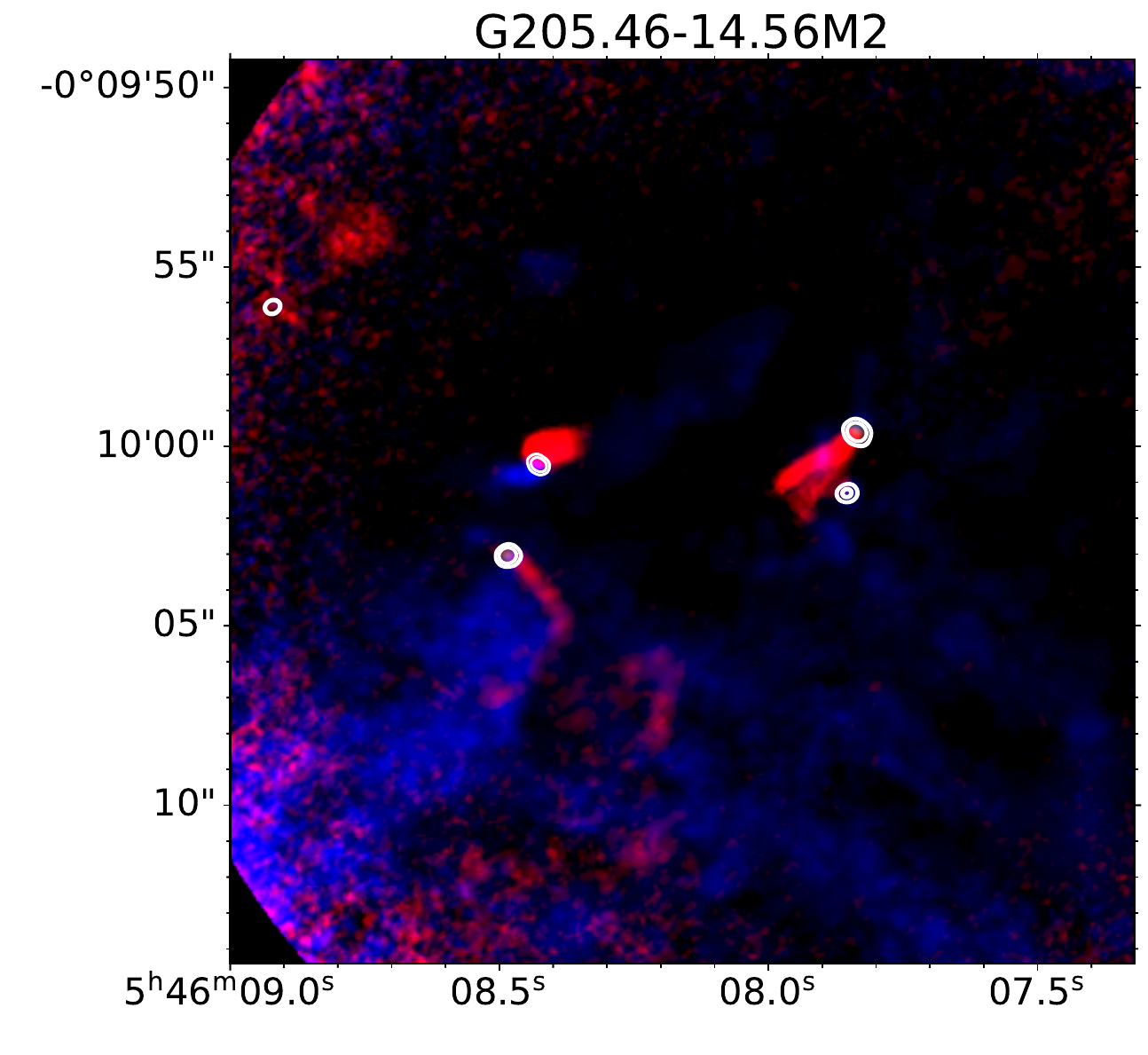}
     
    \end{minipage}
}
{
    \begin{minipage}[b]{0.3\linewidth}
         \centering
         \includegraphics[width=1\linewidth]{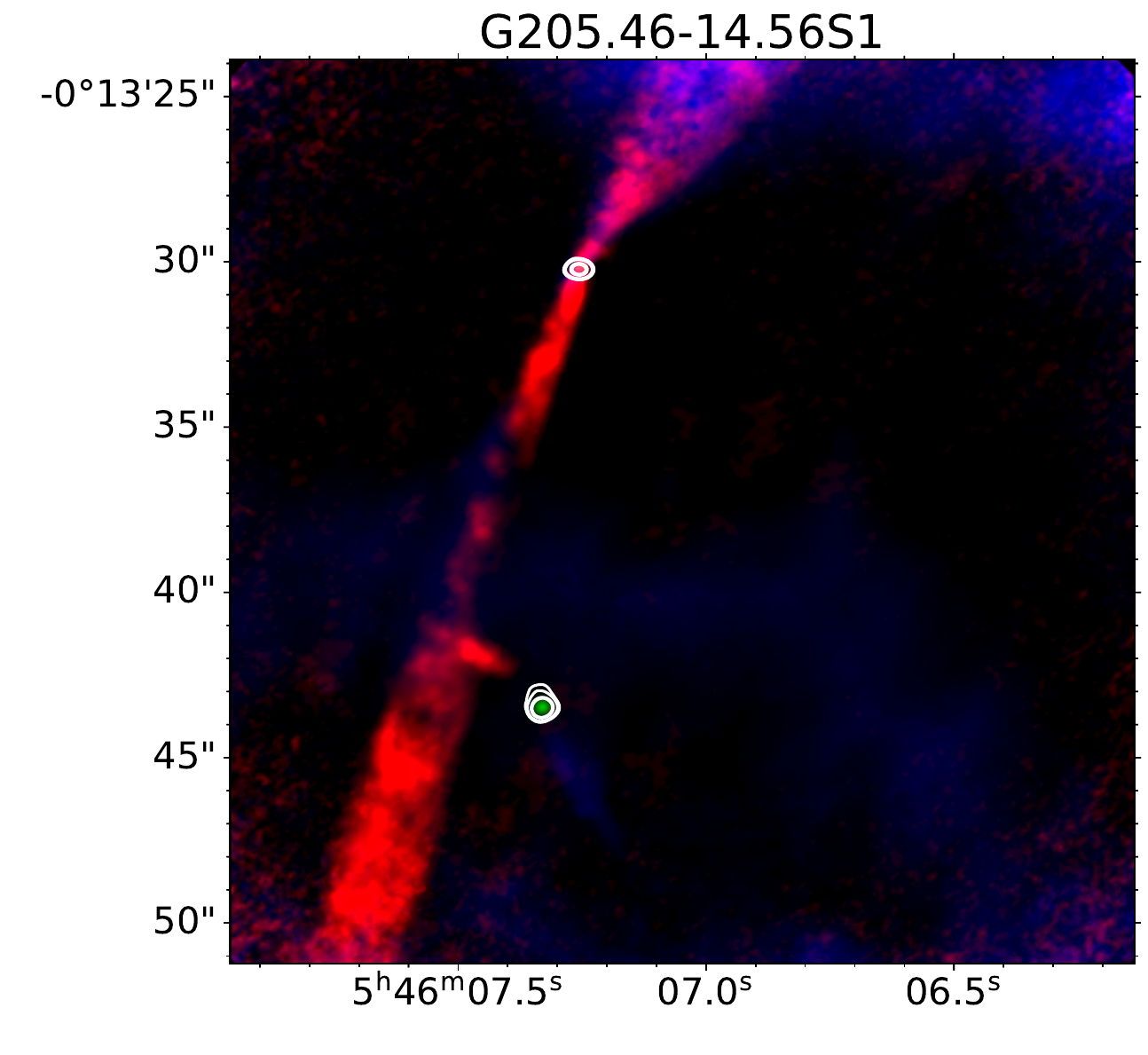}
     
    \end{minipage}
}
{
    \begin{minipage}[b]{0.3\linewidth}
         \centering
         \includegraphics[width=1\linewidth]{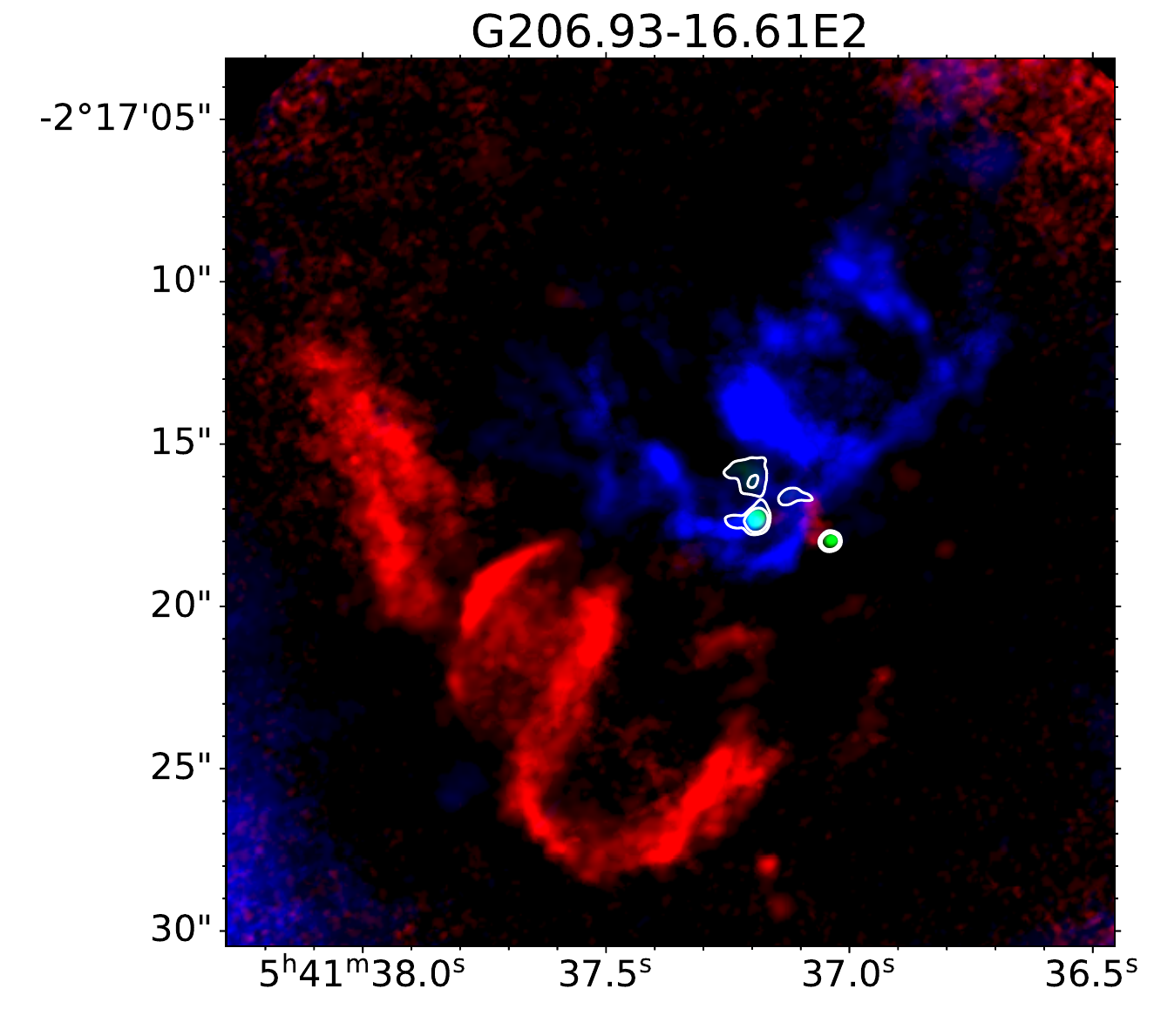}
     
    \end{minipage}
}
{
    \begin{minipage}[b]{0.3\linewidth}
         \centering
         \includegraphics[width=1\linewidth]{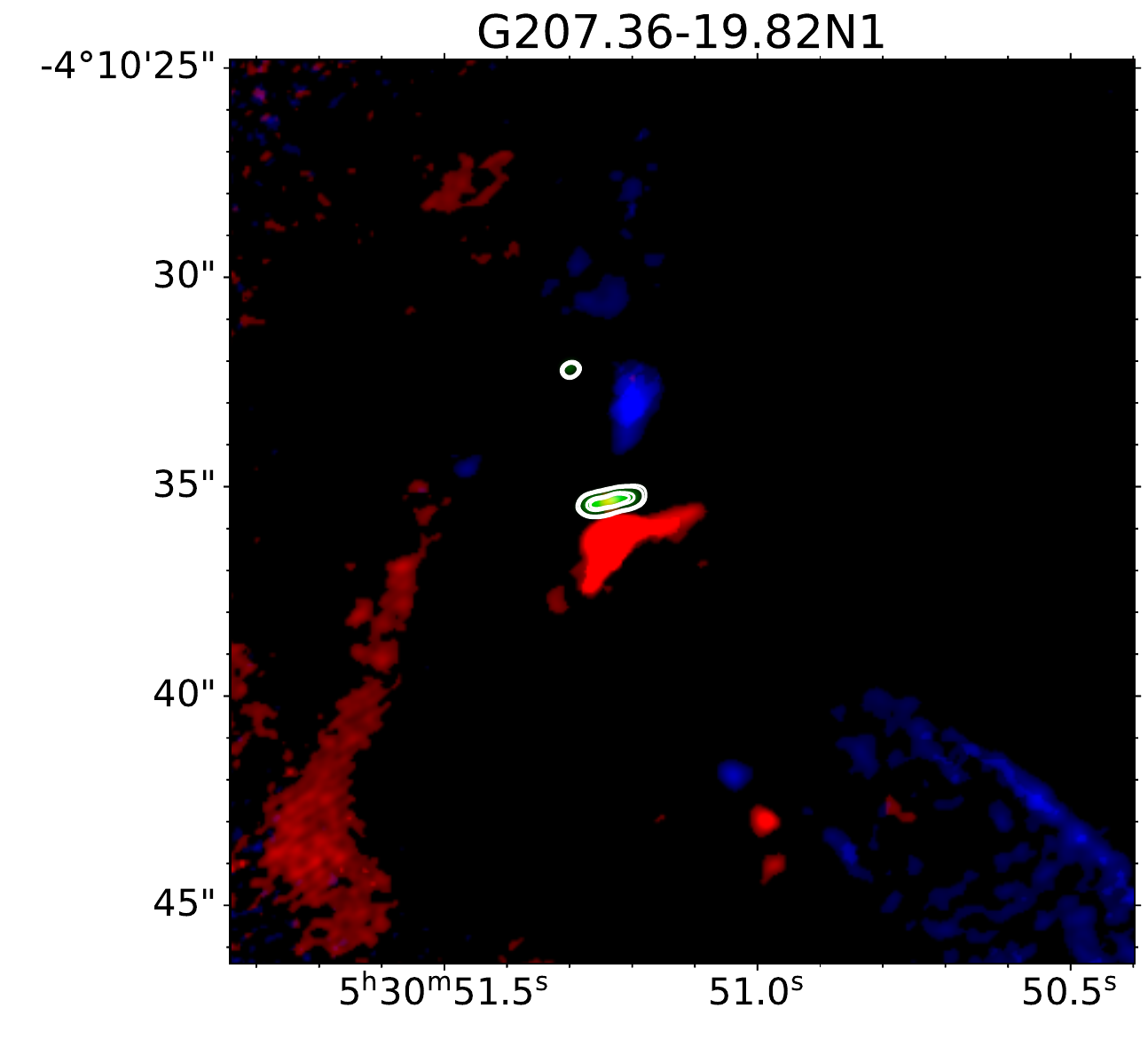}
     
    \end{minipage}
}
{
    \begin{minipage}[b]{0.3\linewidth}
         \centering
         \includegraphics[width=1\linewidth]{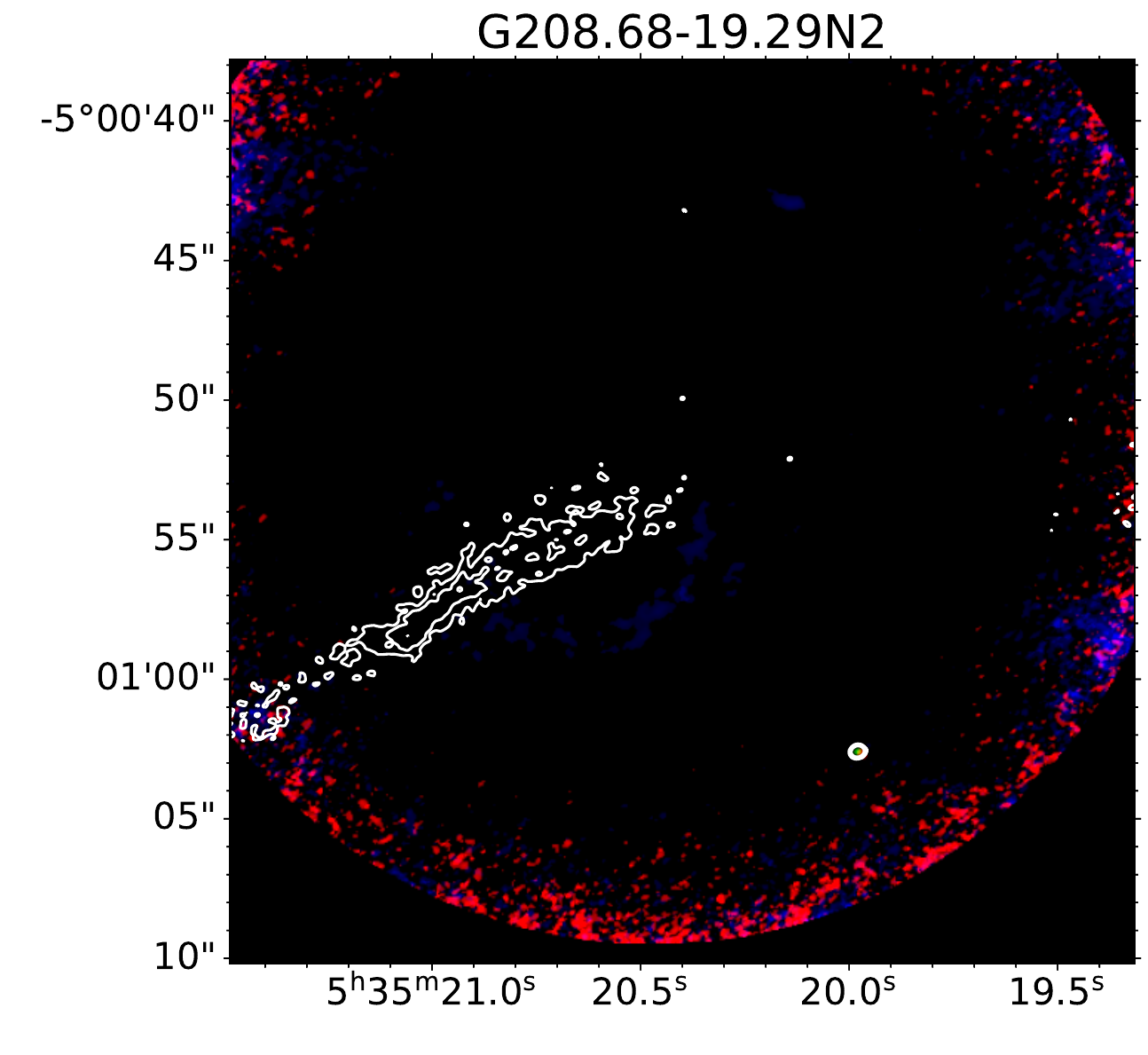}
     
    \end{minipage}
}
{
    \begin{minipage}[b]{0.3\linewidth}
         \centering
         \includegraphics[width=1\linewidth]{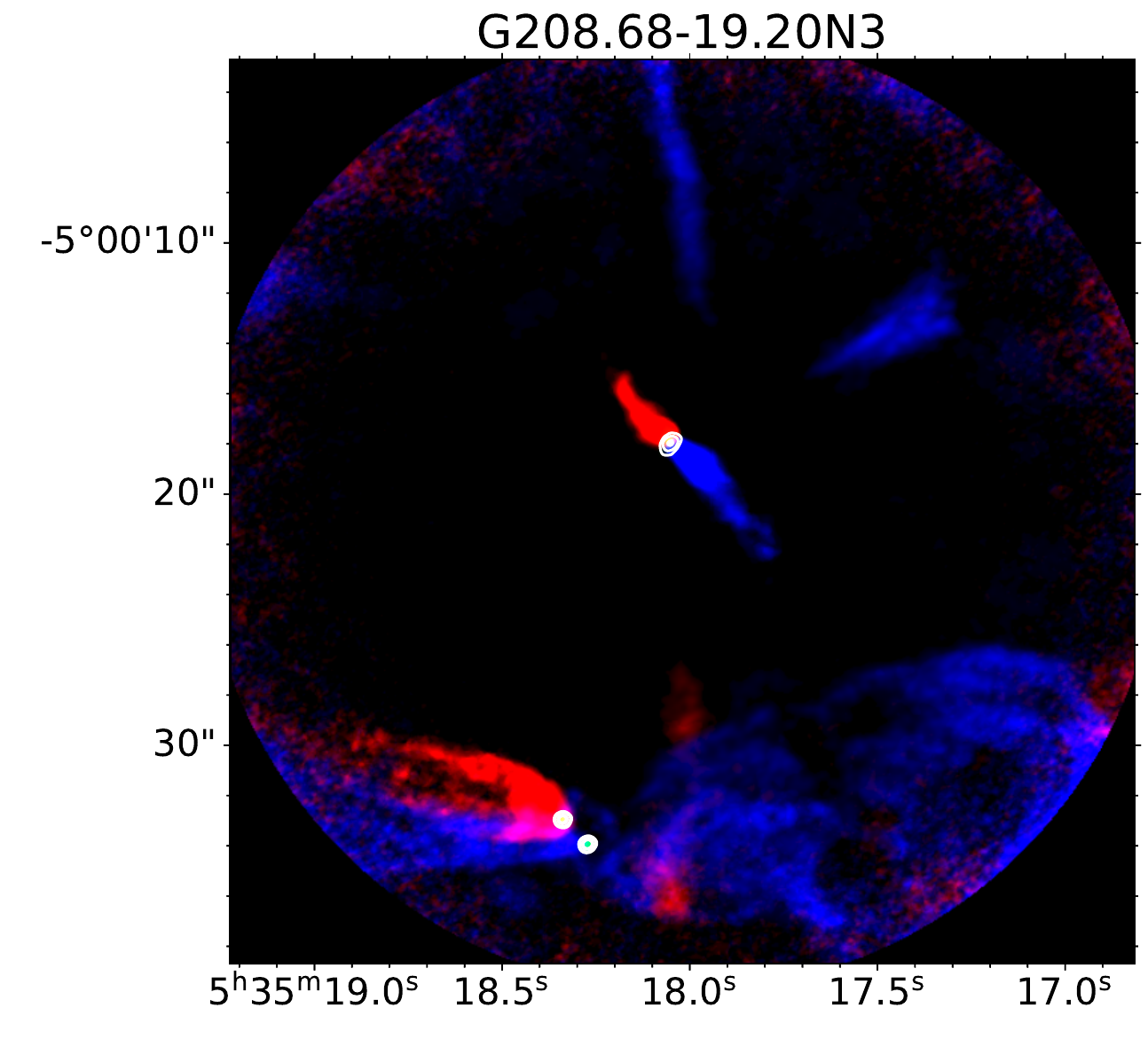}
     
    \end{minipage}
}
{
    \begin{minipage}[b]{0.3\linewidth}
         \centering
         \includegraphics[width=1\linewidth]{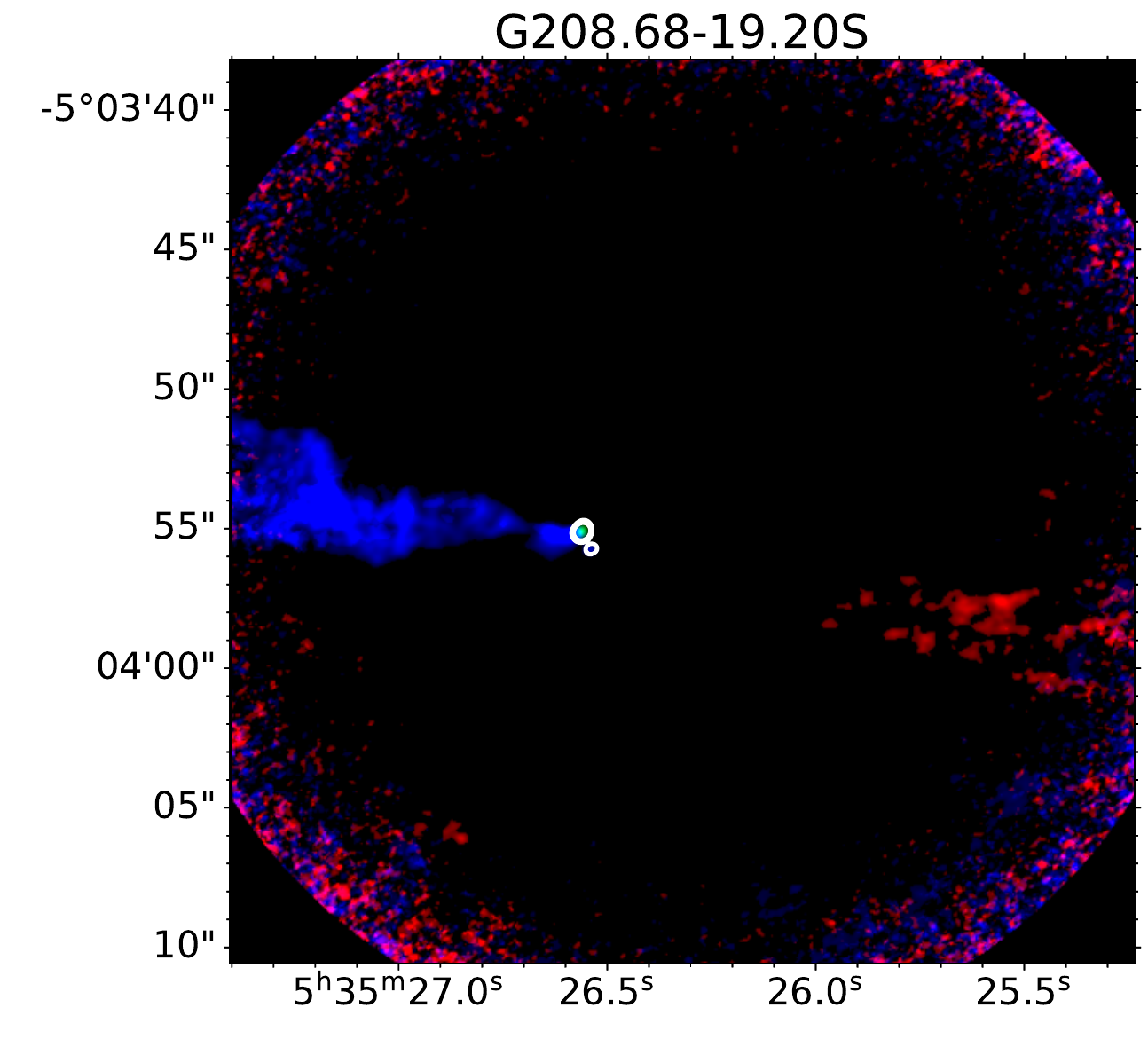}
     
    \end{minipage}
}
  \caption{CO outflows}
\label{figA2}
\end{figure*}

\setcounter{figure}{8}

\begin{figure*}[htbp]
\centering
{
    \begin{minipage}[b]{0.3\linewidth}
         \centering
         \includegraphics[width=1\linewidth]{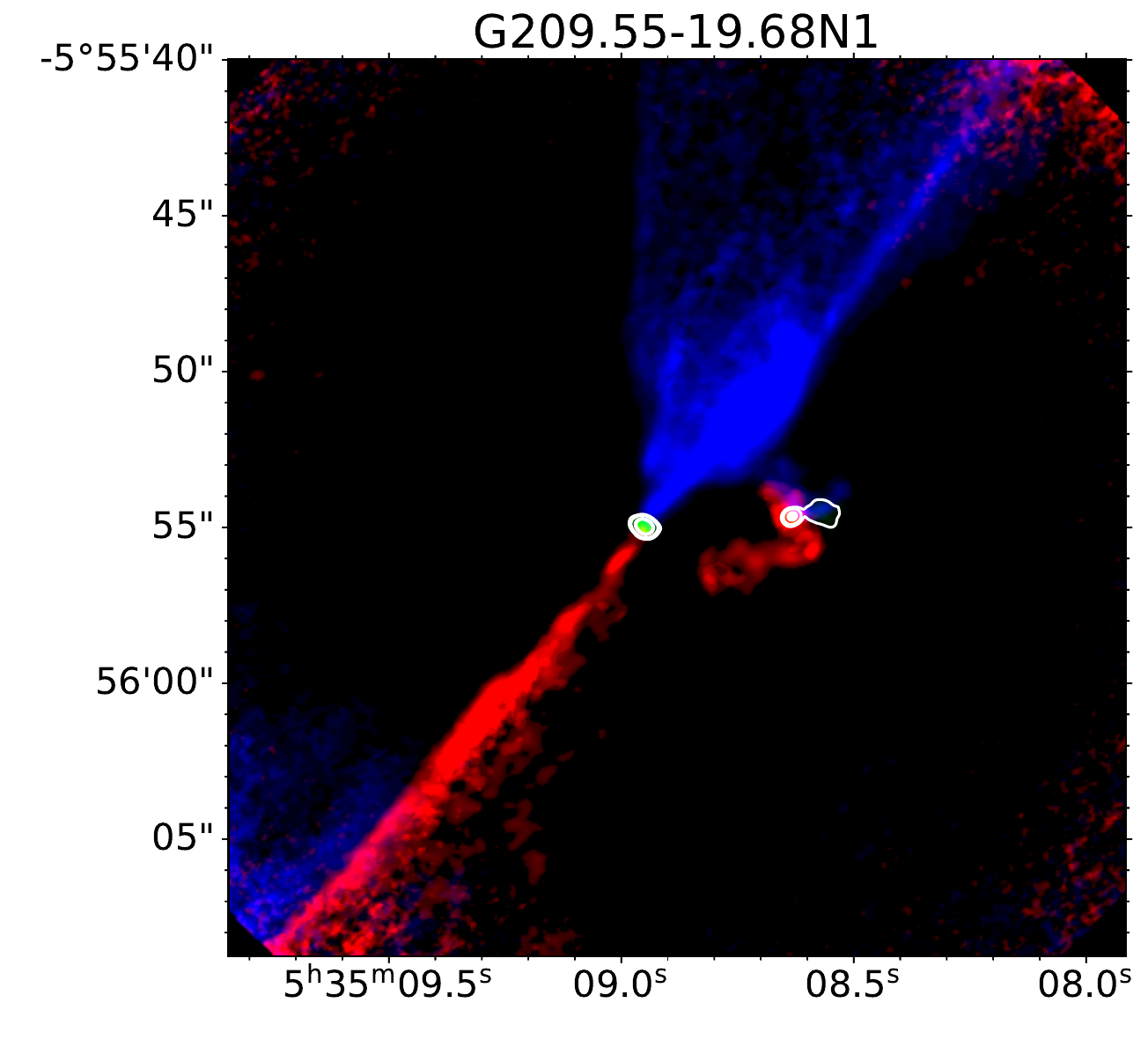}
     
    \end{minipage}
}
{
    \begin{minipage}[b]{0.3\linewidth}
         \centering
         \includegraphics[width=1\linewidth]{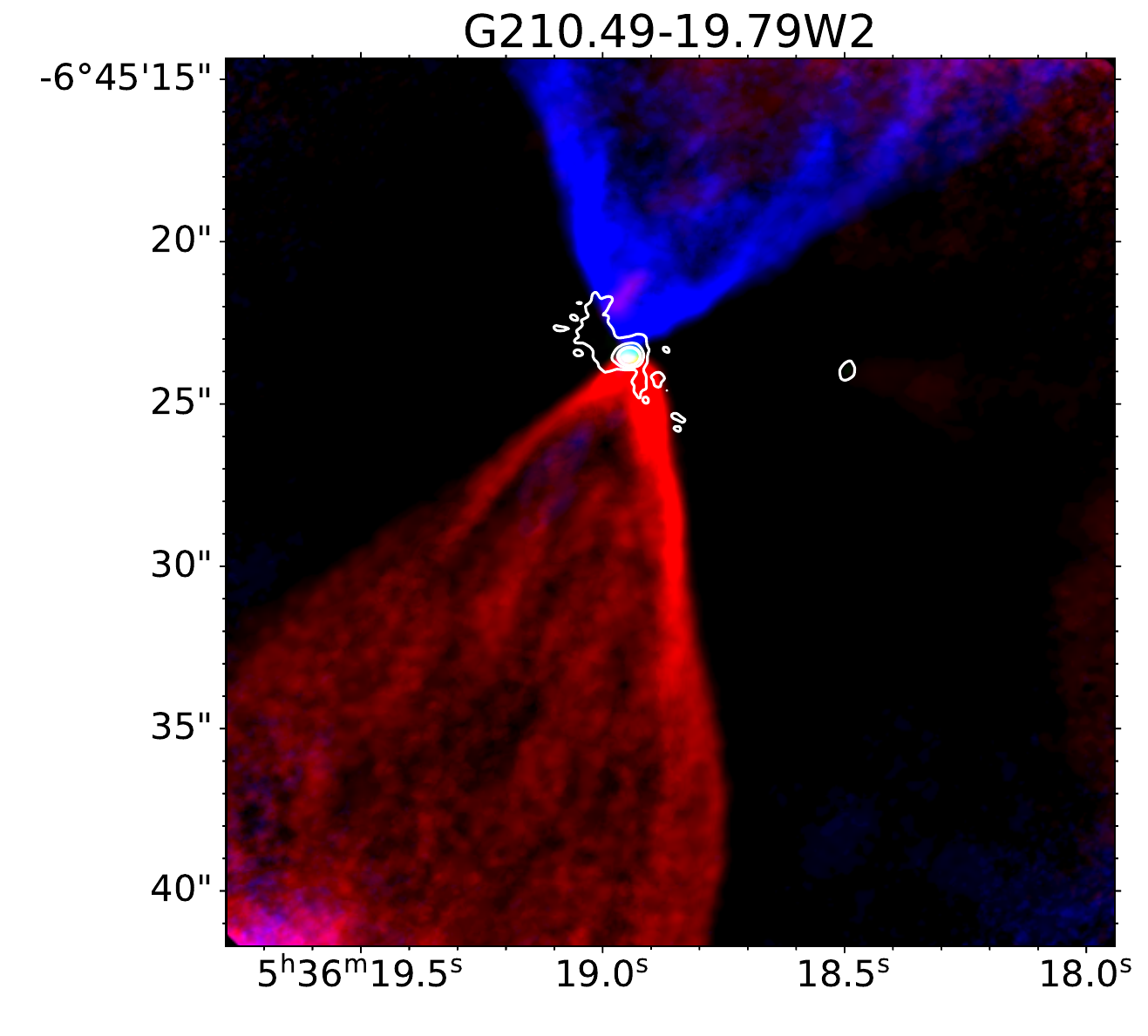}
     
    \end{minipage}
}
{
    \begin{minipage}[b]{0.3\linewidth}
         \centering
         \includegraphics[width=1\linewidth]{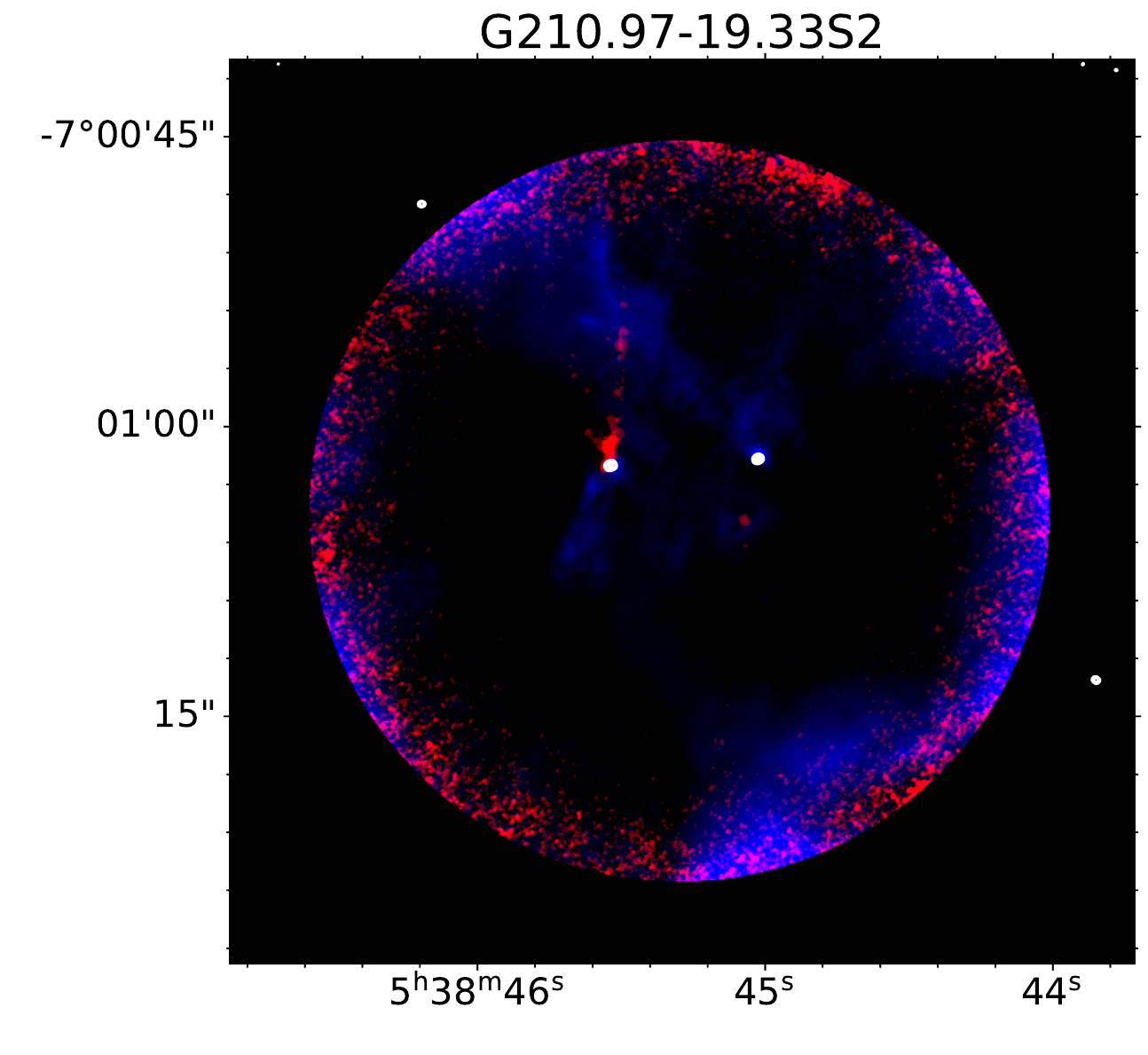}
     
    \end{minipage}
}
{
    \begin{minipage}[b]{0.3\linewidth}
         \centering
         \includegraphics[width=1\linewidth]{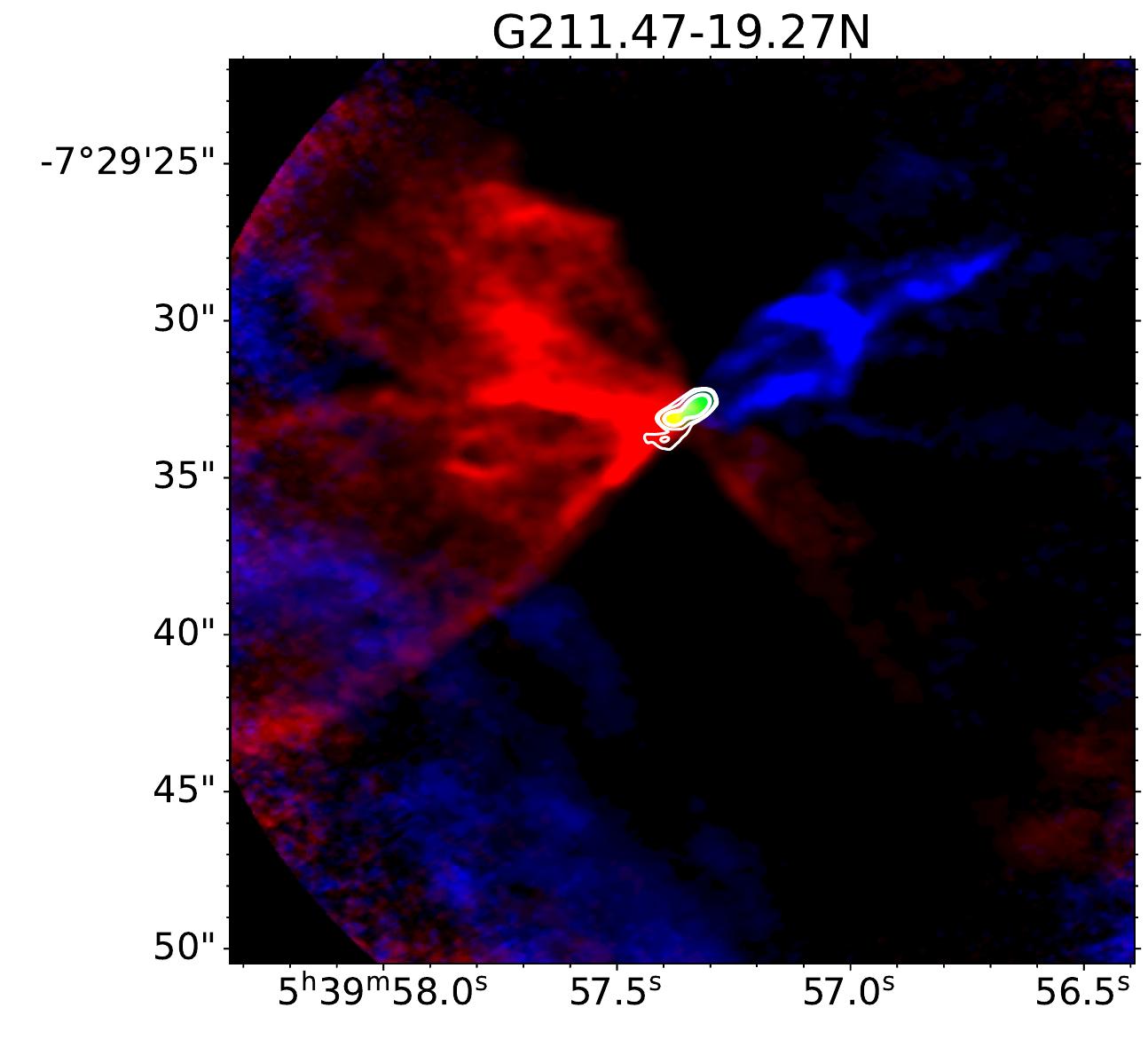}
     
    \end{minipage}
}
{
    \begin{minipage}[b]{0.3\linewidth}
         \centering
         \includegraphics[width=1\linewidth]{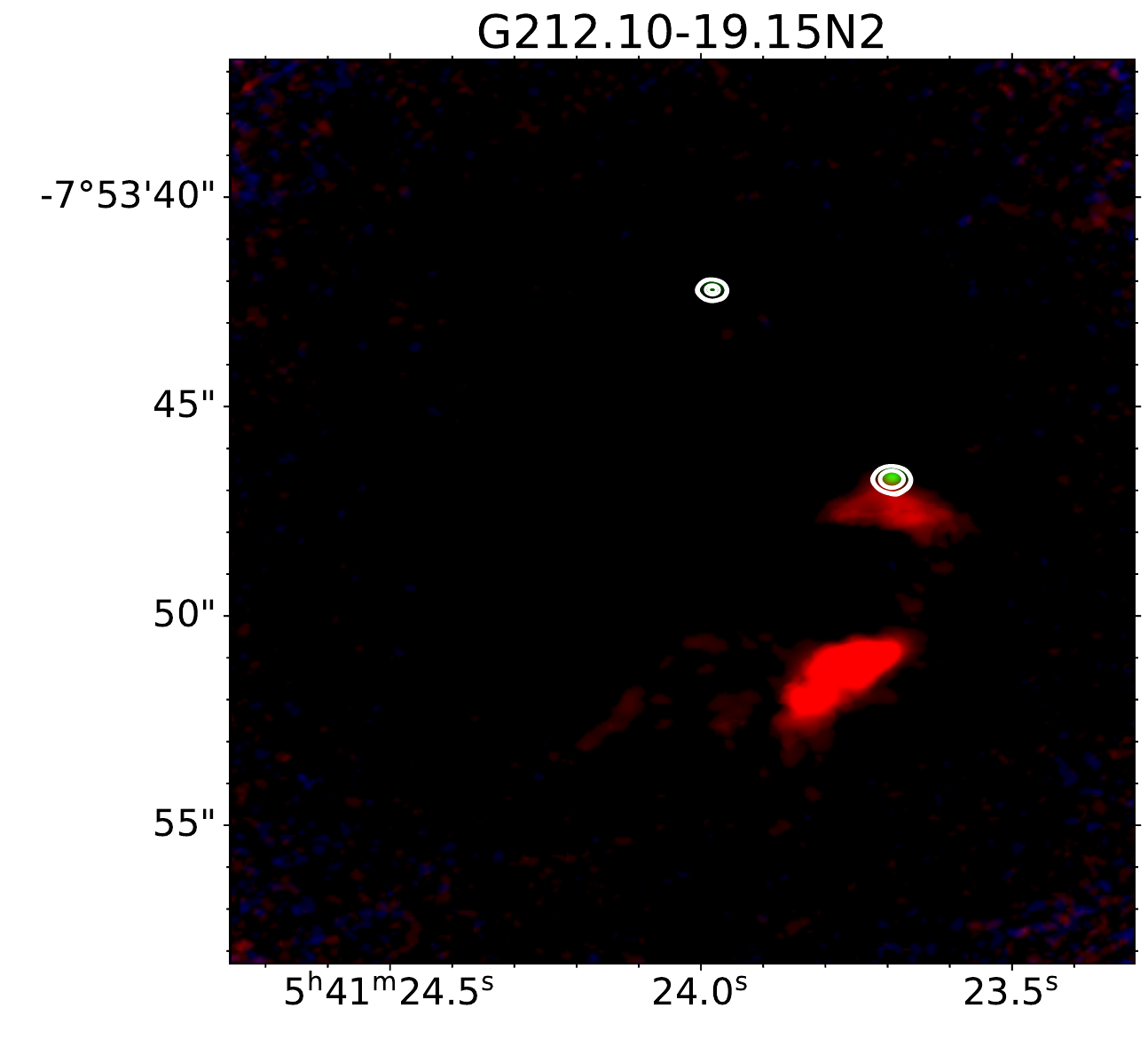}
     
    \end{minipage}
}
  \caption{Continued}
\label{figA2}
\end{figure*}

\clearpage
\bibliography{Main}{}
\bibliographystyle{aasjournal}

\end{CJK*}
\end{document}